\documentclass[
superscriptaddress,
nofootinbib,
amsmath,amssymb,
aps,
prd,
]{revtex4-2}

\usepackage{graphicx}
\usepackage{dcolumn}
\usepackage{bm}
\usepackage{gensymb}
\usepackage{amsmath,siunitx}
\usepackage{hyperref}
\usepackage{booktabs, multirow}
\usepackage{subcaption}
\usepackage{appendix}
\usepackage{csquotes}
\usepackage{booktabs}
\usepackage[justification=raggedright]{caption} 
\usepackage{color}
\usepackage{soul}
\usepackage{afterpage}
\begin{document}
	
	
\title{Orbital Dynamics and Gravitational Wave Signatures of Extreme Mass Ratio Inspirals in Galactic Dark Matter Halos}

\author{Guo-He Li}
\email{liguohe@stu.scu.edu.cn}
\affiliation{College of Physical Science and New Energy, Chongqing University of Technology, Banan, Chongqing 400054, China}
\affiliation{College of Physics, Sichuan University, Chengdu, 610065, China}

\author{Chen-Kai Qiao}
\email{chenkaiqiao@cqut.edu.cn}
\affiliation{College of Physical Science and New Energy, Chongqing University of Technology, Banan, Chongqing 400054, China}
	
\author{Jun Tao}
\email{taojun@scu.edu.cn}
\affiliation{College of Physics, Sichuan University, Chengdu, 610065, China}

	
		
\begin{abstract}
In astrophysics, extreme mass ratio inspiral (EMRI) systems, which consist of a central supermassive black hole and a stellar-mass compact object (SCO), are typically embedded in galactic dark matter (DM) halos. This dark matter environment inevitably affects the orbital dynamics of the SCO and the gravitational wave (GW) signals emitted by the system. In this work, we select two typical dark matter halo profiles---the Navarro-Frenk-White (NFW) and Beta models---to systematically investigate their specific impacts on the long-term orbital evolution of the SCO. By incorporating three dissipative mechanisms---dynamical friction, accretion, and gravitational radiation reaction---our results demonstrate that, compared to a pure vacuum medium, the presence of a dark matter halo significantly alters the trajectories of precessing orbits, the dynamical evolution of orbital parameters, and the waveforms and phases of the emitted gravitational waves. Due to the strong accretion effect within the NFW model, the energy flux exhibits a distinctive ``cusp'' feature, marking a reversal of orbital energy from loss to gain at a specific semi-latus rectum, which is a phenomenon absent in the Beta model. Although short-term observations may not be sufficient to distinguish between the NFW and Beta models, their differences become evident over long-term orbital evolution. The gravitational waveforms computed using the NFW and Beta models exhibit a phase shift, which could be detectable in high-density DM environments. This phase shift becomes even more pronounced for higher eccentric orbits and longer observation times. These results offer a theoretical framework for probing environmental effects on EMRIs across different dark matter models using future space-based gravitational wave observatories.
\end{abstract}
\maketitle
	
\tableofcontents
	
\section{Introduction}

The continued success of ground-based gravitational wave (GW) observatories has established the era of GW astronomy~\cite{LIGOScientific:2016aoc,LIGOScientific:2016emj,LIGOScientific:2018mvr,LIGOScientific:2020ibl,LIGOScientific:2021usb,KAGRA:2021vkt}. To extend the observational window into the millihertz frequency band~\cite{Baibhav:2019rsa,LISA:2022yao,LISA:2022kgy,Karnesis:2022vdp}, space-based gravitational wave missions—represented by the Laser Interferometer Space Antenna (LISA)~\cite{LISA:2017pwj,LISA:2022yao}, Taiji~\cite{Hu:2017mde,Gong:2021gvw}, and TianQin~\cite{TianQin:2015yph,Liu:2020eko}—are currently under active development. As one of the most promising sources for these space-based detectors, extreme mass ratio inspirals (EMRIs) are typically characterized by a stellar-mass compact object (with a mass $m$ ranging from $1M_{\odot}$ to $ 100 M_{\odot}$ approximately), orbiting a supermassive black hole (SMBH) with a mass $M$  spanning from $10^5M_{\odot}$ to $10^8 M_{\odot}$, which generally resides in galactic nuclei~\cite{Babak:2017tow}. Given this extreme mass ratio, 
the stellar-mass compact object can effectively behave as a test particle, tracing out approximately $10^4 \sim 10^5$ orbital cycles within the detector's sensitive frequency band. Crucially, the high precision observations on EMRIs encode rich information, enabling us to decode the astrophysical environments (such as accretion disks, dark matter, and electromagnetic fields)  near the SMBH from GW signatures~\cite{Amaro-Seoane:2007osp,Berry:2019wgg,Seoane:2021kkk,Laghi:2021pqk,McGee:2018qwb,Barausse:2014tra,Cardoso:2022whc,Zi:2025lio}. By analyzing the GW signals, it is possible to directly constrain the influences of these environments. Additionally, EMRIs present exceptional value in probing the quantum gravity effects beyond general relativity~\cite{Yang:2024lmj,Zi:2024jla,Kumar:2025jsi,Battista:2021rlh}.

A vast number of astrophysical observations, including the cosmic microwave background~\cite{Planck:2013pxb}, gravitational lensing and galaxy rotation curves~\cite{Rubin:1970zza,Corbelli:1999af}, consistently indicate that the universe is dominated by dark matter and dark energy. In galactic nuclei, dark matter is expected to form halo structures around the central SMBH, implying that these SMBHs are embedded in dense matter environments~\cite{Cooray:2002dia,Wang:2019ftp}. Extensive numerical simulations and observational studies have motivated various theoretical models to describe these galactic dark matter halos, including the Navarro-Frenk-White (NFW)~\cite{Navarro:1994hi,Navarro:1995iw,Navarro:1996gj}, Beta~\cite{Cavaliere:1976tx}, Moore~\cite{Moore:1999gc}, Einasto~\cite{einasto1965construction,Dutton:2014xda}, Dehnen~\cite{Dehnen:1993uh}, Hernquist~\cite{Hernquist:1990be} and Burkert~\cite{Salucci:2000ps} models. The presence of dark matter halos inevitably exerts profound impacts on gravitational systems in galaxies, such as binary black hole (BBH) coalescences, intermediate mass ratio inspirals (IMRIs) and extreme mass ratio inspirals (EMRIs)~\cite{Dai:2021olt,Cole:2022yzw,Kavanagh:2020cfn,Barsanti:2022ana,Figueiredo:2023gas,Zhou:2024vhk,Zhang:2024ugv,Shadykul:2024ehz,Kadota:2023wlm,Kim:2024rgf,Ding:2025hqf}. Particularly, the dark matter halo influences the orbital dynamics of EMRIs (or IMRIs) through two mechanisms: firstly, the gravitational potential contribute from the dark matter halo modifies the background spacetime geometry; secondly, interactions between the compact object and the surrounding medium give rise to dissipative effects, most notably dynamical friction~\cite{Chandrasekhar:1943ys,Cardoso:2020iji} and accretion~\cite{Bondi:1952ni,Macedo:2013qea,Bondi:1944rnk,Edgar:2004mk}. Over the long-duration inspiral evolution, these environmental effects continuously reduce the orbital energy and angular momentum, causing a cumulative ``dephasing" of the gravitational waveform relative to vacuum predictions. This distinctive phase shift provides a critical observational signature for constraining the dark matter halo models and revealing the matter distribution in galaxies~\cite{Rahman:2023sof,Zhao:2024bpp,Gliorio:2025cbh,Zhao:2026yis}.

The dark matter effects on EMRI systems and their GW signatures can be extracted through the investigations of precessing Keplerian orbits~\cite{Dai:2023cft,Li:2025qtb}, periodic orbits (with zoom-whirl-vertex behaviors)~\cite{Li:2025eln,Alloqulov:2025ucf,Haroon:2025rzx}, chaotic dynamics~\cite{Das:2025vja,Das:2025eiv}, and quasi-normal modes~\cite{Cardoso:2021wlq,Liu:2024xcd,Liu:2023vno,Zhao:2023itk}. Although the influence of dark matter environments on EMRI dynamics has been investigated, distinguishing between different halo density profiles through GW observations remains a significant theoretical challenge. In our previous study, the periodic orbits around an SMBH influenced by various dark matter halo models were investigated for fixed orbital energies or angular momenta (without considering any dissipation mechanisms)~\cite{Li:2025eln}. The periodic orbital shapes and GW waveforms predicted by NFW and Beta halo models exhibit nearly identical characteristics for certain dark matter parameters, making them observationally indistinguishable in such non-dissipative assumptions.  Consequently, analyses limited to conservative dynamics or short-term evolution are insufficient to identify the specific dark matter distribution from GW observations. To resolve this degeneracy, it is necessary to explore the secular orbital evolution of the EMRI system over a large number of orbital periods with dissipative mechanisms taken into account, i.e. gravitational radiation backreaction, dynamical friction. By monitoring the cumulative phase shift induced by gravitational radiation, dynamical friction, and accretion over a long-duration inspiral, it becomes possible to distinguish these halo models.

In this work, we aim to resolve the aforementioned NFW/Beta degeneracy by modeling the long-term adiabatic evolution of EMRIs affected by dark matter environments (described by NFW and Beta density profiles). Specifically, we focus on quantifying how gravitational radiation reaction, accretion, and dark
matter dynamical friction induce the variation of orbital energies, angular momenta, eccentricity, semi-latus rectum, and orbital phase shifts, enabling observational differentiation between these density profiles. We employ the adiabatic approximation~\cite{Hughes:1999bq,Hughes:2001jr,Drasco:2005is,Drasco:2005kz,Sundararajan:2008zm,Isoyama:2021jjd} to evaluate the long-term orbital evolution and use the Numerical Kludge method~\cite{Barack:2003fp,Gair:2005is,Babak:2006uv,Chua:2017ujo} to generate waveforms from these evolved trajectories and calculate the cumulative dephasing over a typical one-year observation timescale, from which a systematic comparison on phase evolution in NFW and Beta halo models can be carried out to magnify their distinguishability. The primary goal is to determine the conditions under which space-based detectors can resolve the specific density structure of the galactic dark matter halo.

This paper is organized as follows: Section~\ref{sec2} introduces the spacetime metric influenced by NFW and Beta dark matter halos. The equations of motion for Keplerian-like orbits are studied in this section. Section~\ref{sec3} outlines the long-term orbital evolution under dissipative forces arising from three mechanisms, including gravitational wave emission, dynamical friction, and accretion. Section~\ref{sec4} presents the numerical results on waveform evolution and analyzes the cumulative phase under different halo densities. Finally, Section~\ref{sec5} summarizes our conclusions and discusses their implications for future GW observations.

\section{Spacetime Metrics and Orbital Motions}
\label{sec2}

In this section, we present the spacetime metrics for black holes immersed in dark matter halos (modeled NFW and Beta profiles). Subsequently, we give an exploration of the Keplerian-like orbits of SCO, which are of great significance in the orbital dynamics and gravitational wave emissions in EMRIs. We study the DM environmental influences by analyzing the observables of these Keplerian-like orbits (e.g., time period, precession angle, and orbital trajectories).
	
\subsection{Spacetime Metric}\label{a1}
 In this work, we consider a supermassive black hole embedded in a spherically symmetric dark matter halo. This type of spherical dark matter halo model is widely used in current galactic dark matter research. Notable examples include the NFW, Beta, Moore, Dehnen, Hernquist, Burkert, and Einasto models, all of which feature spherically symmetric density profiles. Specifically, this study focuses on the NFW and Beta models. The spacetime is described by a static, spherically symmetric metric, written as
\begin{equation}
	ds^{2} = -f(r)dt^{2} + f(r)^{-1}dr^{2} + r^{2}d\Omega^{2}.
\end{equation}
Considering the dark matter halo as an extended mass distribution surrounding the black hole, the total gravitational field can be treated as a combination of the SMBH's gravitational field and the dark matter halo's field. Consequently, the total metric function $f(r)$ is decomposed into the contribution from the central black hole and a correction term $f_{\text{DM}}(r)$ arising from the halo~\cite{Xu:2018wow}\footnote{It can be verified that the spacetime metric constructed via the decomposition in Eqs.~\eqref{eq:f(r)1} and~\eqref{f_DM} satisfies the Einstein field equations, with the $T_{00}$ component of the energy-momentum tensor specified by the dark matter density profile~\cite{Xu:2018wow,Pantig:2022whj,Xu:2020jpv}}:
\begin{equation}
	f(r) = f_{\text{DM}}(r) - \frac{2M}{r}. \label{eq:f(r)1}
\end{equation}
Here, $M$ represents the mass of the central black hole. The halo's contribution $f_{DM}(r)$ is determined by the halo's internal mass distribution, $M_{DM}$. By examining the geodesic equations and considering the Newtonian limit, the tangential velocity $v_{tg}$ of a test particle relates to the enclosed mass via $v_{tg}^2 \approx M_{DM}(r)/r$. Specifically, for a test body moving around the central black hole in a spherically symmetric gravitational field, the tangential velocity can be calculated via the metric function~\cite{Matos:2000ki}:
\begin{equation}
	v_{tg}{}^2(r)=\frac{r}{\sqrt{f(r)}}\cdot\frac{d\sqrt{f(r)}}{dr}=\frac{r(\mathrm{dln}\sqrt{f(r)})}{dr}.
\end{equation}
Solving the ordinary differential equation, one can derive the specific form of the metric correction:
\begin{equation}
	f_{\text{DM}}(r) = \exp\left[ 2 \int \frac{M_{DM}(r)}{r^2} dr \right]. \label{f_DM}
\end{equation}
	
To obtain the explicit metric forms, specific dark matter density profiles $\rho(r)$ are required. We adopt two widely used profiles: the Navarro-Frenk-White (NFW) model  and the Beta model. The density distribution functions for these models are defined as~\cite{Navarro:1994hi,Navarro:1995iw,Navarro:1996gj,Cavaliere:1976tx}:
\begin{subequations}
\begin{align}
	\rho_{\text{NFW}}(r) &= \frac{\rho_{0}}{(r/h)(1+r/h)^{2}} \label{rho_nfw} \\
	\rho_{\text{Beta}}(r) &= \frac{\rho_{0}}{[1+(r/h)^{2}]^{3/2}} \label{rho_beta}
\end{align}
\end{subequations}
where $\rho_0$ denotes the characteristic central density and $h$ is the characteristic radius. For a spherically symmetric distribution, the cumulative dark matter mass $M_{DM}(r)$ enclosed within radius $r$ is calculated by:
\begin{equation}
	M_{\text{DM}}(r) = 4\pi \int_0^r \rho(r')r'^2 dr'. \label{M_DM}
\end{equation}
Substituting Eqs.~(\ref{rho_nfw}) and (\ref{rho_beta}) into Eq.~(\ref{M_DM}) to obtain the respective cumulative masses, and subsequently integrating Eq.~(\ref{f_DM}), we derive the effective metric functions for entire gravitational system~\cite{Qiao:2024ehj,Li:2025eln,Liu:2023xtb},
\begin{subequations}
\begin{align}
	f_{\text{NFW}}(r) &= \left( 1 + \frac{r}{h} \right)^{-\frac{8\pi k}{r}} - \frac{2M}{r}, \label{f_NFW1}\\
	f_{\text{Beta}}(r) &= \exp\left[ -\frac{8\pi k}{r} \sinh^{-1}\left(\frac{r}{h}\right) \right] - \frac{2M}{r},\label{f_Beta1}
\end{align}
\end{subequations}
 where $k=\rho_{0} \cdot h^3$ is related to the dark matter mass. Eqs.~(\ref{f_NFW1}) and (\ref{f_Beta1}) provide a complete analytical description of the spacetime geometry associated with the respective dark matter distributions.

To investigate the environmental effects on EMRIs, we select parameters based on astrophysical estimates. In fact, in most galaxies, the dark matter mass parameter, the scale of the halo, and the mass of the central SMBH typically follow a hierarchical relation (in geometric units): $ M \ll k \ll h $. Guided by these values and this hierarchy, we fix the characteristic radius at $h=10^7 M$ and select a set of mass parameters $k \in \{1000M, 5000M, 10000M, 20000M\}$~\cite{Fukushige:2003xc,Xu:2018wow,Ishihara:2016vdc,Takizawa:2020egm,Li:2020wvn,Ovgun:2018tua,Jusufi:2018jof}. 

\subsection{Keplerian-like orbital motion}
\label{Eccentric motion}
	
\begin{figure}
	\centering
	\includegraphics[width=0.825\textwidth]{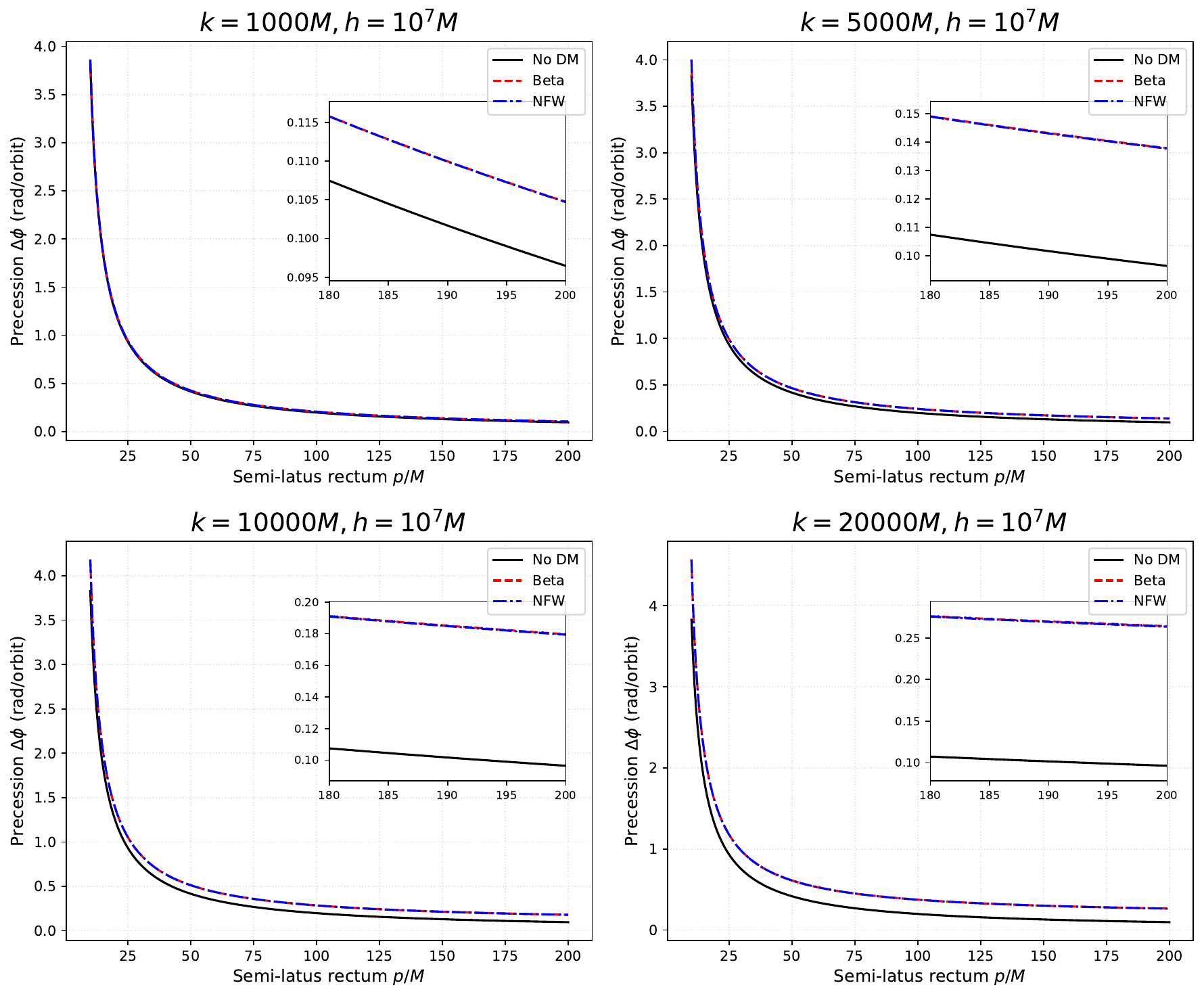}
	\caption{Comparison of the orbital precession $\Delta\phi$ (in rad/orbit) versus $p$ for $e=0.6$ and $p \in [10M, 200M]$. It is observed that the NFW and Beta profiles yield nearly identical results, rendering them indistinguishable even in the high-density scenario.}
	\label{fig:precession_comparison}
\end{figure}

In the framework of general relativity, the motion of a compact object  follows a geodesic in the curved spacetime. The geodesic equation is
\begin{equation}
	\frac{du^\alpha}{d\tau} + \Gamma^\alpha_{\mu\nu} u^\mu u^\nu = 0,
\end{equation}
where $u^\alpha = dx^\alpha/d\tau$ is the four-velocity and $\tau$ denotes the proper time.  Without loss of generality, we restrict the particle's trajectory to the equatorial plane ($\theta = \pi/2$, $\dot{\theta} = 0$) due to the spherical symmetry of the background geometry. For a static and spherically symmetric black hole, the motion admits two conserved quantities per unit mass:  
\begin{subequations}
\begin{align}
	u_t&=-E/\mu=-\varepsilon,\label{u_t}\\
	u_{\phi}&=L/\mu=l,\label{u_phi}
\end{align}
\end{subequations}
where $E$ and $L$ represent the orbital energy and angular momentum of the system, respectively, and $\mu$ is the reduced mass of the system (approximately the mass of the secondary compact object). Utilizing the normalization condition for the four-velocity of a massive object, $g_{\mu\nu}u^{\mu}u^{\nu} = -1$, we derive the radial equation of motion:
\begin{equation}
	\left(\frac{dr}{d\tau}\right)^2 + f(r)\left(1+\frac{l^2}{r^2}\right) = \varepsilon^2.
	\label{radial_motion}
\end{equation}
To describe the Keplerian-like bound orbits (with relativistic precession), we parameterize the trajectory using the semi-latus rectum $p$ and the eccentricity $e$:
\begin{equation}
	r = \frac{p}{1 + e \cos \chi},\label{r(p,e)}
\end{equation}
where $\chi$ is the orbital parameter. The orbital motion is bounded by two turning points: the periapsis $r_p$ and the apoapsis $r_a$, defined as
\begin{equation}
	r_p = \frac{p}{1+e}, \quad r_a = \frac{p}{1-e}.
\end{equation}
At these turning points, the radial velocity vanishes, i.e., $dr/d\tau = 0$. Substituting this condition into the radial equation (\ref{radial_motion}) yields the constraints:
\begin{subequations}
\begin{align}
	\varepsilon^2 &= f(r_p) \left( 1 + \frac{l^2}{r_p^2} \right),\label{eq_rp} \\
	\varepsilon^2 &= f(r_a) \left( 1 + \frac{l^2}{r_a^2} \right).\label{eq_ra}
\end{align}
\end{subequations}
\begin{figure}
	\centering
	\includegraphics[width=0.825\textwidth]{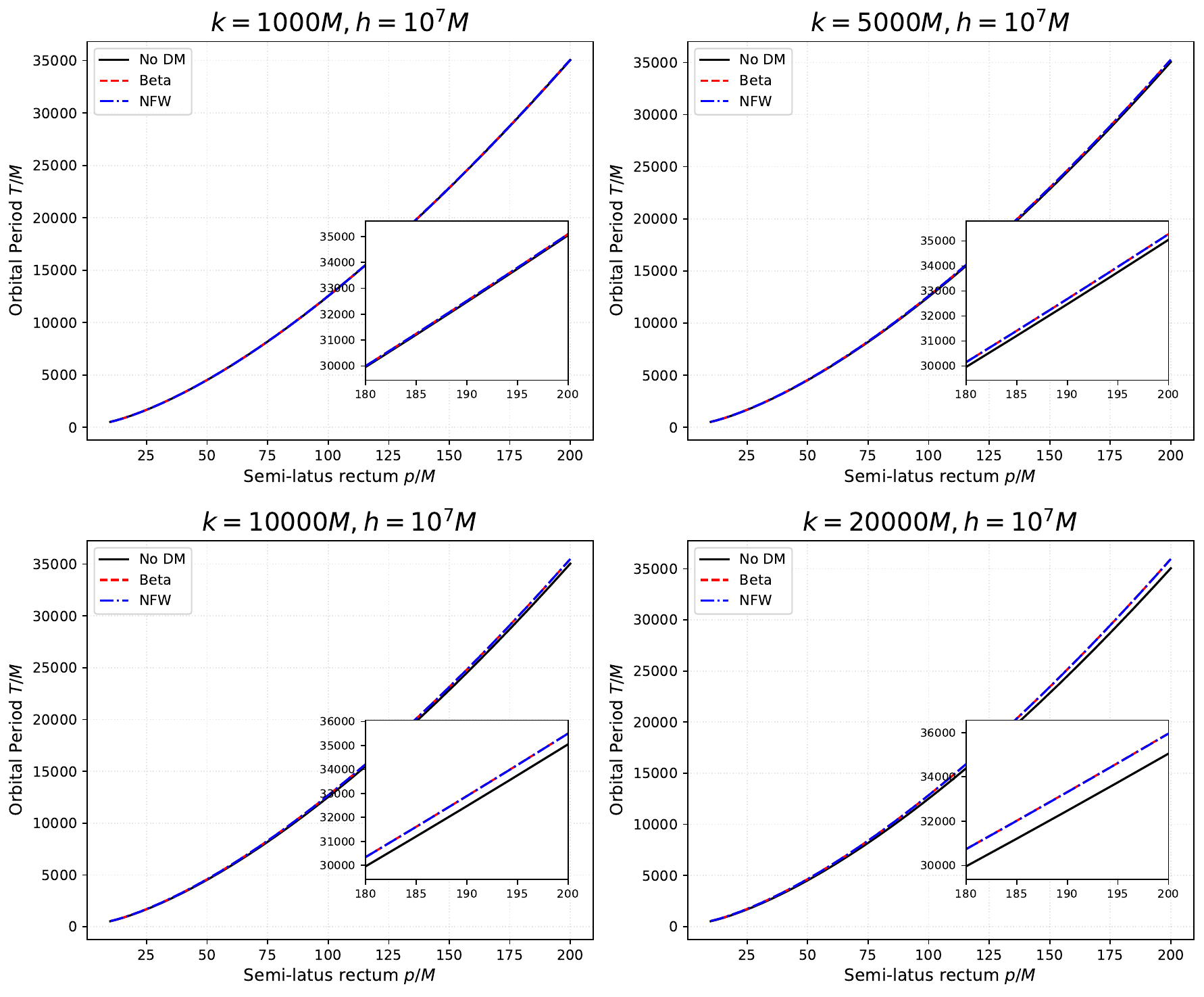}
	\caption{The orbital period $T$ (in hours) as a function of the semi-latus rectum $p$ ranging from $10M$ to $200M$ with $e=0.6$. The panels correspond to different halo mass parameters $k \in \{1000M, 5000M, 10000M, 20000M\}$ with a fixed scale $h=10^7M$. The black solid, red dashed, and blue dash-dotted lines represent the results for the pure black hole case without dark matter, Beta model, and NFW model respectively.}
	\label{fig:period_comparison}
\end{figure}
Solving Eqs. (\ref{eq_rp}) and (\ref{eq_ra}), we express the constants of motion $l^2$ and $\varepsilon^2$ in terms of the turning points:
\begin{subequations}
\begin{align}
	l^2 &= \frac{r_p^2 r_a^2 [f(r_a) - f(r_p)]}{f(r_p) r_a^2 - f(r_a) r_p^2}, \label{l} 
	\\
	\varepsilon^2 &=\frac{f(r_p) f(r_a) (r_a^2 - r_p^2)}{f(r_p) r_a^2 - f(r_a) r_p^2}. \label{varepsilon}
\end{align}
\end{subequations}
From Eqs. (\ref{u_t})-(\ref{r(p,e)}) and applying the chain rule, we obtain:
\begin{subequations}
\begin{align}
  	\frac{d\phi}{d\chi} &= \frac{d\phi}{d\tau}\cdot\frac{d\tau}{d\chi} 
  	=\frac{l}{r^2}\cdot\frac{dr}{d\chi}\cdot\frac{d\tau}{dr} 
  	=\frac{\ell}{r^2}\frac{dr}{d\chi}\left[\varepsilon^{2}-f(r)\left(1+\frac{l^{2}}{r^{2}}\right)\right]^{-1/2}, \label{dphi/dchi}
    \\
  	\frac{dt}{d\chi} &= \frac{dt}{d\tau}\cdot\frac{d\tau}{d\chi}
  	=\frac{\varepsilon}{f(r)}\frac{dr}{d\chi}\left[\varepsilon^{2}-f(r)\left(1+\frac{l^{2}}{r^{2}}\right)\right]^{-1/2}. \label{dt/dchi}
\end{align}
\end{subequations}
The orbital period $T$ and the precession angle $\Delta\phi$ accumulated over one orbital cycle are defined by the following integrals:
\begin{subequations}
\begin{align}
	T &= \int_{0}^{2\pi} \frac{dt}{d\chi} \, d\chi, \label{period_def}
	\\
	\Delta \phi &= \int_{0}^{2\pi} \frac{d\phi}{d\chi} \, d\chi - 2\pi. \label{precession_def}
\end{align}
\end{subequations}
By substituting the effective metric functions for the NFW and Beta profiles (Eqs.~\ref{f_NFW1} and \ref{f_Beta1}) into these differential equations, we can effectively quantify the influence of dark matter on the orbital period and precession angle. To investigate the environmental signatures on Keplerian-like bound orbits of EMRIs, we employ a numerical integration approach to determine the orbital period $T$ and precession angle $\Delta\phi$. We analyze these quantities as functions of the semi-latus rectum $p$ with a fixed eccentricity. In the following part of this section, the eccentricity is selected to be $e=0.6$ as a typical example to investigate the dark matter halo's influences. As illustrated in Fig.~\ref{fig:period_comparison}, the presence of a dark matter halo introduces a cumulative deviation in the orbital period from that given by pure SMBH scenario. Notably, the NFW and Beta profiles remain indistinguishable across all considered values of the halo mass parameter $k$, as evidenced by the complete overlap of their curves in Fig.~\ref{fig:period_comparison}. This indicates that the orbital period is insensitive to the specific density distribution differences between these two models. For orbital precession, as shown in Fig. \ref{fig:precession_comparison}, the precession angle curves for the NFW and Beta models remain indistinguishable even in high-density scenarios ($k=20000M$). Consequently, short-term trajectory observations, specifically the orbital period and precession angle over a single Keplerian cycle, are insufficient to discriminate between these two models. To effectively distinguish them, it is necessary to consider long-term evolutionary effects, such as gravitational wave backreaction, dynamical friction, and accretion.

\begin{figure}
	\centering
	\includegraphics[width=0.9\textwidth]{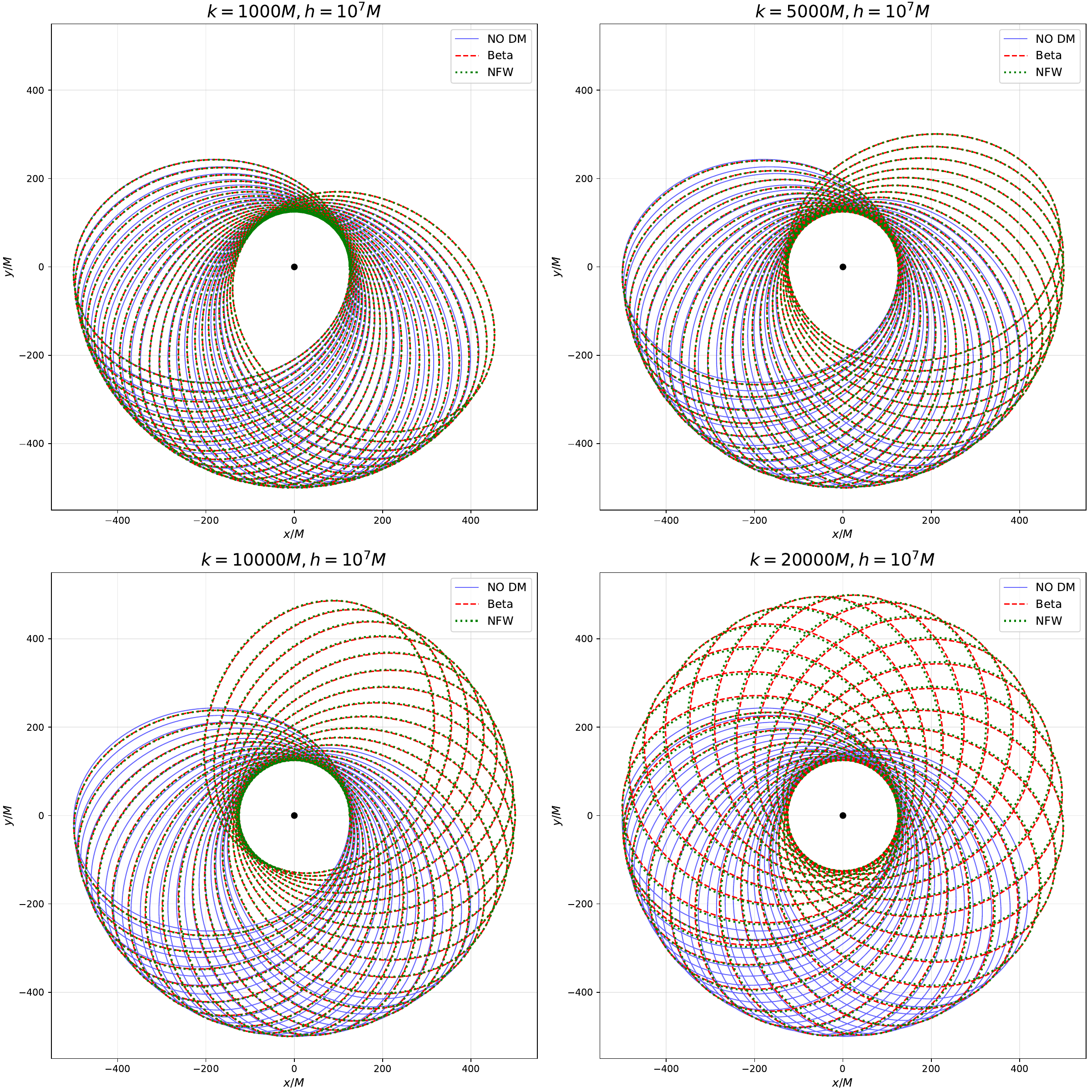}
	\caption{Comparison of orbital trajectories under different dark matter halo profiles. The orbits are computed with fixed parameters $p=200M$ and $e=0.6$. The subplots correspond to different combinations of halo mass $k$ and characteristic radius $h$. The black dot at the center represents the location of SMBH.}
	\label{fig:orbit}
\end{figure}
	
To illustrate the trajectories of SCO, we numerically integrate the geodesic equation $d\phi/d\chi$ to obtain the azimuthal angle $\phi(\chi)$. By combining $\phi(\chi)$ with the radial parameterization $r(\chi)$, the trajectory is determined in polar coordinates $(r, \phi)$ and then mapped into Cartesian coordinates via $x = r \cos\phi$ and $y = r \sin\phi$. Figure \ref{fig:orbit} illustrates the comparison of Keplerian-like bound orbits of EMRIs in galaxies hosting dark matter halos against the pure SMBH scenario. As shown in the figure, the presence of a dark matter halo induces a noticeable cumulative deviation in the trajectory compared to the pure black hole case. Notably, this deviation becomes more pronounced as the parameter $k$ increases, indicating that the dark matter halo's gravitational potential significantly influences orbital trajectories. Furthermore, within the parameter range considered here, the trajectories corresponding to the NFW and Beta profiles are nearly indistinguishable. This is consistent with our findings on periodic orbits in recent study~\cite{Li:2025eln}. It suggests that comparisons based solely on short-term trajectories are insufficient to distinguish between the Beta and NFW models. To effectively differentiate these profiles, one must account for long-term orbital evolution and various dissipative mechanisms, including gravitational radiation reaction, dynamical friction, and accretion.

\section{Orbital Evolution in Dark Matter Environments}
\label{sec3}
	
In the previous section, we analyzed the geodesic motion and Keplerian-like orbits of a small compact object (SCO) in the absence of dissipative effects. However, for a stellar-mass compact object inspiraling into a supermassive black hole embedded in a dark matter halo, the long-term orbital evolution is inevitably governed by various dissipative mechanisms. As the SCO traverses the halo, it interacts with the environment, leading to the accretion of surrounding dark matter and the emergence of dynamical friction~\cite{Chandrasekhar:1943ys,Cardoso:2020iji,Bondi:1952ni,Macedo:2013qea,Bondi:1944rnk,Edgar:2004mk}. Simultaneously, the emission of gravitational radiation continuously extracts energy and angular momentum from the orbit. These dissipative processes have significant impacts on the long-term orbital dynamics of EMRI systems and the resulting phase accumulation of gravitational waves. In this section, assuming the SCO is a black hole, we incorporate the combined effects of dynamical friction, mass accretion, and gravitational wave (GW) radiation reaction to model the evolution of the EMRI system.
	
\subsection{Dissipative Mechanisms }
As the small BH moves through the DM halo, it experiences a gravitational drag force known as dynamical friction. This force arises from the interaction between the BH and the dark matter particles in the halo, leading to a loss of orbital energy and angular momentum. For a collisionless DM medium, the dynamical friction is proportional to the dark matter density and inversely related to the dark matter velocity~\cite{Chandrasekhar:1943ys,Cardoso:2020iji}, via:
\begin{equation}
	\mathbf{f}_{\text{DF}} = - \frac{4\pi \mu^2 \rho_{\text{DM}}(r) \ln\Lambda}{v^3} \mathbf{v},
	\label{eq:F_DF}
\end{equation}
where $\mathbf{v}$ is its velocity, $\rho_{\text{DM}}(r)$ is the local density of the DM halo, and $\ln\Lambda$ is the Coulomb logarithm, typically set to $\ln\Lambda \approx 3$~\cite{Eda:2014kra}. The rates of energy and angular momentum loss induced by dynamical friction are calculated as:
\begin{subequations}
\begin{align}
	\left(\frac{dE}{dt}\right)_{\text{DF}} &= \mathbf{f}_{\text{DF}} \cdot \mathbf{v}, 
	\\
	\left(\frac{dL}{dt}\right)_{\text{DF}} &= \mathbf{r} \times \mathbf{f}_{\text{DF}} 
\end{align}
\end{subequations}
By substituting the orbital parameters defined in Sec. II and the specific density profiles into these equations, we derive the explicit instantaneous loss rates caused by dynamical friction associated with NFW and Beta halo models.

In addition to dynamical friction, the stellar-mass compact object (SCO), assumed to be a black hole, will inevitably capture dark matter (DM) particles from its surroundings. This process is modeled via the Bondi-Hoyle-Lyttleton accretion mechanism. The accretion rate $\dot{\mu}$ depends on the relative velocity of the SCO and the local sound speed $c_s$ of the medium~\cite{Bondi:1952ni,Macedo:2013qea,Bondi:1944rnk,Edgar:2004mk}:
\begin{equation}
	\dot{\mu} = \frac{4\pi \rho_{\text{DM}}(r) \mu^2}{(v^2 + c_s^2)^{3/2}},
	\label{eq:mdot_bondi}
\end{equation}
where $c_s = \sqrt{\delta P / \delta \rho}$ represents the sound speed in the DM fluid. In a typical astrophysical galaxy center near an SMBH, the orbital velocity of the SCO typically far exceeds the sound speed ($v \gg c_s$), allowing us to neglect the $c_s$ term. In this work, we assume an isotropic accretion of dark matter onto the SCO. Under such isotropic accretion from a non-rotating environmental medium, recent studies have demonstrated that no angular momentum is transferred to the SCO~\cite{Hughes:2018qxz}. In accordance with adiabatic invariance, the orbital eccentricity remains constant during this accretion process ($de/dt = 0$), and the orbital angular momentum is conserved ($(dL/dt)_{\text{acc}} = 0$)~\cite{Hughes:2018qxz, Blachier:2023ygh}. The resulting variation in orbital energy due to this mass increase is then given by:
	
\begin{equation}
	\left( \frac{dE}{dt} \right)_{\text{acc}} = \frac{\dot{\mu}}{\mu} E + \mu \dot{\epsilon} = \frac{\dot{\mu}}{\mu} \left( E - \ell \frac{dp}{d\ell} \frac{dE}{dp} \right).
	\label{eq:Edot_acc}
\end{equation}
	
Beyond dynamical friction and accretion, the emission of gravitational waves also causes the loss of the SCO's energy and orbital angular velocity. The averaged rates of energy and angular momentum loss due to GW emission are expressed as~\cite{Peters:1963ux}:
\begin{subequations}
\begin{align}
	\left(\frac{dE}{dt}\right)_{\mathrm{GW}} &= -\frac{1}{5}\ddot{\mathcal{I}}^{jk}\ddot{\mathcal{I}}^{jk},
	\\
	\left(\frac{dL_i}{dt}\right)_{\mathrm{GW}} &= -\frac{2}{5}\epsilon_{ijk}\ddot{\mathcal{I}}^{jl}\ddot{\mathcal{I}}^{kl},
\end{align}
\end{subequations}
where $\mathcal{I}^{jk}$ is the symmetric-traceless part of the mass quadrupole moment of the system. Furthermore, the presence of a dark matter halo introduces other corrections to the GW fluxes. However, it has been pointed out that these direct corrections are proportional to the halo compactness $k/h$~\cite{Dai:2023cft}. For the astrophysical scenarios considered in this work, this compactness is extremely small ($k/h \lesssim 10^{-4}$). In this regime, the direct environmental contributions to the gravitational radiation are negligible compared to the vacuum terms. Therefore, it is reasonable to adopt the vacuum expressions for the energy and angular momentum fluxes, assuming that the dark matter halo influences the GW emission primarily by altering the orbital evolution parameters ($p(t)$, $\mu(t)$, and $e(t)$) rather than the generation of the radiation itself. A similar approach was also adopted in the study~\cite{Ashoorioon:2025ezk}. Under such assumptions, the loss of the SCO's energy and orbital angular momentum caused by GW fluxes is given by~\cite{Peters:1963ux,Peters:1964zz}:
\begin{subequations}
\begin{align}
	\left(\frac{dE}{dt}\right)_{\mathrm{GW}} &= -\frac{32}{5}\frac{\mu^2}{M^2}\left(\frac{M}{p}\right)^5\left(1-e^2\right)^{3/2}\left(1+\frac{73}{24}e^2+\frac{37}{96}e^4\right),
	\\
	\left(\frac{dL}{dt}\right)_{\mathrm{GW}} &= -\frac{32}{5}\frac{\mu^2}{M}\left(\frac{M}{p}\right)^{7/2}\left(1-e^2\right)^{3/2}\left(1+\frac{7}{8}e^2\right),
\end{align}
\end{subequations}

\subsection{Combined Evolution}
\label{subsec:combined_evolution}
The total evolution of the EMRI system is driven by the combined effects of GW radiation reaction, dynamical friction, and accretion. The total rates of change for energy and angular momentum are the sums of the individual contributions:
\begin{subequations}
\begin{align}
	\left\langle \frac{dE}{dt} \right\rangle_{\text{total}} &= \left\langle \frac{dE}{dt} \right\rangle_{\text{GW}} + \left\langle \frac{dE}{dt} \right\rangle_{\text{DF}} + \left\langle \frac{dE}{dt} \right\rangle_{\text{ACC}}, \label{eq:Edot_total} 
	\\
	\left\langle \frac{dL}{dt} \right\rangle_{\text{total}} &= \left\langle \frac{dL}{dt} \right\rangle_{\text{GW}} + \left\langle \frac{dL}{dt} \right\rangle_{\text{DF}}, \label{eq:Ldot_total}
\end{align}
\end{subequations}
where $\langle \dots \rangle$ denotes the time-average over one orbital period $T$.

\begin{figure}
	\centering
	\includegraphics[width=0.625\textwidth]{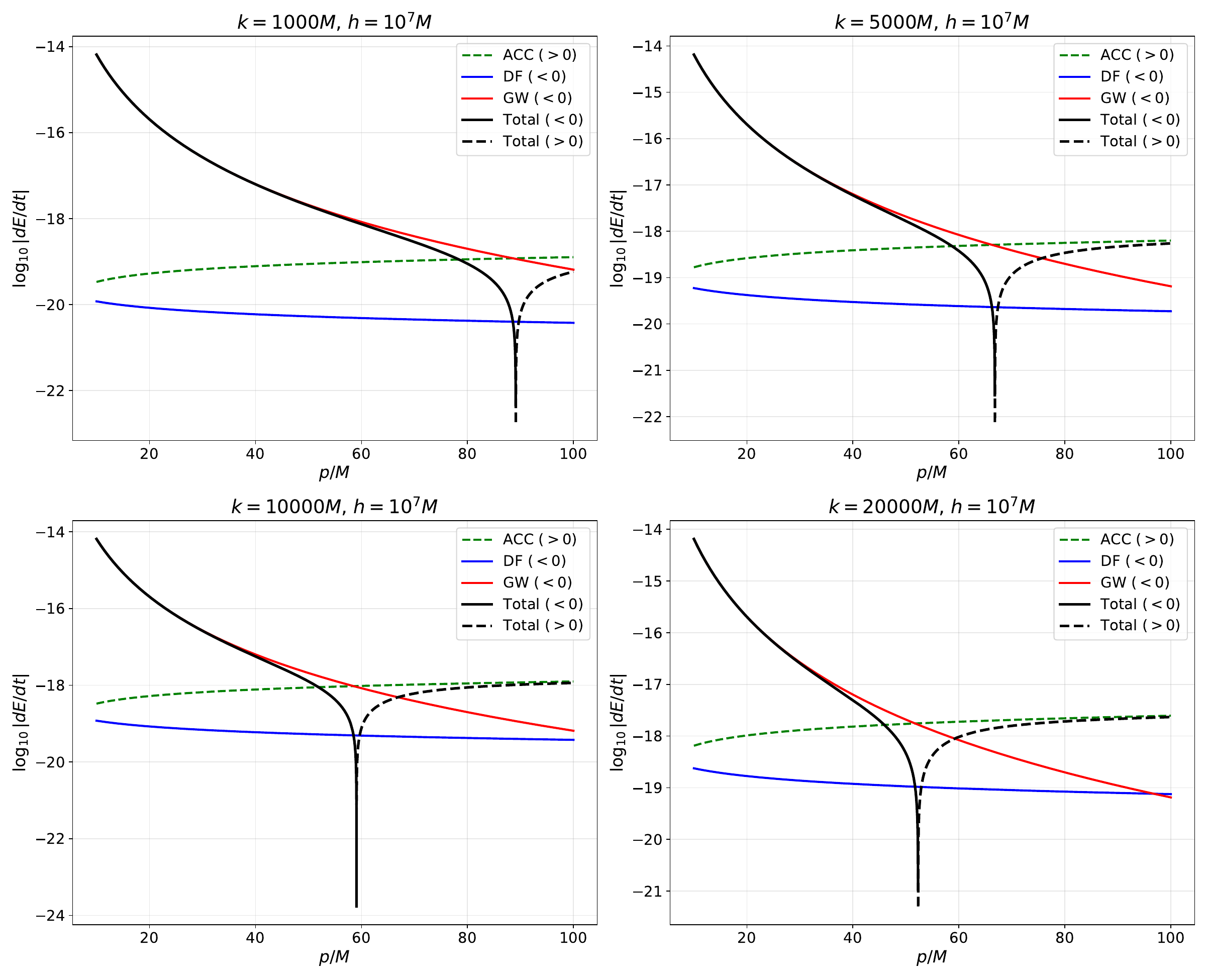} 
	\caption{The absolute values of the time-averaged energy flux $|\langle dE/dt \rangle|$ for the NFW model, plotted on a logarithmic scale. The orbital parameters are $e=0.1$, $M=10^{6}M_{\odot}$, and $\mu=10M_{\odot}$. The total flux is represented by black lines, where the solid segment indicates net energy loss ($dE/dt<0$) and the dashed segment indicates net energy gain ($dE/dt>0$). The sharp cusp in the total flux indicates the competition and balance between the energy injection from accretion and the energy loss from GW radiation and dynamical friction.}
	\label{fig:flux_E_NFW}
\end{figure}

\begin{figure}
	\centering
	\includegraphics[width=0.625\textwidth]{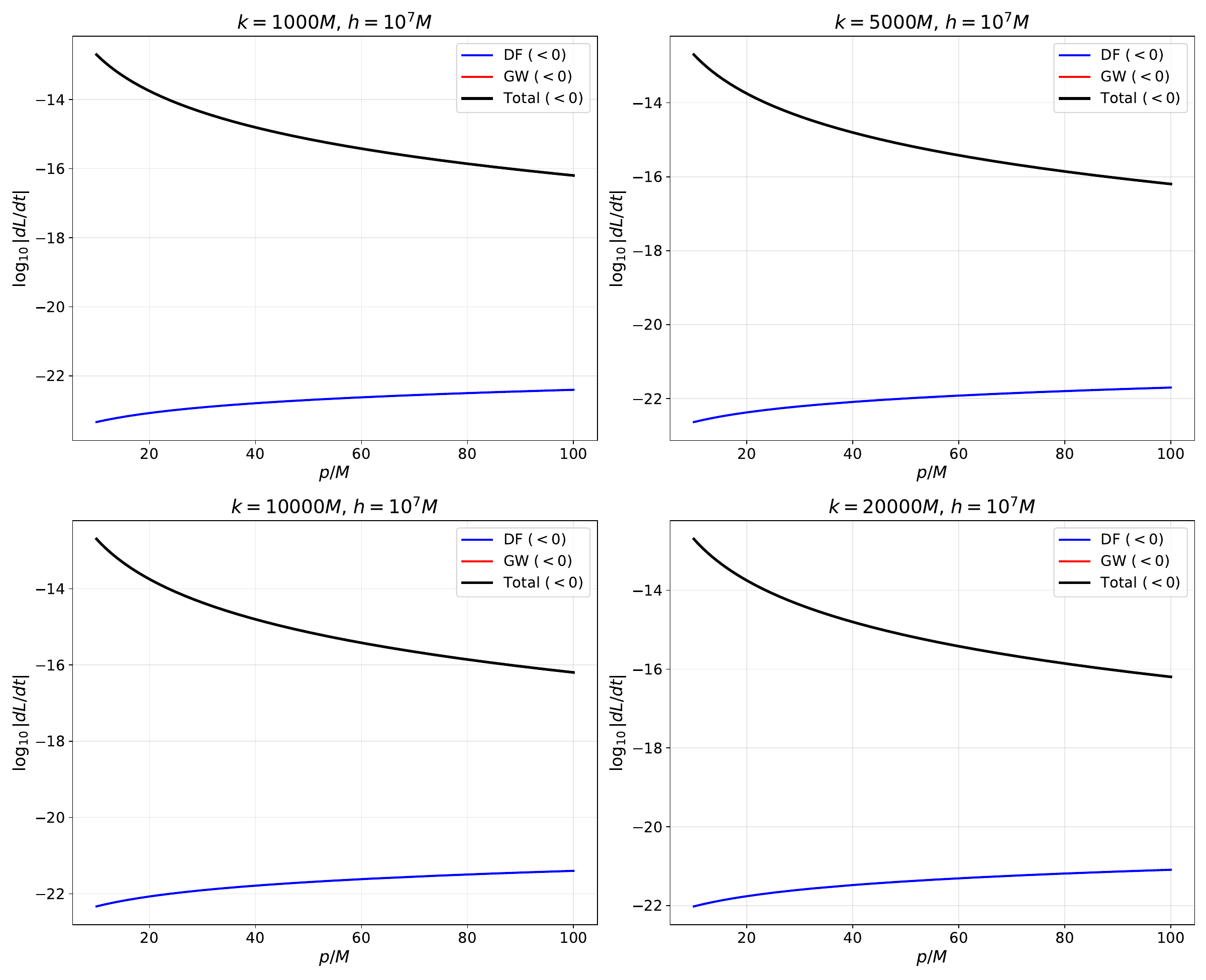}
	\caption{The absolute values of the time-averaged angular momentum flux $|\langle dL/dt \rangle|$ for the NFW model on a logarithmic scale ($e=0.1$, $M=10^{6}M_{\odot}$, $\mu=10M_{\odot}$). The black solid curves (total flux) and red solid curves (GW flux) perfectly overlap. Since accretion is assumed to transfer no angular momentum, the total flux is persistently negative across the entire range of parameter $p$. It is dominated by GW radiation and dynamical friction, and is thus represented by solid lines.}
	\label{fig:flux_L_NFW}
\end{figure}

\begin{figure}
	\centering
	\includegraphics[width=0.625\textwidth]{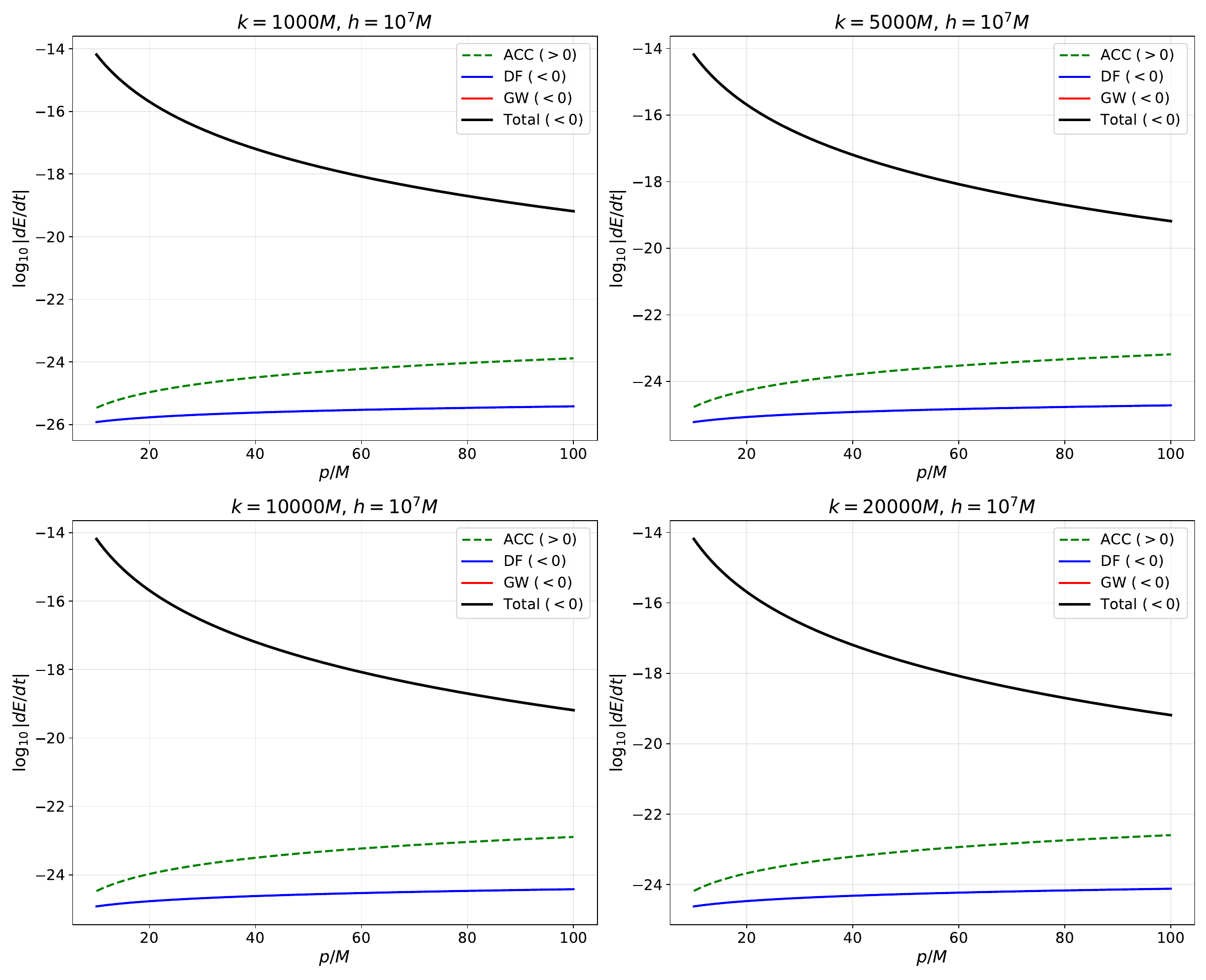}
	\caption{The absolute values of the time-averaged energy flux $|\langle dE/dt \rangle|$ for the Beta model, plotted on a logarithmic scale. The orbital parameters are $e=0.1$, $M=10^{6}M_{\odot}$, and $\mu=10M_{\odot}$. In this figure, the black solid curves indicating the total flux and the red solid curves representing the GW flux perfectly overlap. Unlike the NFW case, the total flux remains strictly negative across the entire range of parameter $p$. This is because the energy injection from accretion is insufficient to overcome the GW dissipation.}
	\label{fig:flux_E_Beta}
\end{figure}

\begin{figure}
	\centering
	\includegraphics[width=0.625\textwidth]{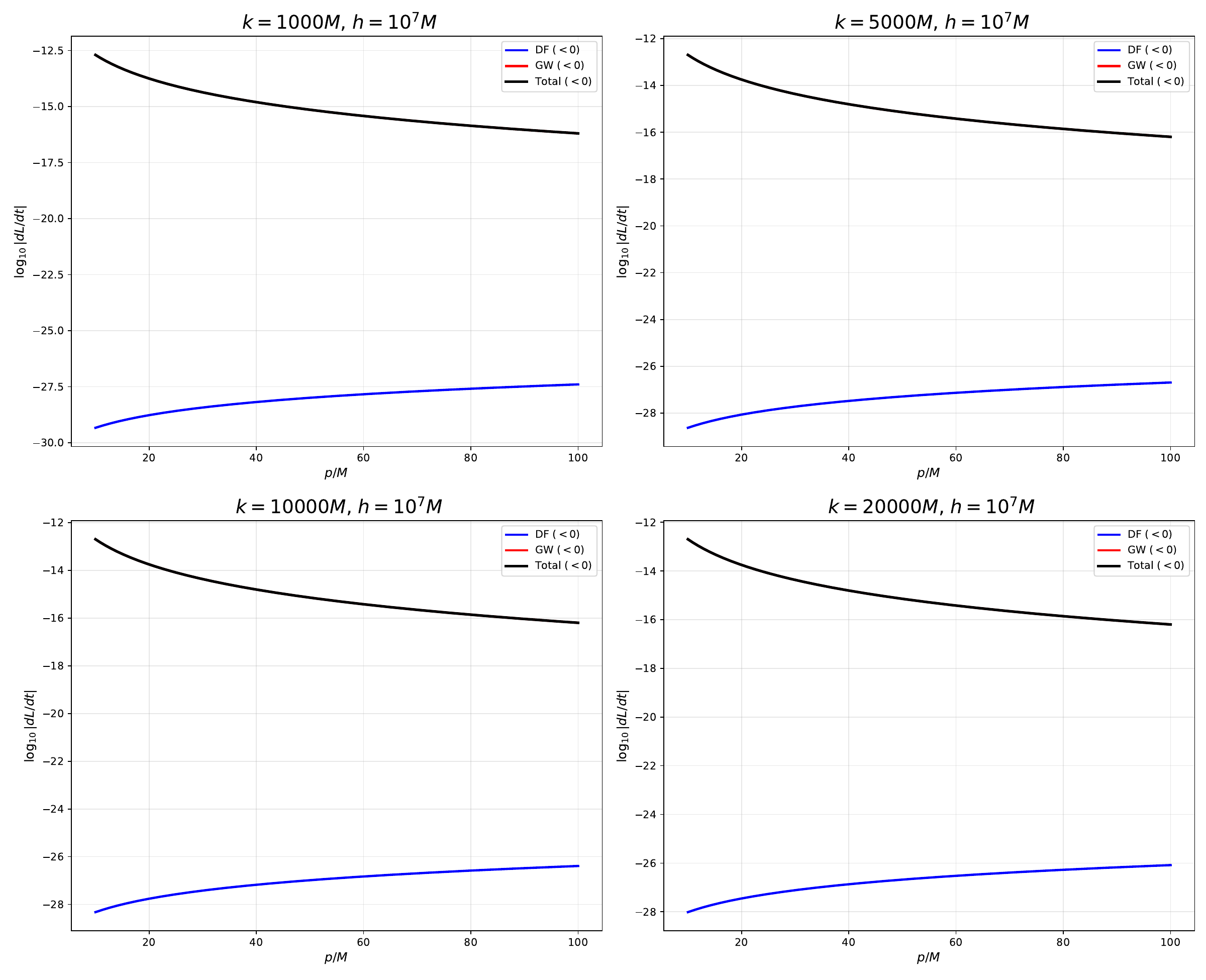}
	\caption{The absolute values of the time-averaged angular momentum flux $|\langle dL/dt \rangle|$ for the Beta model on a logarithmic scale ($e=0.1$, $M=10^{6}M_{\odot}$, $\mu=10M_{\odot}$). The black solid curves (total flux) and the red solid curves (GW flux) perfectly overlap, indicating that the orbital evolution is dominated by GW radiation. Consequently, the total flux remains consistently negative across the entire parameter space, as represented by the solid lines.}
	\label{fig:flux_L_Beta}
\end{figure}

To quantify the impact of different dissipative mechanisms on the orbital evolution, we numerically calculate the time-averaged energy and angular momentum fluxes for both the NFW and Beta models, as presented in Figs.~\ref{fig:flux_E_NFW}--\ref{fig:flux_L_Beta}. In all these cases, the orbital eccentricity is fixed at $e=0.1$ and the mass ratio is set to $\mu/M = 10^{-5}$. We present the absolute values of the time-averaged fluxes on a logarithmic scale to clearly visualize the evolution of energy and angular momentum across several orders of magnitude. In the figures and legends, we explicitly distinguish between the regimes of energy dissipation ($\langle dE/dt \rangle < 0$, denoted by solid lines) and energy injection ($\langle dE/dt \rangle > 0$, denoted by dashed lines). Figs.~\ref{fig:flux_E_NFW} and \ref{fig:flux_L_NFW} present the results for the NFW profile. Since the GW flux drops rapidly as the orbit expands (scaling roughly as $p^{-5}$), it becomes less dominant at larger $p$. In contrast, the DM environmental fluxes (dynamical friction and accretion) change much more slowly with $p$, making them increasingly significant at larger orbital separations. Notably, in Fig.~\ref{fig:flux_E_NFW}, we observe a distinct ``cusp" in the total energy flux. This feature arises from the competition between energy loss (from GWs and dynamical friction) and energy gain from accretion. The cusp marks the point where the energy gained from accretion exactly balances the energy lost due to dissipation. Importantly, the position of this cusp depends on the halo mass. As the halo mass parameter $k$ increases, the local dark matter density rises, which strengthens both dynamical friction and accretion. Consequently, DM environmental effects can compete with GW emission at smaller orbital separations. This is clearly reflected by the shift of the cusp point toward smaller $p$ values (e.g., moving from $p \approx 90M$ for $k=1000M$ to $p \approx 50M$ for $k=20000M$), indicating that massive halos can significantly affect the inspiral dynamics even in the strong-field regime. For angular momentum evolution (Fig.~\ref{fig:flux_L_NFW}), since we assume accretion transfers negligible angular momentum ($\dot{L}_{\text{ACC}} \approx 0$), the total flux of the SCO is determined by the sum of GW radiation and dynamical friction. Both mechanisms reduce the orbital angular momentum, resulting in consistently negative fluxes (solid lines throughout). Unlike the energy flux, no competition or reversal occurs for angular momentum, and the curves in Fig.~\ref{fig:flux_L_NFW} exhibit monotonic behavior across the entire parameter range. For comparison, Figs.~\ref{fig:flux_E_Beta} and \ref{fig:flux_L_Beta} present the results for the Beta model using the same parameter settings. It is evident that the DM environmental contributions from dynamical friction and accretion calculated within the Beta model are orders of magnitude weaker than those in the NFW case. As illustrated in Fig.~\ref{fig:flux_E_Beta}, the total energy flux (black solid lines) virtually coincides with the energy fluxes caused by GW radiation (red solid lines) across the entire range of $p$. The flux remains consistently negative, indicating that accretion is insufficient to trigger the energy injection regime observed in the NFW model. As a result, the characteristic "cusp" feature of the energy flux is absent in the Beta profile.
		

\begin{figure}
	\centering
	\includegraphics[width=0.65\textwidth]{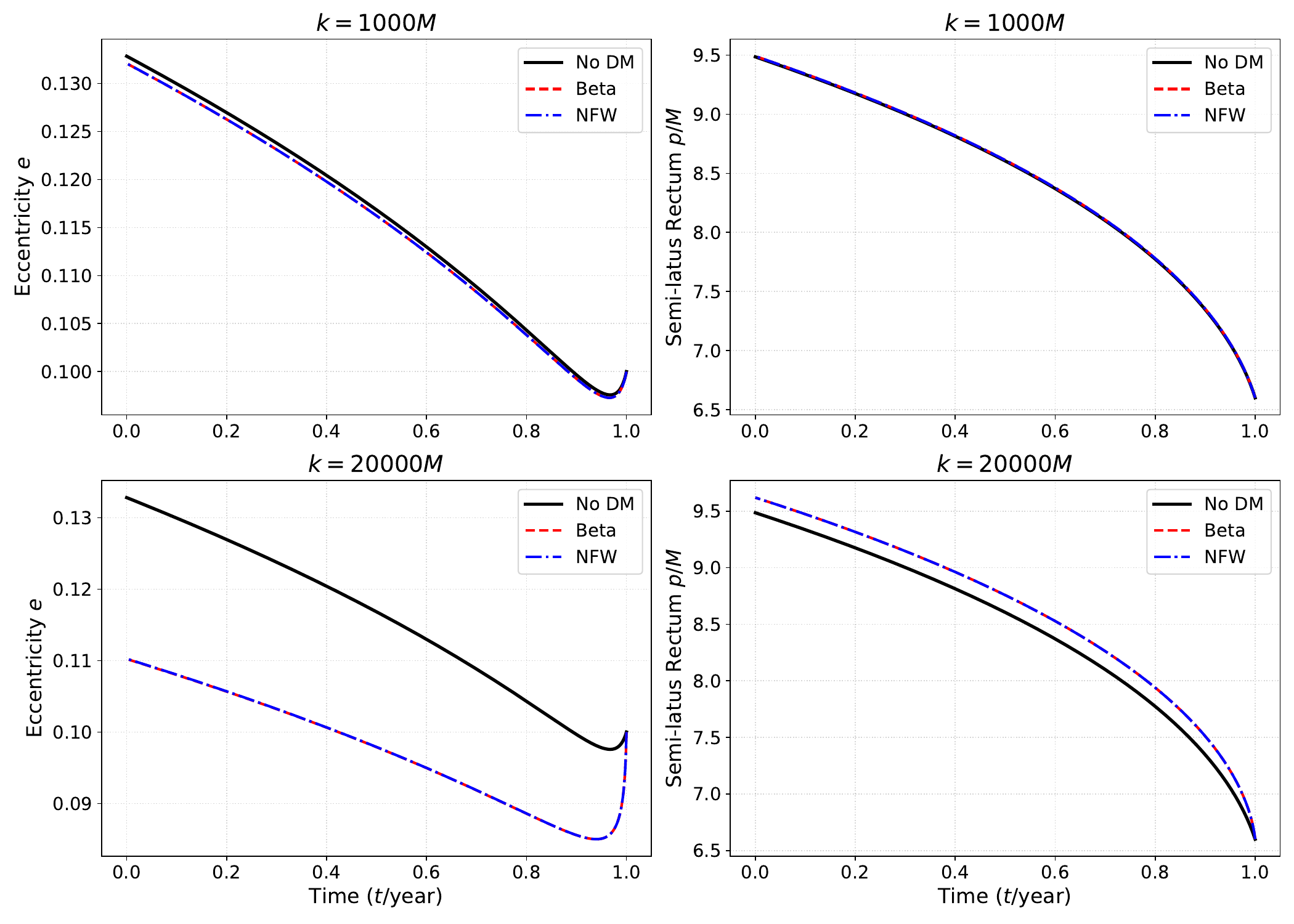} 
	\caption{The backward orbital evolution of the eccentricity $e(t)$ (left panels) and semi-latus rectum $p(t)$ (right panels) over a duration of one year. At the end of this year, the semi-latus rectum and eccentricity of the SCO are set to $p=6.6M$ and $e=0.1$, respectively. The mass of the central SMBH is $M=10^6M_{\odot}$, and the mass of the SCO is $\mu=10M_{\odot}$. The upper panels correspond to a halo density parameter of $k=1000M$, while the lower panels represent a much denser halo with $k=20000M$. For the high-density case (bottom panels), the DM environmental effects significantly accelerate the inspiral, leading to a noticeable deviation from the vacuum trajectory.}
	\label{fig:back_evolution}
\end{figure}
\begin{figure}
	\centering
	\includegraphics[width=0.65\textwidth]{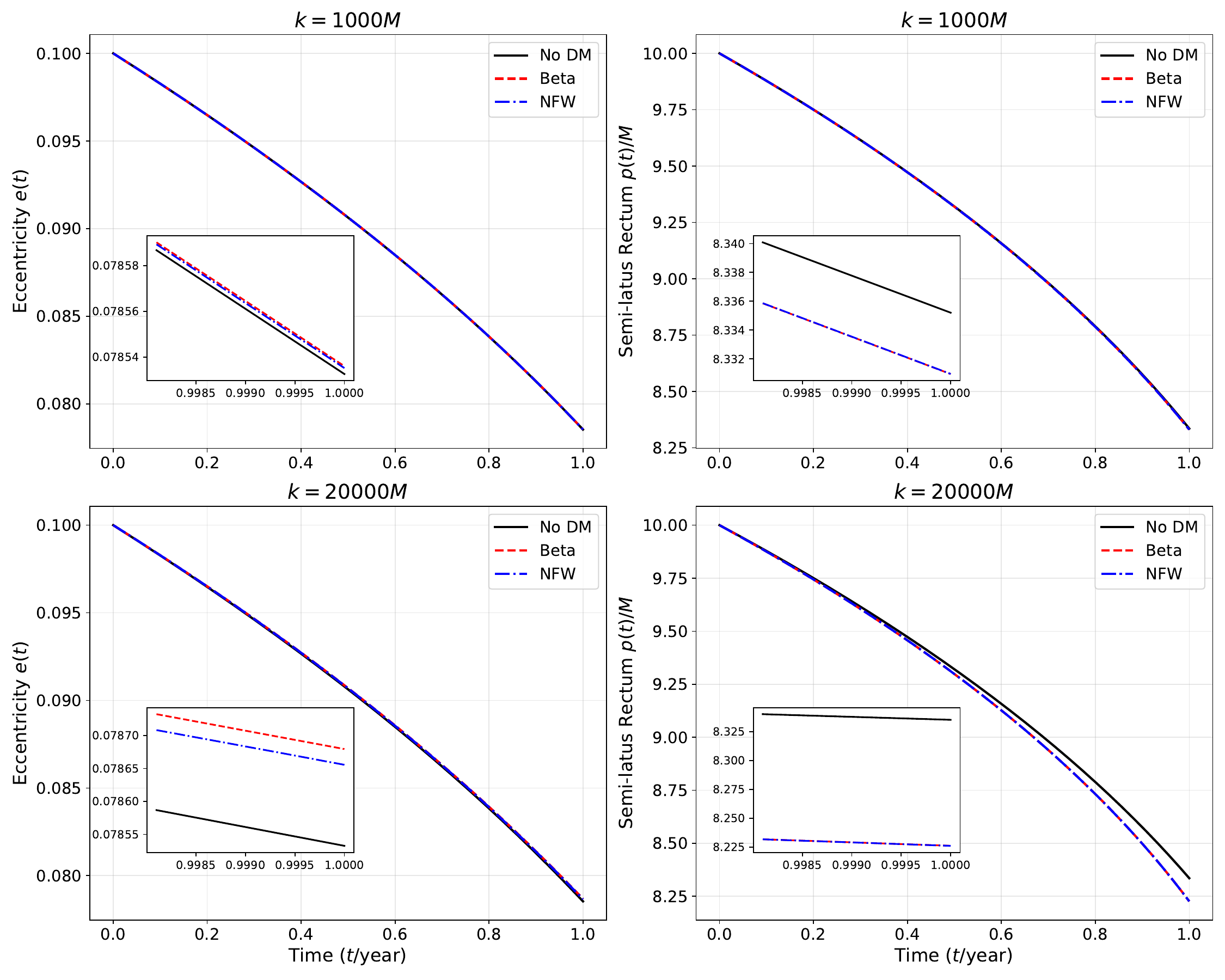}
	\caption{The forward orbital evolution of the eccentricity $e(t)$ (left panels) and semi-latus rectum $p(t)$ (right panels) over a duration of one year, with the SCO's parameters initialized at $p=10.0M$ and $e=0.1$. The upper and lower panels correspond to halo mass parameters $k=1000M$ and $k=20000M$, respectively.}
	\label{fig:forward_evolution}
\end{figure}
	
After comparing the influence of Beta and NFW dark matter halos on the SCO's energy and orbital angular momentum (as a function of the semi-latus rectum $p$), we further investigate and contrast how these two models affect the long-term orbital evolution of EMRI systems. By tracking the variations in eccentricity and semi-latus rectum over long-term evolution, it may be possible to reveal specific signatures from EMRI systems that enable observational discrimination between the Beta and NFW dark matter halo models. To investigate the cumulative impact of DM environmental effects, we performed a backward time integration starting from the SCO's parameters $p=6.6M$ and $e=0.1$ at the end of the year. As illustrated in Fig.~\ref{fig:back_evolution}, dark matter effects are negligible for the low-density case ($k=1000M$). In the high-density scenario ($k=20000M$), both the NFW and Beta models distinctly deviate from the vacuum trajectory, although they remain nearly indistinguishable from each other. The semi-latus rectum $p = 6.6M$ is specifically selected because a critical turning point, characterized by $de/dt = 0$, emerges during the evolution process. In Schwarzschild spacetime, Teukolsky perturbation calculations indicate that this turning point is located at $p = 6.681M$ for the case where $e \approx 0$ \cite{Cutler:1994pb,Gair:2005is}. Consequently, we observe a characteristic reversal in the eccentricity evolution: in contrast to the typical monotonic circularization, the eccentricity undergoes a period of growth when $t > t_{\text{crit}}$ due to environmental influences. In addition, Fig.~\ref{fig:forward_evolution} presents the forward evolution of the SCO's eccentricity and semi-latus rectum over one year, initialized at $p_0=10M$ and $e_0=0.1$. It is evident that, for a large dark matter mass parameter $k=20000M$, the cumulative effect of the dark matter becomes dramatically enhanced: the semi-latus rectum $p(t)$ decreases more rapidly for both NFW and Beta models compared to the vacuum prediction.

\section{Gravitational Waveforms and Phase Accumulation}
\label{sec4}
Having assessed the long-term orbital evolution under different dark matter scenarios, we now focus on the corresponding gravitational wave (GW) signatures. This section presents a comparative analysis of the long-term GW signals from EMRIs within dark matter environments, concentrating on the waveform modulations and phase accumulations affected by the Beta and NFW models. To simulate the GW signals, we employ the Numerical Kludge method~\cite{Babak:2006uv,Chua:2017ujo}. The main strategy of the Kludge method is as follows: first, calculate the orbit of the SCO by solving the geodesic equation, and then use the quadrupole formula of gravitational radiation to get the corresponding gravitational waves\cite{Babak:2006uv,Tu:2023xab}. In the quadrupole approximation, the spatial metric perturbation $h_{ij}$ is determined by the second time derivative of the source's mass quadrupole moment. For a compact object of mass $\mu$ orbiting a central black hole of mass $M$, this perturbation can be expressed in terms of the orbital kinematics as~\cite{Will:2016sgx,Maselli:2021men,Liang:2022gdk}:
\begin{equation}
	h_{ij}=\frac{4\mu M}{D_L}\left(v_iv_j-\frac{M}{r}n_in_j\right),
\end{equation}
where $D_L$ represents the luminosity distance from the EMRI source to the detector, $v_i$ is the velocity vector of the SCO, $r$ represents the radial distance of the SCO from the central SMBH, and $n_i = x_i/r$ is the unit direction vector. To characterize the GW signal as perceived by the detector, we establish a detector-adapted coordinate system $(X, Y, Z)$. The transformation from the source frame is parameterized by the inclination angle $\iota$ and the pericenter longitude $\zeta$. Explicitly, the orthonormal basis vectors for the detector frame are constructed as follows:
\begin{subequations}
\begin{align}
	\mathbf{e}_X &= \begin{pmatrix} \cos \zeta, -\sin \zeta, 0 \end{pmatrix}, 
	\\
	\mathbf{e}_Y &= \begin{pmatrix} \sin \iota \sin \zeta, \cos \iota \cos \zeta, -\sin \iota \end{pmatrix}, 
	\\
	\mathbf{e}_Z &= \begin{pmatrix} \sin \iota \sin \zeta, -\sin \iota \cos \zeta, \cos \iota \end{pmatrix}.
\end{align}
\end{subequations}
In this coordinate system, the gravitational wave polarizations $h_+$ and $h_\times$ observed at the detector are expressed as~\cite{Maselli:2021men,Liang:2022gdk}:
\begin{subequations}
\begin{align}
	h_{+} &= \frac{1}{2}(e_{X}^{i}e_{X}^{j}-e_{Y}^{i}e_{Y}^{j}) h_{ij} \approx -\frac{2\mu M}{D_{L}r}\left(1+\cos^{2}\iota\right)\cos\left(2\phi+2\zeta\right),\label{h_+}
	\\
	h_{\times} &= \frac{1}{2}(e_{X}^{i}e_{Y}^{j}+e_{Y}^{i}e_{X}^{j}) h_{ij} \approx -\frac{4\mu M}{D_{L}r}\cos\iota\sin\left(2\phi+2\zeta\right),\label{h_times}
\end{align}
\end{subequations}
\begin{figure}
	\centering
	\includegraphics[width=0.775\textwidth]{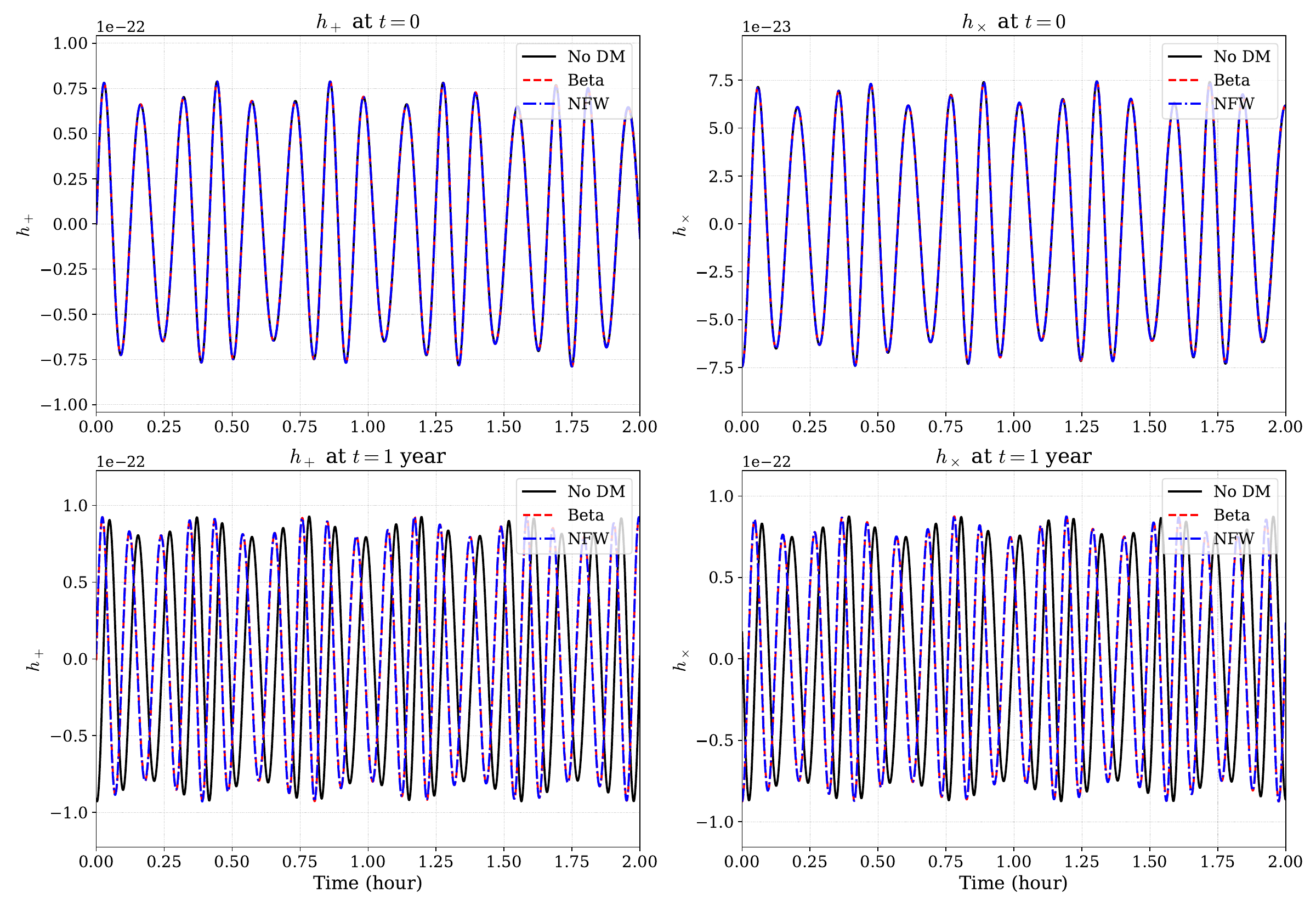}
	\caption{Gravitational waveform comparison for the dark matter mass  $k=1000M$ and halo scale $h=10^7M$.  The top row displays the waveforms at the initial time $t=0$, while the bottom row shows the waveforms after one year of evolution ($t=1$ year).}
	\label{fig:waveform_k_1000}
\end{figure}
\begin{figure}
	\centering
	\includegraphics[width=0.775\textwidth]{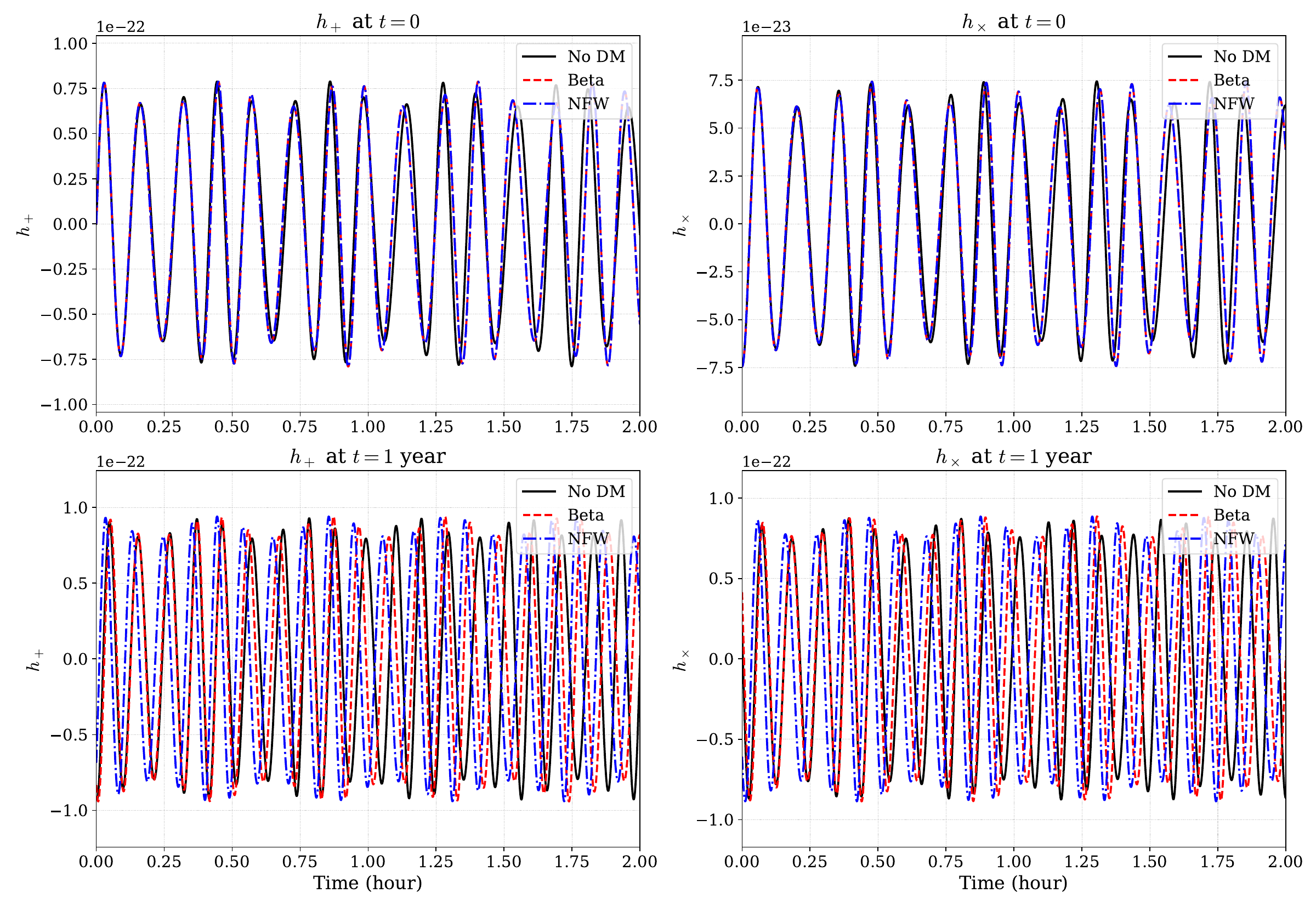}
	\caption{Gravitational waveform comparison for the halo mass $k=20000M$ halo scale $h=10^7M$ The top and bottom rows correspond to the gravitational waveforms at the initial epoch $t=0$ and after an evolutionary period of one year ($t=1$ year), respectively.}
	\label{fig:waveform_k_20000}
\end{figure}
where $\phi$ denotes the orbital phase. For the subsequent waveform calculations, we set the central black hole mass to $M = 10^6 M_{\odot}$ and the compact object mass to $m = 10 M_{\odot}$ (with the reduced mass approximated as $\mu \approx m$). We assume a luminosity distance of $D_L = 2$ Gpc, and fix both the inclination angle $\iota$ and the pericenter phase $\zeta$ at $\pi/4$. To investigate the detectability of dark matter (DM) halos and the distinguishability of different density profiles, we compute the gravitational waveforms at the initial stage ($t=0$) and after one year of evolution ($t=1$ year) using Eqs.~(\ref{h_+}) and (\ref{h_times}), whose results are plotted in Figs~\ref{fig:waveform_k_1000}-\ref{fig:waveform_k_20000}. Fig.~\ref{fig:waveform_k_1000} illustrates the waveforms for a low-density halo ($k = 1000M$). At the beginning of the evolution, the DM environmental effects are not visible, and all waveforms overlap perfectly. After one year of inspiral, a slight dephasing emerges between the DM environments and the vacuum case, but the waveforms calculated using the NFW and Beta models remain close to each other, exhibiting only minor deviations. This suggests that distinguishing between specific density profiles for a sparse dark matter halo (with a low mass density) is challenging. A different trend is observed in Fig.~\ref{fig:waveform_k_20000} for a dense DM environment ($k=20000M$). The strong gravitational potential of the halo is sufficient to produce an immediate frequency shift. As shown in the top panels, the gravitational waveforms emitted in DM halos readily deviate from the vacuum baseline shortly after the start of the observation. While these waveforms differ clearly from the vacuum case, the NFW and Beta models remain nearly identical at $t = 0$. The crucial distinction between the two profiles emerges only through long-term phase accumulation. After one year of evolution, the cumulative phase shift becomes clearly visible, providing a reliable signature to distinguish these density models in sufficiently dense halos.

In the context of EMRI gravitational wave detection, phase accumulation plays a crucial role in observations. It not only helps analyze the physical properties of gravitational sources but also provides an effective means to test and distinguish between different theoretical models~\cite{Berti:2004bd,Kavanagh:2020cfn,Barsanti:2022ana}. To quantitatively evaluate the cumulative effects of the dark matter halo on the orbital evolution and gravitational wave phase, we calculate the accumulated orbital phase $\mathcal{N}$ over a one-year observation duration ($t_2 - t_1 = 1$ year):
\begin{equation}
	\mathcal{N} = \int_{t_1}^{t_2} \dot{\phi}(t) \, dt.
\end{equation}
By evaluating the dephasing $\Delta \mathcal{N}$, defined as the difference in accumulated phase between a DM environment and a vacuum scenario or among distinct DM models, we can quantify the detectability of halo signatures in gravitational wave observations. To ensure that the gravitational wave signals in these scenarios are evaluated at an identical initial frequency, we adopt the same specific initial orbital period as the evolutionary starting point for all three models.

\begin{figure} 
	\centering
	\begin{subfigure}[b]{0.48\textwidth}
		\includegraphics[width=\textwidth]{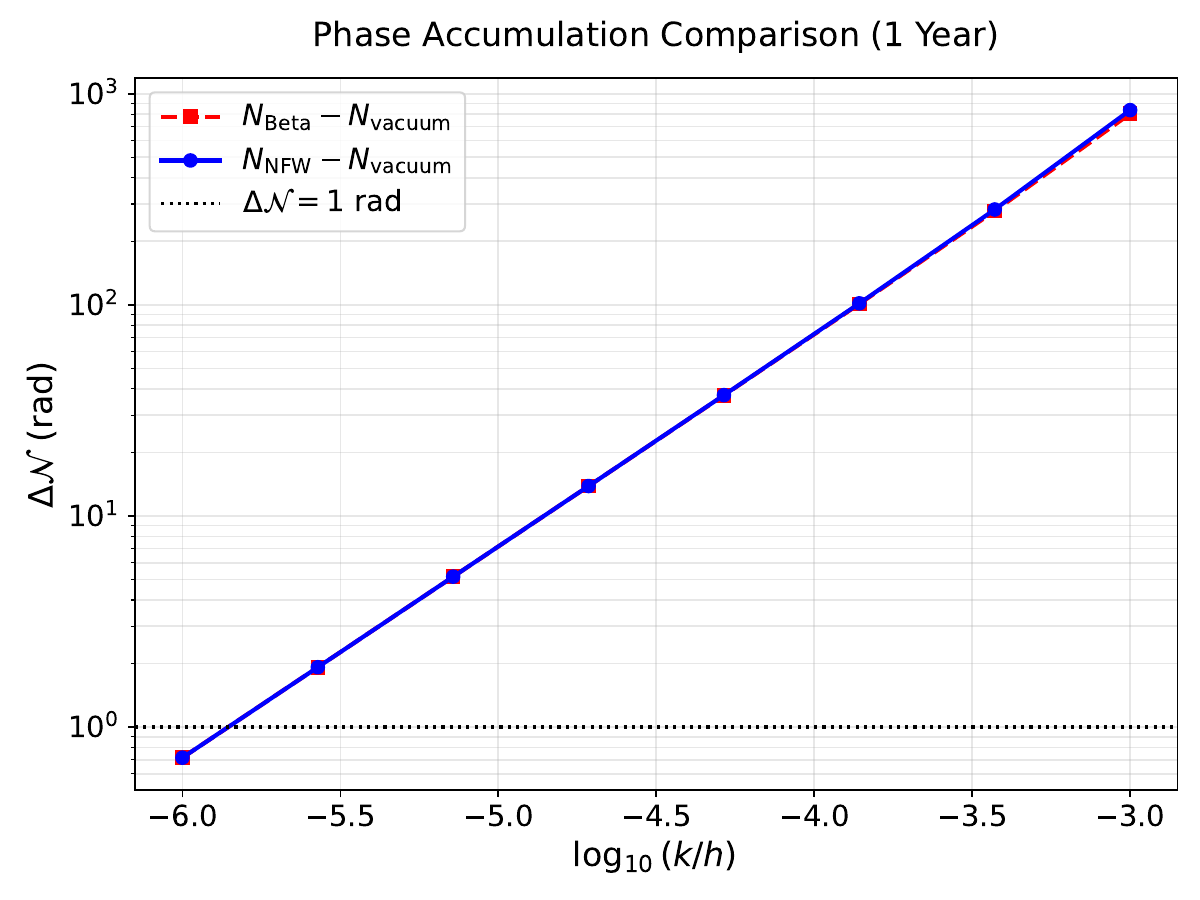}
		\caption{} 
		\label{fig:phase_vac}
	\end{subfigure}
	\hfill 
	\begin{subfigure}[b]{0.48\textwidth}
		\includegraphics[width=\textwidth]{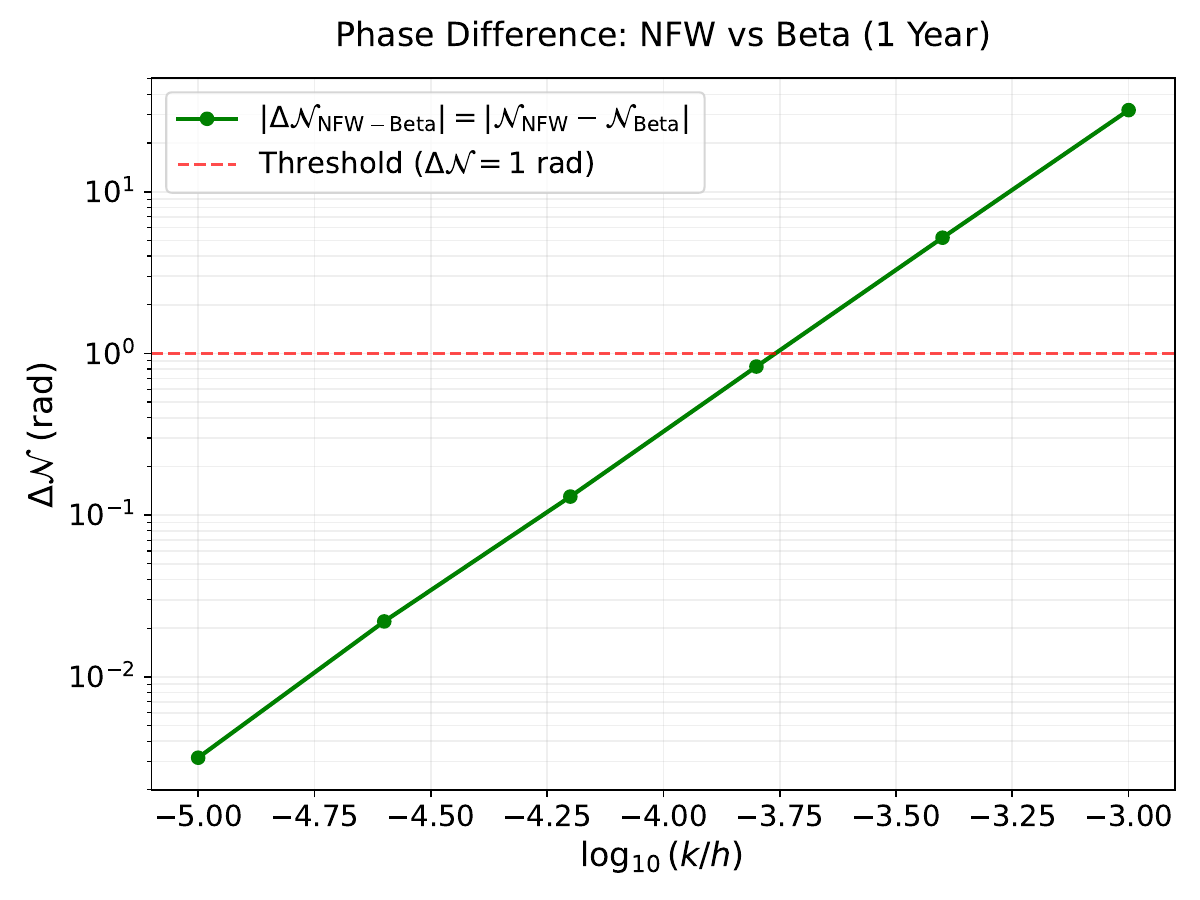}
		\caption{} 
		\label{fig:phase_diff}
	\end{subfigure}
	
	\caption{
		(a) The accumulated dephasing $\Delta \mathcal{N}$ relative to the vacuum background over a one-year observation period. 
		(b) The absolute phase difference $|\Delta \mathcal{N}_{\rm NFW} - \Delta \mathcal{N}_{\rm Beta}|$ between the NFW and Beta models. 
		We adopt the system parameters $e_0=0.1$,  $M=10^6 M_\odot$, $\mu=10 M_\odot$, and $k=100M$. The initial semi-latus rectum $p_0$ is set to correspond to an initial orbital period $T = 2\pi\sqrt{(10M)^3/M}$.
	}
	
	\label{fig:combined_phase}
\end{figure}

\begin{figure*}[htbp]
	\centering
	
	\begin{subfigure}{0.48\textwidth}
		\includegraphics[width=\linewidth]{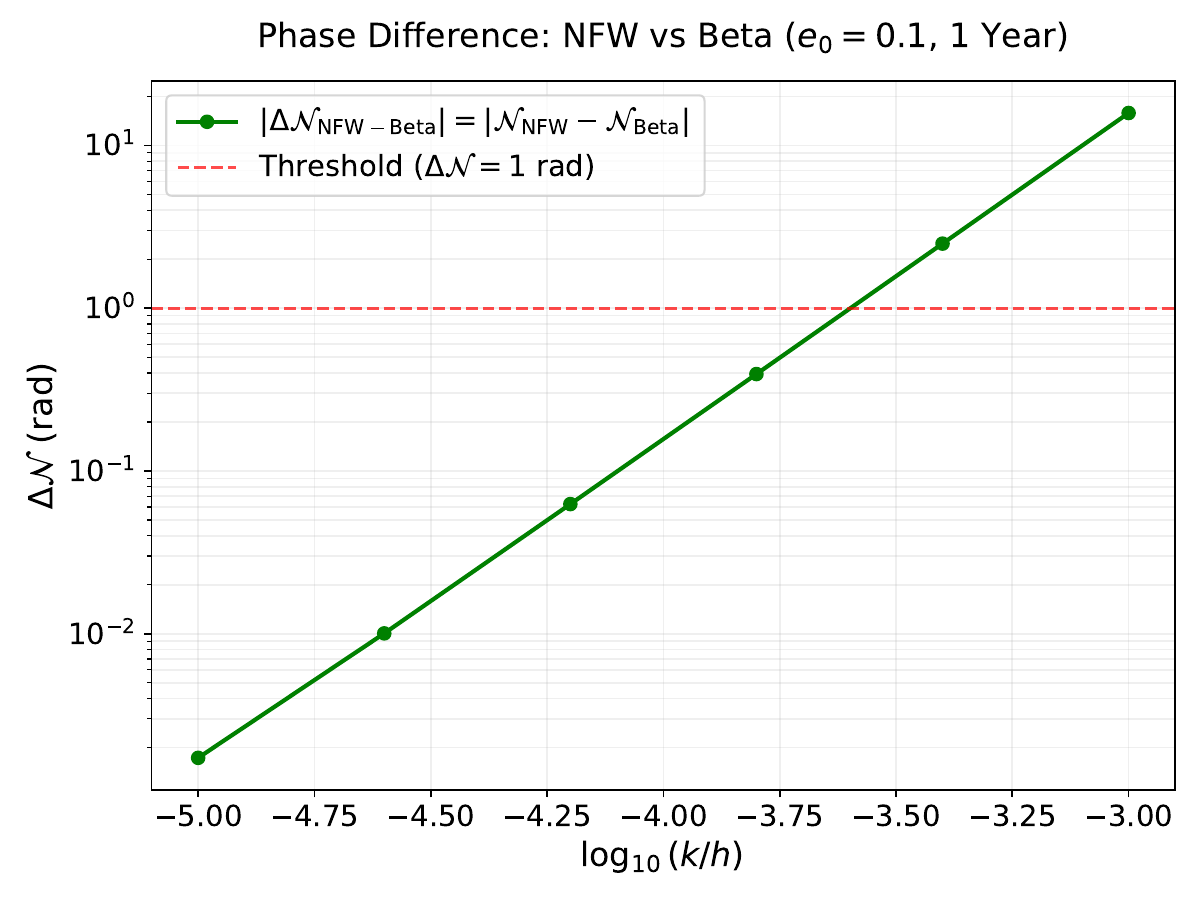}
		\label{fig:e01_1yr}
	\end{subfigure}
	\hfill
	\begin{subfigure}{0.48\textwidth}
		\includegraphics[width=\linewidth]{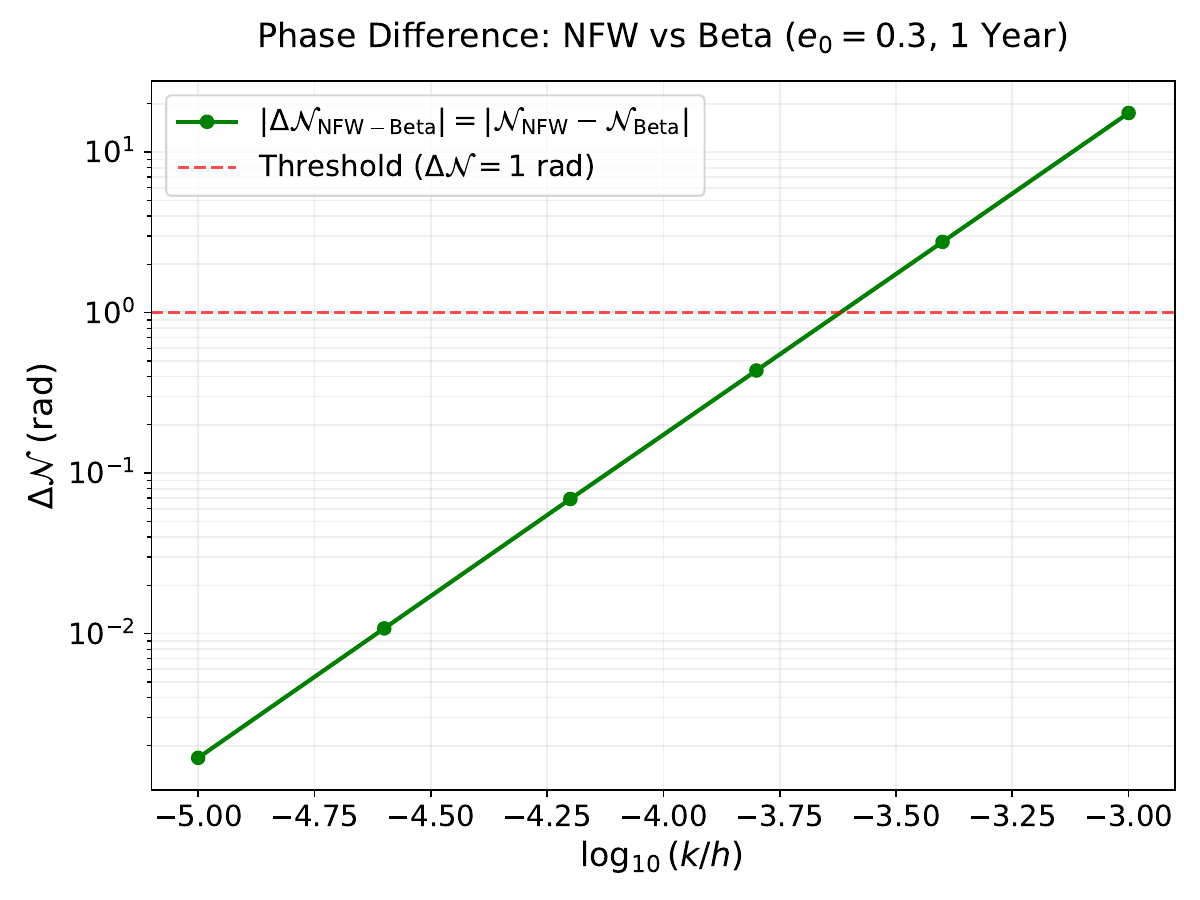}
		\label{fig:e02_1yr}
	\end{subfigure}
	
	
	\begin{subfigure}{0.48\textwidth}
		\includegraphics[width=\linewidth]{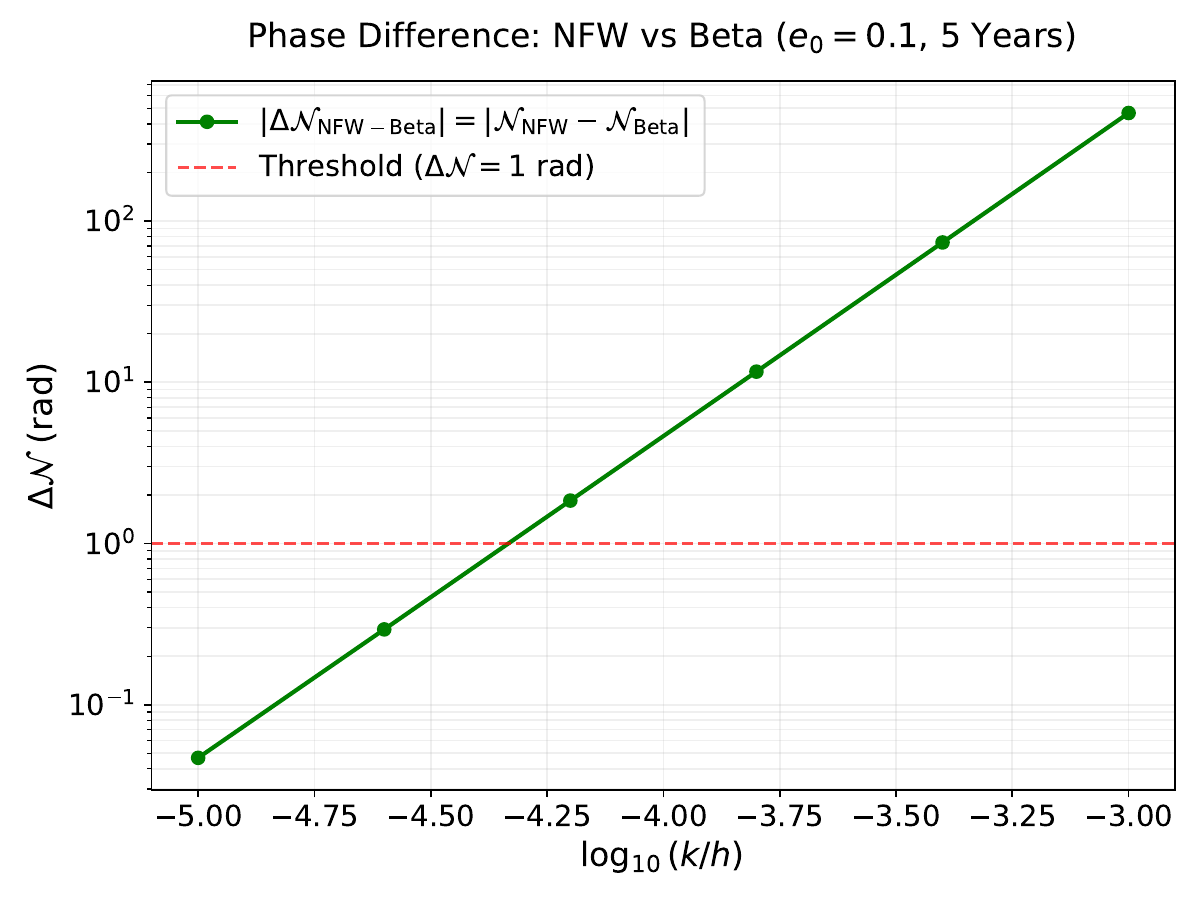}
		\label{fig:e01_5yr}
	\end{subfigure}
	\hfill
	\begin{subfigure}{0.48\textwidth}
		\includegraphics[width=\linewidth]{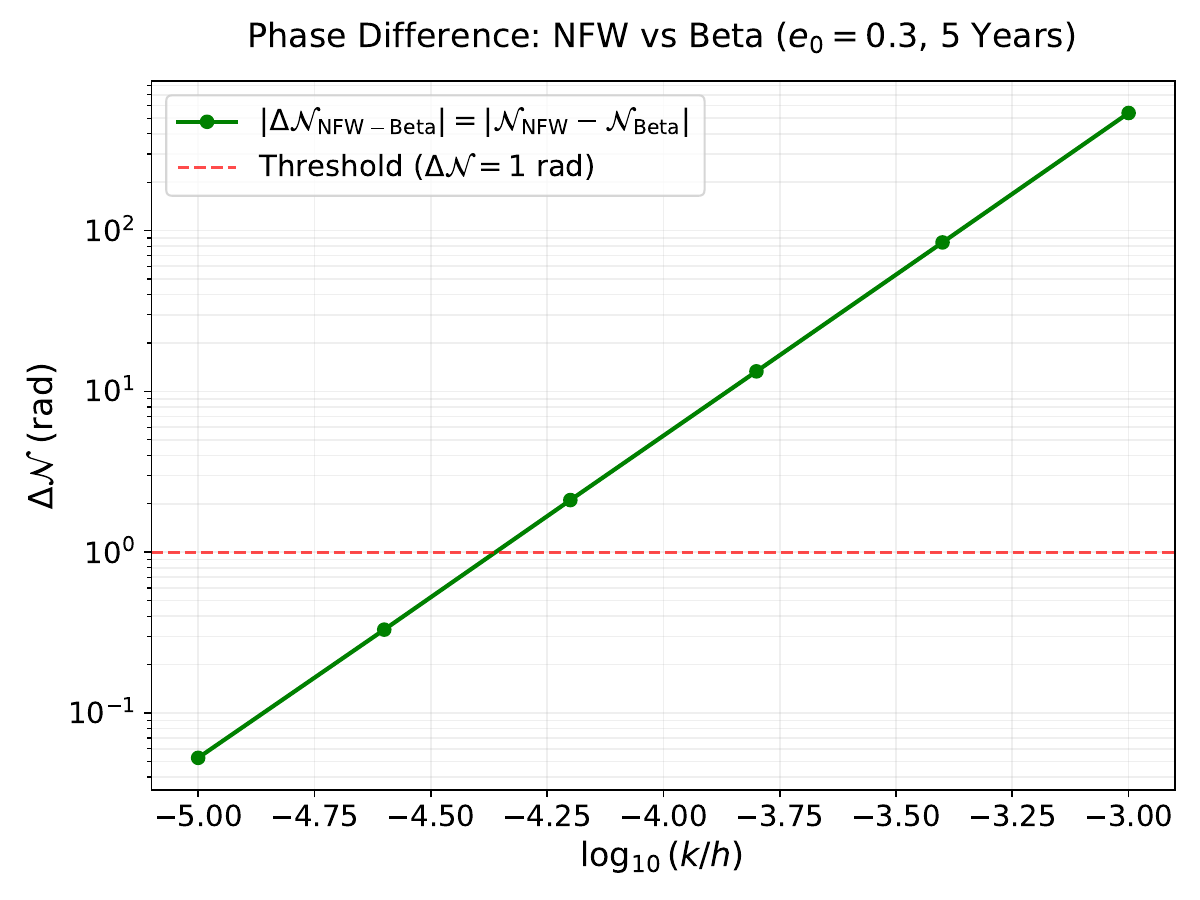}
		\label{fig:e02_5yr}
	\end{subfigure}
	
	
	\begin{subfigure}{0.48\textwidth}
		\includegraphics[width=\linewidth]{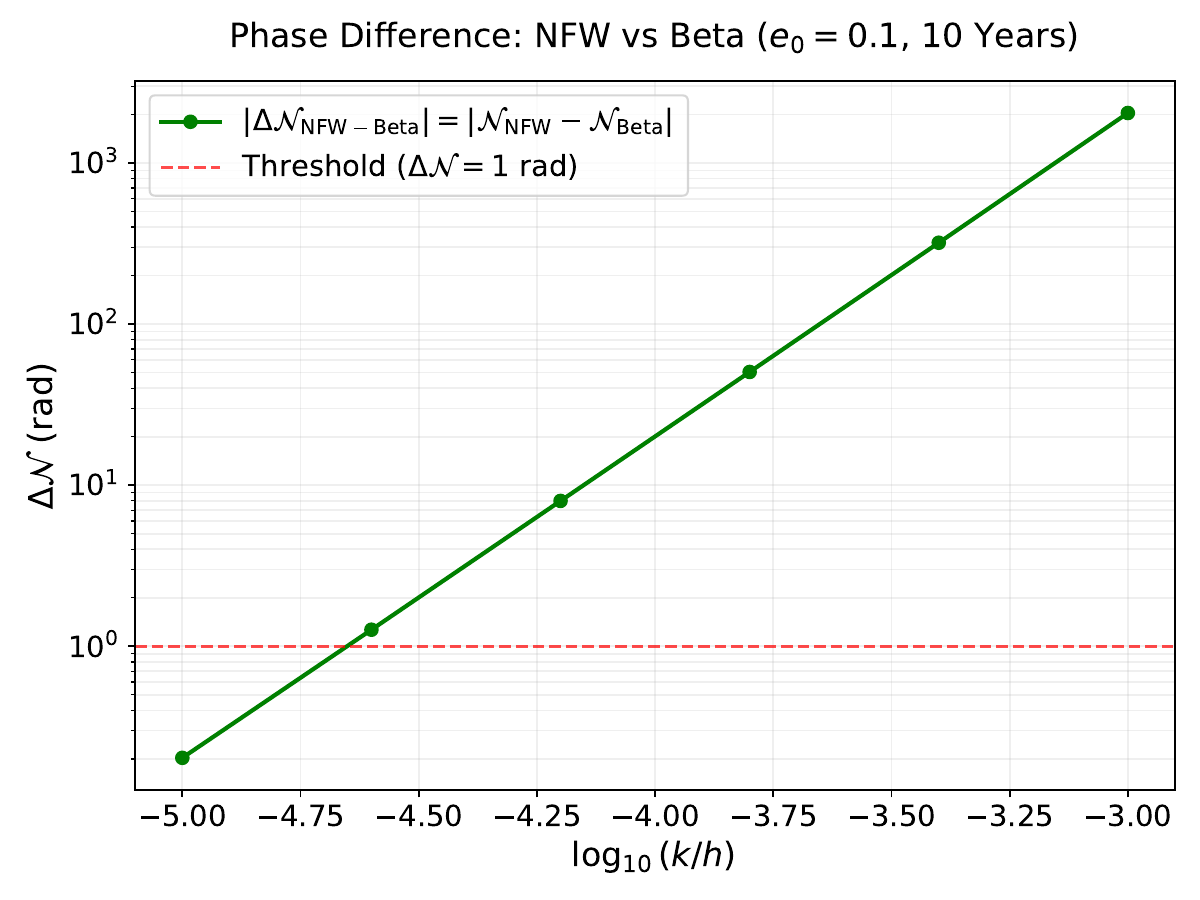}
		\label{fig:e01_10yr}
	\end{subfigure}
	\hfill
	\begin{subfigure}{0.48\textwidth}
		\includegraphics[width=\linewidth]{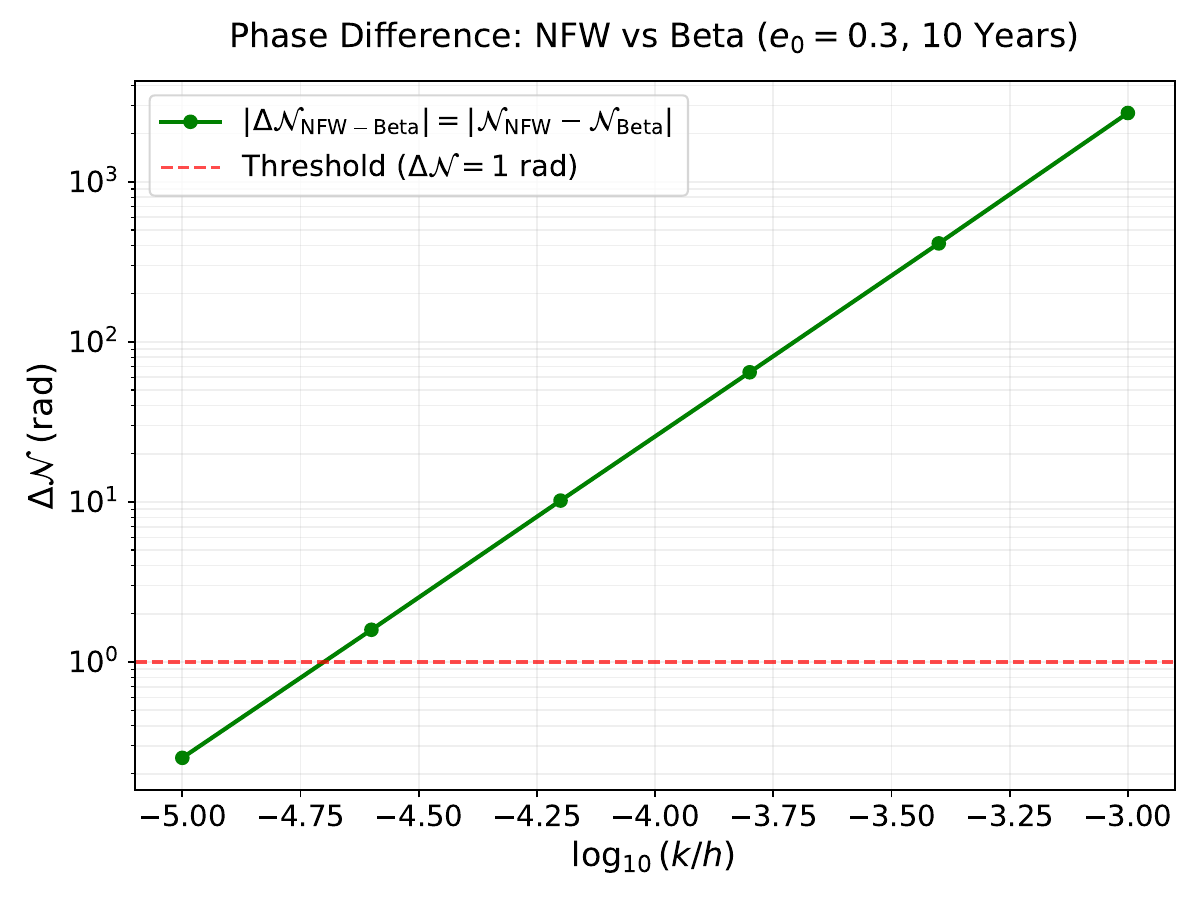} 
		\label{fig:e02_10yr}
	\end{subfigure}
	
	\caption{The absolute phase difference $|\Delta \mathcal{N}|$ between the NFW and Beta dark matter profiles as a function of the halo compactness $\log_{10}(k/h)$. The left and right columns illustrate the evolution for initial eccentricities of $e_0=0.1$ and $e_0=0.3$, respectively. From top to bottom, the panels display the accumulated dephasing over an observational span of 1, 5, and 10 years. The horizontal dashed line denotes the detectability threshold of $\Delta \mathcal{N} = 1$ rad. Throughout the simulations, the masses of the central black hole and the inspiraling compact object are fixed at $M=10^6 M_\odot$ and $\mu=10 M_\odot$, with the dark matter mass chosen as $k=100M$. The initial semi-latus rectum $p_0$ is determined by setting the initial orbital period to $T = 2\pi\sqrt{(20M)^3/M}$.} 
	\label{fig:phase_diff_all}
\end{figure*}

Fig.~\ref{fig:phase_vac} illustrates the dephasing of the NFW and Beta profiles relative to the vacuum case as a function of the halo compactness $\log_{10}(k/h)$. Both density profiles manifest a significant phase shift driven by dynamical friction and accretion. For both ground-based and space-based gravitational wave detectors, two signals are distinguishable by the detector if their phase difference exceeds a specific threshold. Specifically, we adopt a threshold of $\Delta \mathcal{N} \gtrsim 1$ rad as the requirement for a detectable deviation from the vacuum case~\cite{Maselli:2020zgv}. We find that the halo-induced dephasing becomes distinguishable from the vacuum background over a wide range of halo parameters. While the presence of a halo is readily detectable, distinguishing between different density profiles is more computationally and observationally demanding. As shown in Fig.~\ref{fig:phase_diff}, the absolute phase difference between the NFW and Beta models, $|\Delta \mathcal{N}_{\text{NFW-Beta}}| = |\mathcal{N}_{\text{NFW}} - \mathcal{N}_{\text{Beta}}|$, grows with increasing density but remains below the detectability threshold for a wider parameter range. Consequently, although the existence of a DM halo can be identified at a small halo compactness $k/h$ (with low dark matter densities), characterizing the specific profile requires a significantly denser environment. Although this dephasing $|\Delta \mathcal{N}_{\text{NFW-Beta}}|$ increases with $k/h$, it remains below $1$ rad for a larger portion of the parameter space compared to the vacuum-relative phase shift $|\mathcal{N}_{\text{DM}} - \mathcal{N}_{\text{vacuum}}|$. Specifically, the NFW and Beta models only become observationally distinct in dense environments where $log_{10}(k/h) \gtrsim -3.75$. This comparison suggests that while gravitational wave observations can easily discriminate a DM environment from a vacuum medium, higher halo densities are necessary to break the degeneracy between different density profiles.

To further explore the observational distinguishability between NFW and Beta profiles, Fig.~\ref{fig:phase_diff_all} presents the phase shift between these two models across different observation timescales and initial eccentricities. Consistent with expectations, a longer observation duration significantly amplifies the total dephasing. As the observation time increases from 1 to 10 years, the phase difference grows accordingly, shifting the critical point (where $\Delta N = 1$) toward much lower compactness values within the dark matter parameter space. Furthermore, a comparison between the left panels ($e_0 = 0.1$) and the right panels ($e_0 = 0.3$) shows that higher orbital eccentricity promotes a faster accumulation of phase deviation. Specifically, a highly eccentric orbit brings the compact object to a periapsis much closer to the central region. In this innermost area, the difference in dark matter density between the two models becomes significantly pronounced: as indicated by Eqs.~\eqref{rho_nfw} and \eqref{rho_beta}, the density profile of the NFW model exhibits a characteristic ``cusp" feature as $r \to 0$, whereas the Beta model does not possess this feature, instead presenting a flat core. By probing these regions where the dark matter density differs, highly eccentric orbits significantly accelerate the dephasing $|\Delta \mathcal{N}_{\text{NFW-Beta}}|$. Therefore, the combination of extended observation periods and high orbital eccentricities offers a highly favorable scenario for constraining the specific dark matter density distribution.

\section{Conclusion}
\label{sec5}

In this work, we have presented a systematic investigation of the orbital dynamics and gravitational wave signatures of Extreme Mass Ratio Inspirals (EMRIs) in galactic dark matter halos within the framework of general relativity. To effectively discriminate between the Navarro-Frenk-White (NFW) and Beta dark matter halo models via EMRIs, it is necessary to consider the long-term evolution of the SCO's parameters, gravitational waveforms, and accumulated dephasing in the presence of dissipative mechanisms. In addition to gravitational wave back-reaction, we also incorporate environmental dissipative effects, specifically dynamical friction and Bondi-Hoyle accretion of dark matter particles. By comparing the numerical results calculated for the NFW and Beta density profiles, we comprehensively explore the effects of the dark matter halo on orbital evolution and gravitational wave emission. Our results reported in this work would be potentially helpful for distinguishing different dark matter models using future space-based gravitational wave detectors.

In a galactic dark matter environment, the background spacetime geometry is significantly influenced by the gravitational potential of the dark matter halo, causing modifications to the spacetime metric through dark matter density profiles (predicted by NFW and Beta halo models). Our results indicate that the presence of a dark matter halo significantly influences the geodesic motion of test particles by altering the effective potential of the spacetime, inducing a noticeable cumulative deviation in the trajectory compared to the pure black hole case. Furthermore, we find that while the halo potential introduces a cumulative deviation from vacuum predictions, short-term observables—specifically orbital periods and precession angles—remain insensitive to the specific density distribution differences between the NFW and Beta models, even in high-density scenarios. Consequently, distinguishing these two density profiles via short-term dynamical observations remains challenging and insufficient. 

To efficiently probe the imprints from NFW and Beta models in EMRI systems, we further examined the long-term orbital evolution driven by dissipative mechanisms (dynamical friction, accretion, and gravitational radiation reaction). Our analysis reveals that incorporating these dissipative mechanisms can enhance the distinct environmental effects on EMRIs' dynamical evolution from different halo models. In the NFW profile, energy injection from accretion counteracts dissipation from gravitational waves and dynamical friction, bringing a distinct sharp cusp in the total energy flux at a specific semi-latus rectum. This cusp shifts toward a smaller semi-latus rectum region as the halo mass increases. In contrast, for the Beta model, environmental dissipation is orders of magnitude weaker than in the NFW case, and the orbital evolution is dominated by gravitational wave emission, lacking such distinct flux features because accretion is insufficient to trigger an energy injection regime. Importantly, the presence of a high-density halo (e.g., $k=20000M$) accelerates the inspiral, driving the evolutions of both the NFW and Beta models in the eccentricity-semilatus rectum ($e-p$) plane significantly away from the vacuum case. Regarding gravitational wave emission, we find that waveforms for NFW and Beta profiles remain nearly identical during the initial observation, even in massive halos ($k=20000M$) where they distinctly deviate from the vacuum baseline. However, cumulative phase differences driven by distinct dissipative dynamics gradually emerge over long-term adiabatic evolution, providing a reliable signature to distinguish these density models. Quantitatively, our dephasing analysis confirms that the presence of a dark matter halo is highly detectable relative to a vacuum background across a broad parameter space. The absolute phase difference between the NFW and Beta models generally exceeds the detection threshold of $\Delta\mathcal{N} \ge 1$ rad in high-density halo environments. Furthermore, we also demonstrated that extending the observation duration (e.g., from 1 year to 10 years) and targeting SCOs with higher orbital eccentricities (e.g., $e_0 = 0.3$) can significantly mitigate this limitation. By bringing the compact object closer to the galaxy center, these highly eccentric orbits efficiently probe the inner region where distinct density features (cusp vs flat) of different halo models become dominant, accelerating the dephasing $|\Delta \mathcal{N}_{\text{NFW-Beta}}|$ and lowering the required halo compactness for the detection threshold. This emphasizes the necessity of high-precision, long-duration observations to effectively probe the environmental effects on EMRIs' gravitational waveforms and phase shifts, constraining the density profiles in galactic dark matter halos.

In summary, this study highlights the critical role of dissipation mechanisms (dynamical friction, accretion, and gravitational radiation reaction) in distinguishing dark matter halo profiles in EMRI observations. Although the NFW and Beta models are indistinguishable through short-term geodesic orbital observations, their cumulative effects produce measurable differences in the long-term evolution of the SCO, especially in the energy fluxes caused by accretion and dynamical friction, as well as in the dephasing of gravitational waveforms. These results provide theoretical support and guidance for probing EMRI gravitational wave signatures in galactic dark matter environments, offering potential applications for next-generation space-based gravitational wave detectors, such as LISA, Taiji, and TianQin.

\begin{acknowledgments}
This research is supported by the Natural Science Foundation of China (Grant No. 12175212), Natural Science Foundation of Chongqing Municipality (Grant No. CSTB2022NSCQ-MSX0932), the Scientific and Technological Research Program of Chongqing Municipal Education Commission (Grant No. KJQN202201126).
\end{acknowledgments}

\nocite{*}
\bibliography{./ref}

\begin{thebibliography}{110}%
\makeatletter
\providecommand \@ifxundefined [1]{%
 \@ifx{#1\undefined}
}%
\providecommand \@ifnum [1]{%
 \ifnum #1\expandafter \@firstoftwo
 \else \expandafter \@secondoftwo
 \fi
}%
\providecommand \@ifx [1]{%
 \ifx #1\expandafter \@firstoftwo
 \else \expandafter \@secondoftwo
 \fi
}%
\providecommand \natexlab [1]{#1}%
\providecommand \enquote  [1]{``#1''}%
\providecommand \bibnamefont  [1]{#1}%
\providecommand \bibfnamefont [1]{#1}%
\providecommand \citenamefont [1]{#1}%
\providecommand \href@noop [0]{\@secondoftwo}%
\providecommand \href [0]{\begingroup \@sanitize@url \@href}%
\providecommand \@href[1]{\@@startlink{#1}\@@href}%
\providecommand \@@href[1]{\endgroup#1\@@endlink}%
\providecommand \@sanitize@url [0]{\catcode `\\12\catcode `\$12\catcode
  `\&12\catcode `\#12\catcode `\^12\catcode `\_12\catcode `\%12\relax}%
\providecommand \@@startlink[1]{}%
\providecommand \@@endlink[0]{}%
\providecommand \url  [0]{\begingroup\@sanitize@url \@url }%
\providecommand \@url [1]{\endgroup\@href {#1}{\urlprefix }}%
\providecommand \urlprefix  [0]{URL }%
\providecommand \Eprint [0]{\href }%
\providecommand \doibase [0]{https://doi.org/}%
\providecommand \selectlanguage [0]{\@gobble}%
\providecommand \bibinfo  [0]{\@secondoftwo}%
\providecommand \bibfield  [0]{\@secondoftwo}%
\providecommand \translation [1]{[#1]}%
\providecommand \BibitemOpen [0]{}%
\providecommand \bibitemStop [0]{}%
\providecommand \bibitemNoStop [0]{.\EOS\space}%
\providecommand \EOS [0]{\spacefactor3000\relax}%
\providecommand \BibitemShut  [1]{\csname bibitem#1\endcsname}%
\let\auto@bib@innerbib\@empty
\bibitem [{\citenamefont {Abbott}\ \emph
  {et~al.}(2016{\natexlab{a}})\citenamefont {Abbott} \emph
  {et~al.}}]{LIGOScientific:2016aoc}%
  \BibitemOpen
  \bibfield  {author} {\bibinfo {author} {\bibfnamefont {B.~P.}\ \bibnamefont
  {Abbott}} \emph {et~al.} (\bibinfo {collaboration} {LIGO Scientific,
  Virgo}),\ }\bibfield  {title} {\bibinfo {title} {{Observation of
  Gravitational Waves from a Binary Black Hole Merger}},\ }\href
  {https://doi.org/10.1103/PhysRevLett.116.061102} {\bibfield  {journal}
  {\bibinfo  {journal} {Phys. Rev. Lett.}\ }\textbf {\bibinfo {volume} {116}},\
  \bibinfo {pages} {061102} (\bibinfo {year} {2016}{\natexlab{a}})},\ \Eprint
  {https://arxiv.org/abs/1602.03837} {arXiv:1602.03837 [gr-qc]} \BibitemShut
  {NoStop}%
\bibitem [{\citenamefont {Abbott}\ \emph
  {et~al.}(2016{\natexlab{b}})\citenamefont {Abbott} \emph
  {et~al.}}]{LIGOScientific:2016emj}%
  \BibitemOpen
  \bibfield  {author} {\bibinfo {author} {\bibfnamefont {B.~P.}\ \bibnamefont
  {Abbott}} \emph {et~al.} (\bibinfo {collaboration} {LIGO Scientific,
  Virgo}),\ }\bibfield  {title} {\bibinfo {title} {{GW150914: The Advanced LIGO
  Detectors in the Era of First Discoveries}},\ }\href
  {https://doi.org/10.1103/PhysRevLett.116.131103} {\bibfield  {journal}
  {\bibinfo  {journal} {Phys. Rev. Lett.}\ }\textbf {\bibinfo {volume} {116}},\
  \bibinfo {pages} {131103} (\bibinfo {year} {2016}{\natexlab{b}})},\ \Eprint
  {https://arxiv.org/abs/1602.03838} {arXiv:1602.03838 [gr-qc]} \BibitemShut
  {NoStop}%
\bibitem [{\citenamefont {Abbott}\ \emph {et~al.}(2019)\citenamefont {Abbott}
  \emph {et~al.}}]{LIGOScientific:2018mvr}%
  \BibitemOpen
  \bibfield  {author} {\bibinfo {author} {\bibfnamefont {B.~P.}\ \bibnamefont
  {Abbott}} \emph {et~al.} (\bibinfo {collaboration} {LIGO Scientific,
  Virgo}),\ }\bibfield  {title} {\bibinfo {title} {{GWTC-1: A
  Gravitational-Wave Transient Catalog of Compact Binary Mergers Observed by
  LIGO and Virgo during the First and Second Observing Runs}},\ }\href
  {https://doi.org/10.1103/PhysRevX.9.031040} {\bibfield  {journal} {\bibinfo
  {journal} {Phys. Rev. X}\ }\textbf {\bibinfo {volume} {9}},\ \bibinfo {pages}
  {031040} (\bibinfo {year} {2019})},\ \Eprint
  {https://arxiv.org/abs/1811.12907} {arXiv:1811.12907 [astro-ph.HE]}
  \BibitemShut {NoStop}%
\bibitem [{\citenamefont {Abbott}\ \emph {et~al.}(2021)\citenamefont {Abbott}
  \emph {et~al.}}]{LIGOScientific:2020ibl}%
  \BibitemOpen
  \bibfield  {author} {\bibinfo {author} {\bibfnamefont {R.}~\bibnamefont
  {Abbott}} \emph {et~al.} (\bibinfo {collaboration} {LIGO Scientific,
  Virgo}),\ }\bibfield  {title} {\bibinfo {title} {{GWTC-2: Compact Binary
  Coalescences Observed by LIGO and Virgo During the First Half of the Third
  Observing Run}},\ }\href {https://doi.org/10.1103/PhysRevX.11.021053}
  {\bibfield  {journal} {\bibinfo  {journal} {Phys. Rev. X}\ }\textbf {\bibinfo
  {volume} {11}},\ \bibinfo {pages} {021053} (\bibinfo {year} {2021})},\
  \Eprint {https://arxiv.org/abs/2010.14527} {arXiv:2010.14527 [gr-qc]}
  \BibitemShut {NoStop}%
\bibitem [{\citenamefont {Abbott}\ \emph {et~al.}(2024)\citenamefont {Abbott}
  \emph {et~al.}}]{LIGOScientific:2021usb}%
  \BibitemOpen
  \bibfield  {author} {\bibinfo {author} {\bibfnamefont {R.}~\bibnamefont
  {Abbott}} \emph {et~al.} (\bibinfo {collaboration} {LIGO Scientific,
  VIRGO}),\ }\bibfield  {title} {\bibinfo {title} {{GWTC-2.1: Deep extended
  catalog of compact binary coalescences observed by LIGO and Virgo during the
  first half of the third observing run}},\ }\href
  {https://doi.org/10.1103/PhysRevD.109.022001} {\bibfield  {journal} {\bibinfo
   {journal} {Phys. Rev. D}\ }\textbf {\bibinfo {volume} {109}},\ \bibinfo
  {pages} {022001} (\bibinfo {year} {2024})},\ \Eprint
  {https://arxiv.org/abs/2108.01045} {arXiv:2108.01045 [gr-qc]} \BibitemShut
  {NoStop}%
\bibitem [{\citenamefont {Abbott}\ \emph {et~al.}(2023)\citenamefont {Abbott}
  \emph {et~al.}}]{KAGRA:2021vkt}%
  \BibitemOpen
  \bibfield  {author} {\bibinfo {author} {\bibfnamefont {R.}~\bibnamefont
  {Abbott}} \emph {et~al.} (\bibinfo {collaboration} {KAGRA, VIRGO, LIGO
  Scientific}),\ }\bibfield  {title} {\bibinfo {title} {{GWTC-3: Compact Binary
  Coalescences Observed by LIGO and Virgo during the Second Part of the Third
  Observing Run}},\ }\href {https://doi.org/10.1103/PhysRevX.13.041039}
  {\bibfield  {journal} {\bibinfo  {journal} {Phys. Rev. X}\ }\textbf {\bibinfo
  {volume} {13}},\ \bibinfo {pages} {041039} (\bibinfo {year} {2023})},\
  \Eprint {https://arxiv.org/abs/2111.03606} {arXiv:2111.03606 [gr-qc]}
  \BibitemShut {NoStop}%
\bibitem [{\citenamefont {Baibhav}\ \emph {et~al.}(2021)\citenamefont {Baibhav}
  \emph {et~al.}}]{Baibhav:2019rsa}%
  \BibitemOpen
  \bibfield  {author} {\bibinfo {author} {\bibfnamefont {V.}~\bibnamefont
  {Baibhav}} \emph {et~al.},\ }\bibfield  {title} {\bibinfo {title} {{Probing
  the nature of black holes: Deep in the mHz gravitational-wave sky}},\ }\href
  {https://doi.org/10.1007/s10686-021-09741-9} {\bibfield  {journal} {\bibinfo
  {journal} {Exper. Astron.}\ }\textbf {\bibinfo {volume} {51}},\ \bibinfo
  {pages} {1385} (\bibinfo {year} {2021})},\ \Eprint
  {https://arxiv.org/abs/1908.11390} {arXiv:1908.11390 [astro-ph.HE]}
  \BibitemShut {NoStop}%
\bibitem [{\citenamefont {Seoane}\ \emph {et~al.}(2023)\citenamefont {Seoane}
  \emph {et~al.}}]{LISA:2022yao}%
  \BibitemOpen
  \bibfield  {author} {\bibinfo {author} {\bibfnamefont {P.~A.}\ \bibnamefont
  {Seoane}} \emph {et~al.} (\bibinfo {collaboration} {LISA}),\ }\bibfield
  {title} {\bibinfo {title} {{Astrophysics with the Laser Interferometer Space
  Antenna}},\ }\href {https://doi.org/10.1007/s41114-022-00041-y} {\bibfield
  {journal} {\bibinfo  {journal} {Living Rev. Rel.}\ }\textbf {\bibinfo
  {volume} {26}},\ \bibinfo {pages} {2} (\bibinfo {year} {2023})},\ \Eprint
  {https://arxiv.org/abs/2203.06016} {arXiv:2203.06016 [gr-qc]} \BibitemShut
  {NoStop}%
\bibitem [{\citenamefont {Arun}\ \emph {et~al.}(2022)\citenamefont {Arun} \emph
  {et~al.}}]{LISA:2022kgy}%
  \BibitemOpen
  \bibfield  {author} {\bibinfo {author} {\bibfnamefont {K.~G.}\ \bibnamefont
  {Arun}} \emph {et~al.} (\bibinfo {collaboration} {LISA}),\ }\bibfield
  {title} {\bibinfo {title} {{New horizons for fundamental physics with
  LISA}},\ }\href {https://doi.org/10.1007/s41114-022-00036-9} {\bibfield
  {journal} {\bibinfo  {journal} {Living Rev. Rel.}\ }\textbf {\bibinfo
  {volume} {25}},\ \bibinfo {pages} {4} (\bibinfo {year} {2022})},\ \Eprint
  {https://arxiv.org/abs/2205.01597} {arXiv:2205.01597 [gr-qc]} \BibitemShut
  {NoStop}%
\bibitem [{\citenamefont {Karnesis}\ \emph {et~al.}(2024)\citenamefont
  {Karnesis} \emph {et~al.}}]{Karnesis:2022vdp}%
  \BibitemOpen
  \bibfield  {author} {\bibinfo {author} {\bibfnamefont {N.}~\bibnamefont
  {Karnesis}} \emph {et~al.},\ }\bibfield  {title} {\bibinfo {title} {{The
  Laser Interferometer Space Antenna mission in Greece White Paper}},\ }\href
  {https://doi.org/10.1142/S0218271824500275} {\bibfield  {journal} {\bibinfo
  {journal} {Int. J. Mod. Phys. D}\ }\textbf {\bibinfo {volume} {33}},\
  \bibinfo {pages} {2450027} (\bibinfo {year} {2024})},\ \Eprint
  {https://arxiv.org/abs/2209.04358} {arXiv:2209.04358 [gr-qc]} \BibitemShut
  {NoStop}%
\bibitem [{\citenamefont {Amaro-Seoane}\ \emph {et~al.}(2017)\citenamefont
  {Amaro-Seoane} \emph {et~al.}}]{LISA:2017pwj}%
  \BibitemOpen
  \bibfield  {author} {\bibinfo {author} {\bibfnamefont {P.}~\bibnamefont
  {Amaro-Seoane}} \emph {et~al.} (\bibinfo {collaboration} {LISA}),\ }\bibfield
   {title} {\bibinfo {title} {{Laser Interferometer Space Antenna}},\
  }\href@noop {} {\  (\bibinfo {year} {2017})},\ \Eprint
  {https://arxiv.org/abs/1702.00786} {arXiv:1702.00786 [astro-ph.IM]}
  \BibitemShut {NoStop}%
\bibitem [{\citenamefont {Hu}\ and\ \citenamefont {Wu}(2017)}]{Hu:2017mde}%
  \BibitemOpen
  \bibfield  {author} {\bibinfo {author} {\bibfnamefont {W.-R.}\ \bibnamefont
  {Hu}}\ and\ \bibinfo {author} {\bibfnamefont {Y.-L.}\ \bibnamefont {Wu}},\
  }\bibfield  {title} {\bibinfo {title} {{The Taiji Program in Space for
  gravitational wave physics and the nature of gravity}},\ }\href
  {https://doi.org/10.1093/nsr/nwx116} {\bibfield  {journal} {\bibinfo
  {journal} {Natl. Sci. Rev.}\ }\textbf {\bibinfo {volume} {4}},\ \bibinfo
  {pages} {685} (\bibinfo {year} {2017})}\BibitemShut {NoStop}%
\bibitem [{\citenamefont {Gong}\ \emph {et~al.}(2021)\citenamefont {Gong},
  \citenamefont {Luo},\ and\ \citenamefont {Wang}}]{Gong:2021gvw}%
  \BibitemOpen
  \bibfield  {author} {\bibinfo {author} {\bibfnamefont {Y.}~\bibnamefont
  {Gong}}, \bibinfo {author} {\bibfnamefont {J.}~\bibnamefont {Luo}},\ and\
  \bibinfo {author} {\bibfnamefont {B.}~\bibnamefont {Wang}},\ }\bibfield
  {title} {\bibinfo {title} {{Concepts and status of Chinese space
  gravitational wave detection projects}},\ }\href
  {https://doi.org/10.1038/s41550-021-01480-3} {\bibfield  {journal} {\bibinfo
  {journal} {Nature Astron.}\ }\textbf {\bibinfo {volume} {5}},\ \bibinfo
  {pages} {881} (\bibinfo {year} {2021})},\ \Eprint
  {https://arxiv.org/abs/2109.07442} {arXiv:2109.07442 [astro-ph.IM]}
  \BibitemShut {NoStop}%
\bibitem [{\citenamefont {Luo}\ \emph {et~al.}(2016)\citenamefont {Luo} \emph
  {et~al.}}]{TianQin:2015yph}%
  \BibitemOpen
  \bibfield  {author} {\bibinfo {author} {\bibfnamefont {J.}~\bibnamefont
  {Luo}} \emph {et~al.} (\bibinfo {collaboration} {TianQin}),\ }\bibfield
  {title} {\bibinfo {title} {{TianQin: a space-borne gravitational wave
  detector}},\ }\href {https://doi.org/10.1088/0264-9381/33/3/035010}
  {\bibfield  {journal} {\bibinfo  {journal} {Class. Quant. Grav.}\ }\textbf
  {\bibinfo {volume} {33}},\ \bibinfo {pages} {035010} (\bibinfo {year}
  {2016})},\ \Eprint {https://arxiv.org/abs/1512.02076} {arXiv:1512.02076
  [astro-ph.IM]} \BibitemShut {NoStop}%
\bibitem [{\citenamefont {Liu}\ \emph {et~al.}(2020)\citenamefont {Liu},
  \citenamefont {Hu}, \citenamefont {Zhang},\ and\ \citenamefont
  {Mei}}]{Liu:2020eko}%
  \BibitemOpen
  \bibfield  {author} {\bibinfo {author} {\bibfnamefont {S.}~\bibnamefont
  {Liu}}, \bibinfo {author} {\bibfnamefont {Y.-M.}\ \bibnamefont {Hu}},
  \bibinfo {author} {\bibfnamefont {J.-d.}\ \bibnamefont {Zhang}},\ and\
  \bibinfo {author} {\bibfnamefont {J.}~\bibnamefont {Mei}},\ }\bibfield
  {title} {\bibinfo {title} {{Science with the TianQin observatory: Preliminary
  results on stellar-mass binary black holes}},\ }\href
  {https://doi.org/10.1103/PhysRevD.101.103027} {\bibfield  {journal} {\bibinfo
   {journal} {Phys. Rev. D}\ }\textbf {\bibinfo {volume} {101}},\ \bibinfo
  {pages} {103027} (\bibinfo {year} {2020})},\ \Eprint
  {https://arxiv.org/abs/2004.14242} {arXiv:2004.14242 [astro-ph.HE]}
  \BibitemShut {NoStop}%
\bibitem [{\citenamefont {Babak}\ \emph {et~al.}(2017)\citenamefont {Babak},
  \citenamefont {Gair}, \citenamefont {Sesana}, \citenamefont {Barausse},
  \citenamefont {Sopuerta}, \citenamefont {Berry}, \citenamefont {Berti},
  \citenamefont {Amaro-Seoane}, \citenamefont {Petiteau},\ and\ \citenamefont
  {Klein}}]{Babak:2017tow}%
  \BibitemOpen
  \bibfield  {author} {\bibinfo {author} {\bibfnamefont {S.}~\bibnamefont
  {Babak}}, \bibinfo {author} {\bibfnamefont {J.}~\bibnamefont {Gair}},
  \bibinfo {author} {\bibfnamefont {A.}~\bibnamefont {Sesana}}, \bibinfo
  {author} {\bibfnamefont {E.}~\bibnamefont {Barausse}}, \bibinfo {author}
  {\bibfnamefont {C.~F.}\ \bibnamefont {Sopuerta}}, \bibinfo {author}
  {\bibfnamefont {C.~P.~L.}\ \bibnamefont {Berry}}, \bibinfo {author}
  {\bibfnamefont {E.}~\bibnamefont {Berti}}, \bibinfo {author} {\bibfnamefont
  {P.}~\bibnamefont {Amaro-Seoane}}, \bibinfo {author} {\bibfnamefont
  {A.}~\bibnamefont {Petiteau}},\ and\ \bibinfo {author} {\bibfnamefont
  {A.}~\bibnamefont {Klein}},\ }\bibfield  {title} {\bibinfo {title} {{Science
  with the space-based interferometer LISA. V: Extreme mass-ratio inspirals}},\
  }\href {https://doi.org/10.1103/PhysRevD.95.103012} {\bibfield  {journal}
  {\bibinfo  {journal} {Phys. Rev. D}\ }\textbf {\bibinfo {volume} {95}},\
  \bibinfo {pages} {103012} (\bibinfo {year} {2017})},\ \Eprint
  {https://arxiv.org/abs/1703.09722} {arXiv:1703.09722 [gr-qc]} \BibitemShut
  {NoStop}%
\bibitem [{\citenamefont {Amaro-Seoane}\ \emph {et~al.}(2007)\citenamefont
  {Amaro-Seoane}, \citenamefont {Gair}, \citenamefont {Freitag}, \citenamefont
  {Coleman~Miller}, \citenamefont {Mandel}, \citenamefont {Cutler},\ and\
  \citenamefont {Babak}}]{Amaro-Seoane:2007osp}%
  \BibitemOpen
  \bibfield  {author} {\bibinfo {author} {\bibfnamefont {P.}~\bibnamefont
  {Amaro-Seoane}}, \bibinfo {author} {\bibfnamefont {J.~R.}\ \bibnamefont
  {Gair}}, \bibinfo {author} {\bibfnamefont {M.}~\bibnamefont {Freitag}},
  \bibinfo {author} {\bibfnamefont {M.}~\bibnamefont {Coleman~Miller}},
  \bibinfo {author} {\bibfnamefont {I.}~\bibnamefont {Mandel}}, \bibinfo
  {author} {\bibfnamefont {C.~J.}\ \bibnamefont {Cutler}},\ and\ \bibinfo
  {author} {\bibfnamefont {S.}~\bibnamefont {Babak}},\ }\bibfield  {title}
  {\bibinfo {title} {{Astrophysics, detection and science applications of
  intermediate- and extreme mass-ratio inspirals}},\ }\href
  {https://doi.org/10.1088/0264-9381/24/17/R01} {\bibfield  {journal} {\bibinfo
   {journal} {Class. Quant. Grav.}\ }\textbf {\bibinfo {volume} {24}},\
  \bibinfo {pages} {R113} (\bibinfo {year} {2007})},\ \Eprint
  {https://arxiv.org/abs/astro-ph/0703495} {arXiv:astro-ph/0703495}
  \BibitemShut {NoStop}%
\bibitem [{\citenamefont {Berry}\ \emph {et~al.}(2019)\citenamefont {Berry},
  \citenamefont {Hughes}, \citenamefont {Sopuerta}, \citenamefont {Chua},
  \citenamefont {Heffernan}, \citenamefont {Holley-Bockelmann}, \citenamefont
  {Mihaylov}, \citenamefont {Miller},\ and\ \citenamefont
  {Sesana}}]{Berry:2019wgg}%
  \BibitemOpen
  \bibfield  {author} {\bibinfo {author} {\bibfnamefont {C.~P.~L.}\
  \bibnamefont {Berry}}, \bibinfo {author} {\bibfnamefont {S.~A.}\ \bibnamefont
  {Hughes}}, \bibinfo {author} {\bibfnamefont {C.~F.}\ \bibnamefont
  {Sopuerta}}, \bibinfo {author} {\bibfnamefont {A.~J.~K.}\ \bibnamefont
  {Chua}}, \bibinfo {author} {\bibfnamefont {A.}~\bibnamefont {Heffernan}},
  \bibinfo {author} {\bibfnamefont {K.}~\bibnamefont {Holley-Bockelmann}},
  \bibinfo {author} {\bibfnamefont {D.~P.}\ \bibnamefont {Mihaylov}}, \bibinfo
  {author} {\bibfnamefont {M.~C.}\ \bibnamefont {Miller}},\ and\ \bibinfo
  {author} {\bibfnamefont {A.}~\bibnamefont {Sesana}},\ }\bibfield  {title}
  {\bibinfo {title} {{The unique potential of extreme mass-ratio inspirals for
  gravitational-wave astronomy}},\ }\href@noop {} {\bibfield  {journal}
  {\bibinfo  {journal} {Bull. Am. Astron. Soc.}\ }\textbf {\bibinfo {volume}
  {51}},\ \bibinfo {pages} {42} (\bibinfo {year} {2019})},\ \Eprint
  {https://arxiv.org/abs/1903.03686} {arXiv:1903.03686 [astro-ph.HE]}
  \BibitemShut {NoStop}%
\bibitem [{\citenamefont {Seoane}\ \emph {et~al.}(2022)\citenamefont {Seoane}
  \emph {et~al.}}]{Seoane:2021kkk}%
  \BibitemOpen
  \bibfield  {author} {\bibinfo {author} {\bibfnamefont {P.~A.}\ \bibnamefont
  {Seoane}} \emph {et~al.},\ }\bibfield  {title} {\bibinfo {title} {{The effect
  of mission duration on LISA science objectives}},\ }\href
  {https://doi.org/10.1007/s10714-021-02889-x} {\bibfield  {journal} {\bibinfo
  {journal} {Gen. Rel. Grav.}\ }\textbf {\bibinfo {volume} {54}},\ \bibinfo
  {pages} {3} (\bibinfo {year} {2022})},\ \Eprint
  {https://arxiv.org/abs/2107.09665} {arXiv:2107.09665 [astro-ph.IM]}
  \BibitemShut {NoStop}%
\bibitem [{\citenamefont {Laghi}\ \emph {et~al.}(2021)\citenamefont {Laghi},
  \citenamefont {Tamanini}, \citenamefont {Del~Pozzo}, \citenamefont {Sesana},
  \citenamefont {Gair}, \citenamefont {Babak},\ and\ \citenamefont
  {Izquierdo-Villalba}}]{Laghi:2021pqk}%
  \BibitemOpen
  \bibfield  {author} {\bibinfo {author} {\bibfnamefont {D.}~\bibnamefont
  {Laghi}}, \bibinfo {author} {\bibfnamefont {N.}~\bibnamefont {Tamanini}},
  \bibinfo {author} {\bibfnamefont {W.}~\bibnamefont {Del~Pozzo}}, \bibinfo
  {author} {\bibfnamefont {A.}~\bibnamefont {Sesana}}, \bibinfo {author}
  {\bibfnamefont {J.}~\bibnamefont {Gair}}, \bibinfo {author} {\bibfnamefont
  {S.}~\bibnamefont {Babak}},\ and\ \bibinfo {author} {\bibfnamefont
  {D.}~\bibnamefont {Izquierdo-Villalba}},\ }\bibfield  {title} {\bibinfo
  {title} {{Gravitational-wave cosmology with extreme mass-ratio inspirals}},\
  }\href {https://doi.org/10.1093/mnras/stab2741} {\bibfield  {journal}
  {\bibinfo  {journal} {Mon. Not. Roy. Astron. Soc.}\ }\textbf {\bibinfo
  {volume} {508}},\ \bibinfo {pages} {4512} (\bibinfo {year} {2021})},\ \Eprint
  {https://arxiv.org/abs/2102.01708} {arXiv:2102.01708 [astro-ph.CO]}
  \BibitemShut {NoStop}%
\bibitem [{\citenamefont {McGee}\ \emph {et~al.}(2020)\citenamefont {McGee},
  \citenamefont {Sesana},\ and\ \citenamefont {Vecchio}}]{McGee:2018qwb}%
  \BibitemOpen
  \bibfield  {author} {\bibinfo {author} {\bibfnamefont {S.}~\bibnamefont
  {McGee}}, \bibinfo {author} {\bibfnamefont {A.}~\bibnamefont {Sesana}},\ and\
  \bibinfo {author} {\bibfnamefont {A.}~\bibnamefont {Vecchio}},\ }\bibfield
  {title} {\bibinfo {title} {{Linking gravitational waves and X-ray phenomena
  with joint LISA and Athena observations}},\ }\href
  {https://doi.org/10.1038/s41550-019-0969-7} {\bibfield  {journal} {\bibinfo
  {journal} {Nature Astron.}\ }\textbf {\bibinfo {volume} {4}},\ \bibinfo
  {pages} {26} (\bibinfo {year} {2020})},\ \Eprint
  {https://arxiv.org/abs/1811.00050} {arXiv:1811.00050 [astro-ph.HE]}
  \BibitemShut {NoStop}%
\bibitem [{\citenamefont {Barausse}\ \emph {et~al.}(2014)\citenamefont
  {Barausse}, \citenamefont {Cardoso},\ and\ \citenamefont
  {Pani}}]{Barausse:2014tra}%
  \BibitemOpen
  \bibfield  {author} {\bibinfo {author} {\bibfnamefont {E.}~\bibnamefont
  {Barausse}}, \bibinfo {author} {\bibfnamefont {V.}~\bibnamefont {Cardoso}},\
  and\ \bibinfo {author} {\bibfnamefont {P.}~\bibnamefont {Pani}},\ }\bibfield
  {title} {\bibinfo {title} {{Can environmental effects spoil precision
  gravitational-wave astrophysics?}},\ }\href
  {https://doi.org/10.1103/PhysRevD.89.104059} {\bibfield  {journal} {\bibinfo
  {journal} {Phys. Rev. D}\ }\textbf {\bibinfo {volume} {89}},\ \bibinfo
  {pages} {104059} (\bibinfo {year} {2014})},\ \Eprint
  {https://arxiv.org/abs/1404.7149} {arXiv:1404.7149 [gr-qc]} \BibitemShut
  {NoStop}%
\bibitem [{\citenamefont {Cardoso}\ \emph
  {et~al.}(2022{\natexlab{a}})\citenamefont {Cardoso}, \citenamefont
  {Destounis}, \citenamefont {Duque}, \citenamefont {Panosso~Macedo},\ and\
  \citenamefont {Maselli}}]{Cardoso:2022whc}%
  \BibitemOpen
  \bibfield  {author} {\bibinfo {author} {\bibfnamefont {V.}~\bibnamefont
  {Cardoso}}, \bibinfo {author} {\bibfnamefont {K.}~\bibnamefont {Destounis}},
  \bibinfo {author} {\bibfnamefont {F.}~\bibnamefont {Duque}}, \bibinfo
  {author} {\bibfnamefont {R.}~\bibnamefont {Panosso~Macedo}},\ and\ \bibinfo
  {author} {\bibfnamefont {A.}~\bibnamefont {Maselli}},\ }\bibfield  {title}
  {\bibinfo {title} {{Gravitational Waves from Extreme-Mass-Ratio Systems in
  Astrophysical Environments}},\ }\href
  {https://doi.org/10.1103/PhysRevLett.129.241103} {\bibfield  {journal}
  {\bibinfo  {journal} {Phys. Rev. Lett.}\ }\textbf {\bibinfo {volume} {129}},\
  \bibinfo {pages} {241103} (\bibinfo {year} {2022}{\natexlab{a}})},\ \Eprint
  {https://arxiv.org/abs/2210.01133} {arXiv:2210.01133 [gr-qc]} \BibitemShut
  {NoStop}%
\bibitem [{\citenamefont {Zi}\ and\ \citenamefont
  {Kumar}(2025{\natexlab{a}})}]{Zi:2025lio}%
  \BibitemOpen
  \bibfield  {author} {\bibinfo {author} {\bibfnamefont {T.}~\bibnamefont
  {Zi}}\ and\ \bibinfo {author} {\bibfnamefont {S.}~\bibnamefont {Kumar}},\
  }\bibfield  {title} {\bibinfo {title} {{Probing scalar field with generic
  extreme mass-ratio inspirals around Kerr black holes}},\ }\href@noop {} {\
  (\bibinfo {year} {2025}{\natexlab{a}})},\ \Eprint
  {https://arxiv.org/abs/2508.00516} {arXiv:2508.00516 [gr-qc]} \BibitemShut
  {NoStop}%
\bibitem [{\citenamefont {Yang}\ \emph {et~al.}(2025)\citenamefont {Yang},
  \citenamefont {Zhang}, \citenamefont {Zhu}, \citenamefont {Zhao},\ and\
  \citenamefont {Liu}}]{Yang:2024lmj}%
  \BibitemOpen
  \bibfield  {author} {\bibinfo {author} {\bibfnamefont {S.}~\bibnamefont
  {Yang}}, \bibinfo {author} {\bibfnamefont {Y.-P.}\ \bibnamefont {Zhang}},
  \bibinfo {author} {\bibfnamefont {T.}~\bibnamefont {Zhu}}, \bibinfo {author}
  {\bibfnamefont {L.}~\bibnamefont {Zhao}},\ and\ \bibinfo {author}
  {\bibfnamefont {Y.-X.}\ \bibnamefont {Liu}},\ }\bibfield  {title} {\bibinfo
  {title} {{Gravitational waveforms from periodic orbits around a
  quantum-corrected black hole}},\ }\href
  {https://doi.org/10.1088/1475-7516/2025/01/091} {\bibfield  {journal}
  {\bibinfo  {journal} {JCAP}\ }\textbf {\bibinfo {volume} {01}},\ \bibinfo
  {pages} {091}},\ \Eprint {https://arxiv.org/abs/2407.00283} {arXiv:2407.00283
  [gr-qc]} \BibitemShut {NoStop}%
\bibitem [{\citenamefont {Zi}\ and\ \citenamefont
  {Kumar}(2025{\natexlab{b}})}]{Zi:2024jla}%
  \BibitemOpen
  \bibfield  {author} {\bibinfo {author} {\bibfnamefont {T.}~\bibnamefont
  {Zi}}\ and\ \bibinfo {author} {\bibfnamefont {S.}~\bibnamefont {Kumar}},\
  }\bibfield  {title} {\bibinfo {title} {{Eccentric extreme mass-ratio
  inspirals: a gateway to probe quantum gravity effects}},\ }\href
  {https://doi.org/10.1140/epjc/s10052-025-14330-7} {\bibfield  {journal}
  {\bibinfo  {journal} {Eur. Phys. J. C}\ }\textbf {\bibinfo {volume} {85}},\
  \bibinfo {pages} {592} (\bibinfo {year} {2025}{\natexlab{b}})},\ \Eprint
  {https://arxiv.org/abs/2409.17765} {arXiv:2409.17765 [gr-qc]} \BibitemShut
  {NoStop}%
\bibitem [{\citenamefont {Kumar}\ \emph {et~al.}(2025)\citenamefont {Kumar},
  \citenamefont {Zi},\ and\ \citenamefont {Bhattacharyya}}]{Kumar:2025jsi}%
  \BibitemOpen
  \bibfield  {author} {\bibinfo {author} {\bibfnamefont {S.}~\bibnamefont
  {Kumar}}, \bibinfo {author} {\bibfnamefont {T.}~\bibnamefont {Zi}},\ and\
  \bibinfo {author} {\bibfnamefont {A.}~\bibnamefont {Bhattacharyya}},\
  }\bibfield  {title} {\bibinfo {title} {{Extreme mass-ratio inspirals and
  extra dimensions: Insights from modified Teukolsky framework}},\ }\href@noop
  {} {\  (\bibinfo {year} {2025})},\ \Eprint {https://arxiv.org/abs/2507.03380}
  {arXiv:2507.03380 [gr-qc]} \BibitemShut {NoStop}%
\bibitem [{\citenamefont {Battista}\ and\ \citenamefont
  {De~Falco}(2021)}]{Battista:2021rlh}%
  \BibitemOpen
  \bibfield  {author} {\bibinfo {author} {\bibfnamefont {E.}~\bibnamefont
  {Battista}}\ and\ \bibinfo {author} {\bibfnamefont {V.}~\bibnamefont
  {De~Falco}},\ }\bibfield  {title} {\bibinfo {title} {{First post-Newtonian
  generation of gravitational waves in Einstein-Cartan theory}},\ }\href
  {https://doi.org/10.1103/PhysRevD.104.084067} {\bibfield  {journal} {\bibinfo
   {journal} {Phys. Rev. D}\ }\textbf {\bibinfo {volume} {104}},\ \bibinfo
  {pages} {084067} (\bibinfo {year} {2021})},\ \Eprint
  {https://arxiv.org/abs/2109.01384} {arXiv:2109.01384 [gr-qc]} \BibitemShut
  {NoStop}%
\bibitem [{\citenamefont {Ade}\ \emph {et~al.}(2014)\citenamefont {Ade} \emph
  {et~al.}}]{Planck:2013pxb}%
  \BibitemOpen
  \bibfield  {author} {\bibinfo {author} {\bibfnamefont {P.~A.~R.}\
  \bibnamefont {Ade}} \emph {et~al.} (\bibinfo {collaboration} {Planck}),\
  }\bibfield  {title} {\bibinfo {title} {{Planck 2013 results. XVI.
  Cosmological parameters}},\ }\href
  {https://doi.org/10.1051/0004-6361/201321591} {\bibfield  {journal} {\bibinfo
   {journal} {Astron. Astrophys.}\ }\textbf {\bibinfo {volume} {571}},\
  \bibinfo {pages} {A16} (\bibinfo {year} {2014})},\ \Eprint
  {https://arxiv.org/abs/1303.5076} {arXiv:1303.5076 [astro-ph.CO]}
  \BibitemShut {NoStop}%
\bibitem [{\citenamefont {Rubin}\ and\ \citenamefont
  {Ford}(1970)}]{Rubin:1970zza}%
  \BibitemOpen
  \bibfield  {author} {\bibinfo {author} {\bibfnamefont {V.~C.}\ \bibnamefont
  {Rubin}}\ and\ \bibinfo {author} {\bibfnamefont {W.~K.}\ \bibnamefont {Ford},
  \bibfnamefont {Jr.}},\ }\bibfield  {title} {\bibinfo {title} {{Rotation of
  the Andromeda Nebula from a Spectroscopic Survey of Emission Regions}},\
  }\href {https://doi.org/10.1086/150317} {\bibfield  {journal} {\bibinfo
  {journal} {Astrophys. J.}\ }\textbf {\bibinfo {volume} {159}},\ \bibinfo
  {pages} {379} (\bibinfo {year} {1970})}\BibitemShut {NoStop}%
\bibitem [{\citenamefont {Corbelli}\ and\ \citenamefont
  {Salucci}(2000)}]{Corbelli:1999af}%
  \BibitemOpen
  \bibfield  {author} {\bibinfo {author} {\bibfnamefont {E.}~\bibnamefont
  {Corbelli}}\ and\ \bibinfo {author} {\bibfnamefont {P.}~\bibnamefont
  {Salucci}},\ }\bibfield  {title} {\bibinfo {title} {{The Extended Rotation
  Curve and the Dark Matter Halo of M33}},\ }\href
  {https://doi.org/10.1046/j.1365-8711.2000.03075.x} {\bibfield  {journal}
  {\bibinfo  {journal} {Mon. Not. Roy. Astron. Soc.}\ }\textbf {\bibinfo
  {volume} {311}},\ \bibinfo {pages} {441} (\bibinfo {year} {2000})},\ \Eprint
  {https://arxiv.org/abs/astro-ph/9909252} {arXiv:astro-ph/9909252}
  \BibitemShut {NoStop}%
\bibitem [{\citenamefont {Cooray}\ and\ \citenamefont
  {Sheth}(2002)}]{Cooray:2002dia}%
  \BibitemOpen
  \bibfield  {author} {\bibinfo {author} {\bibfnamefont {A.}~\bibnamefont
  {Cooray}}\ and\ \bibinfo {author} {\bibfnamefont {R.~K.}\ \bibnamefont
  {Sheth}},\ }\bibfield  {title} {\bibinfo {title} {{Halo Models of Large Scale
  Structure}},\ }\href {https://doi.org/10.1016/S0370-1573(02)00276-4}
  {\bibfield  {journal} {\bibinfo  {journal} {Phys. Rept.}\ }\textbf {\bibinfo
  {volume} {372}},\ \bibinfo {pages} {1} (\bibinfo {year} {2002})},\ \Eprint
  {https://arxiv.org/abs/astro-ph/0206508} {arXiv:astro-ph/0206508}
  \BibitemShut {NoStop}%
\bibitem [{\citenamefont {Wang}\ \emph {et~al.}(2020)\citenamefont {Wang},
  \citenamefont {Bose}, \citenamefont {Frenk}, \citenamefont {Gao},
  \citenamefont {Jenkins}, \citenamefont {Springel},\ and\ \citenamefont
  {White}}]{Wang:2019ftp}%
  \BibitemOpen
  \bibfield  {author} {\bibinfo {author} {\bibfnamefont {J.}~\bibnamefont
  {Wang}}, \bibinfo {author} {\bibfnamefont {S.}~\bibnamefont {Bose}}, \bibinfo
  {author} {\bibfnamefont {C.~S.}\ \bibnamefont {Frenk}}, \bibinfo {author}
  {\bibfnamefont {L.}~\bibnamefont {Gao}}, \bibinfo {author} {\bibfnamefont
  {A.}~\bibnamefont {Jenkins}}, \bibinfo {author} {\bibfnamefont
  {V.}~\bibnamefont {Springel}},\ and\ \bibinfo {author} {\bibfnamefont
  {S.~D.~M.}\ \bibnamefont {White}},\ }\bibfield  {title} {\bibinfo {title}
  {{Universal structure of dark matter haloes over a mass range of 20 orders of
  magnitude}},\ }\href {https://doi.org/10.1038/s41586-020-2642-9} {\bibfield
  {journal} {\bibinfo  {journal} {Nature}\ }\textbf {\bibinfo {volume} {585}},\
  \bibinfo {pages} {39} (\bibinfo {year} {2020})},\ \Eprint
  {https://arxiv.org/abs/1911.09720} {arXiv:1911.09720 [astro-ph.CO]}
  \BibitemShut {NoStop}%
\bibitem [{\citenamefont {Navarro}\ \emph {et~al.}(1995)\citenamefont
  {Navarro}, \citenamefont {Frenk},\ and\ \citenamefont
  {White}}]{Navarro:1994hi}%
  \BibitemOpen
  \bibfield  {author} {\bibinfo {author} {\bibfnamefont {J.~F.}\ \bibnamefont
  {Navarro}}, \bibinfo {author} {\bibfnamefont {C.~S.}\ \bibnamefont {Frenk}},\
  and\ \bibinfo {author} {\bibfnamefont {S.~D.~M.}\ \bibnamefont {White}},\
  }\bibfield  {title} {\bibinfo {title} {{Simulations of x-ray clusters}},\
  }\href {https://doi.org/10.1093/mnras/275.3.720} {\bibfield  {journal}
  {\bibinfo  {journal} {Mon. Not. Roy. Astron. Soc.}\ }\textbf {\bibinfo
  {volume} {275}},\ \bibinfo {pages} {720} (\bibinfo {year} {1995})},\ \Eprint
  {https://arxiv.org/abs/astro-ph/9408069} {arXiv:astro-ph/9408069}
  \BibitemShut {NoStop}%
\bibitem [{\citenamefont {Navarro}\ \emph {et~al.}(1996)\citenamefont
  {Navarro}, \citenamefont {Frenk},\ and\ \citenamefont
  {White}}]{Navarro:1995iw}%
  \BibitemOpen
  \bibfield  {author} {\bibinfo {author} {\bibfnamefont {J.~F.}\ \bibnamefont
  {Navarro}}, \bibinfo {author} {\bibfnamefont {C.~S.}\ \bibnamefont {Frenk}},\
  and\ \bibinfo {author} {\bibfnamefont {S.~D.~M.}\ \bibnamefont {White}},\
  }\bibfield  {title} {\bibinfo {title} {{The Structure of cold dark matter
  halos}},\ }\href {https://doi.org/10.1086/177173} {\bibfield  {journal}
  {\bibinfo  {journal} {Astrophys. J.}\ }\textbf {\bibinfo {volume} {462}},\
  \bibinfo {pages} {563} (\bibinfo {year} {1996})},\ \Eprint
  {https://arxiv.org/abs/astro-ph/9508025} {arXiv:astro-ph/9508025}
  \BibitemShut {NoStop}%
\bibitem [{\citenamefont {Navarro}\ \emph {et~al.}(1997)\citenamefont
  {Navarro}, \citenamefont {Frenk},\ and\ \citenamefont
  {White}}]{Navarro:1996gj}%
  \BibitemOpen
  \bibfield  {author} {\bibinfo {author} {\bibfnamefont {J.~F.}\ \bibnamefont
  {Navarro}}, \bibinfo {author} {\bibfnamefont {C.~S.}\ \bibnamefont {Frenk}},\
  and\ \bibinfo {author} {\bibfnamefont {S.~D.~M.}\ \bibnamefont {White}},\
  }\bibfield  {title} {\bibinfo {title} {{A Universal density profile from
  hierarchical clustering}},\ }\href {https://doi.org/10.1086/304888}
  {\bibfield  {journal} {\bibinfo  {journal} {Astrophys. J.}\ }\textbf
  {\bibinfo {volume} {490}},\ \bibinfo {pages} {493} (\bibinfo {year}
  {1997})},\ \Eprint {https://arxiv.org/abs/astro-ph/9611107}
  {arXiv:astro-ph/9611107} \BibitemShut {NoStop}%
\bibitem [{\citenamefont {Cavaliere}\ and\ \citenamefont
  {Fusco-Femiano}(1976)}]{Cavaliere:1976tx}%
  \BibitemOpen
  \bibfield  {author} {\bibinfo {author} {\bibfnamefont {A.}~\bibnamefont
  {Cavaliere}}\ and\ \bibinfo {author} {\bibfnamefont {R.}~\bibnamefont
  {Fusco-Femiano}},\ }\bibfield  {title} {\bibinfo {title} {{X-rays from hot
  plasma in clusters of galaxies}},\ }\href@noop {} {\bibfield  {journal}
  {\bibinfo  {journal} {Astron. Astrophys.}\ }\textbf {\bibinfo {volume}
  {49}},\ \bibinfo {pages} {137} (\bibinfo {year} {1976})}\BibitemShut
  {NoStop}%
\bibitem [{\citenamefont {Moore}\ \emph {et~al.}(1999)\citenamefont {Moore},
  \citenamefont {Quinn}, \citenamefont {Governato}, \citenamefont {Stadel},\
  and\ \citenamefont {Lake}}]{Moore:1999gc}%
  \BibitemOpen
  \bibfield  {author} {\bibinfo {author} {\bibfnamefont {B.}~\bibnamefont
  {Moore}}, \bibinfo {author} {\bibfnamefont {T.~R.}\ \bibnamefont {Quinn}},
  \bibinfo {author} {\bibfnamefont {F.}~\bibnamefont {Governato}}, \bibinfo
  {author} {\bibfnamefont {J.}~\bibnamefont {Stadel}},\ and\ \bibinfo {author}
  {\bibfnamefont {G.}~\bibnamefont {Lake}},\ }\bibfield  {title} {\bibinfo
  {title} {{Cold collapse and the core catastrophe}},\ }\href
  {https://doi.org/10.1046/j.1365-8711.1999.03039.x} {\bibfield  {journal}
  {\bibinfo  {journal} {Mon. Not. Roy. Astron. Soc.}\ }\textbf {\bibinfo
  {volume} {310}},\ \bibinfo {pages} {1147} (\bibinfo {year} {1999})},\ \Eprint
  {https://arxiv.org/abs/astro-ph/9903164} {arXiv:astro-ph/9903164}
  \BibitemShut {NoStop}%
\bibitem [{\citenamefont {Einasto}(1965)}]{einasto1965construction}%
  \BibitemOpen
  \bibfield  {author} {\bibinfo {author} {\bibfnamefont {J.}~\bibnamefont
  {Einasto}},\ }\bibfield  {title} {\bibinfo {title} {On the construction of a
  composite model for the galaxy and on the determination of the system of
  galactic parameters},\ }\href@noop {} {\bibfield  {journal} {\bibinfo
  {journal} {Trudy Astrofizicheskogo Instituta Alma-Ata, Vol. 5, p. 87-100,
  1965}\ }\textbf {\bibinfo {volume} {5}},\ \bibinfo {pages} {87} (\bibinfo
  {year} {1965})}\BibitemShut {NoStop}%
\bibitem [{\citenamefont {Dutton}\ and\ \citenamefont
  {Macci{\`o}}(2014)}]{Dutton:2014xda}%
  \BibitemOpen
  \bibfield  {author} {\bibinfo {author} {\bibfnamefont {A.~A.}\ \bibnamefont
  {Dutton}}\ and\ \bibinfo {author} {\bibfnamefont {A.~V.}\ \bibnamefont
  {Macci{\`o}}},\ }\bibfield  {title} {\bibinfo {title} {{Cold dark matter
  haloes in the Planck era: evolution of structural parameters for Einasto and
  NFW profiles}},\ }\href {https://doi.org/10.1093/mnras/stu742} {\bibfield
  {journal} {\bibinfo  {journal} {Mon. Not. Roy. Astron. Soc.}\ }\textbf
  {\bibinfo {volume} {441}},\ \bibinfo {pages} {3359} (\bibinfo {year}
  {2014})},\ \Eprint {https://arxiv.org/abs/1402.7073} {arXiv:1402.7073
  [astro-ph.CO]} \BibitemShut {NoStop}%
\bibitem [{\citenamefont {Dehnen}(1993)}]{Dehnen:1993uh}%
  \BibitemOpen
  \bibfield  {author} {\bibinfo {author} {\bibfnamefont {W.}~\bibnamefont
  {Dehnen}},\ }\bibfield  {title} {\bibinfo {title} {{A Family of
  Potential-Density Pairs for Spherical Galaxies and Bulges}},\ }\href@noop {}
  {\bibfield  {journal} {\bibinfo  {journal} {Mon. Not. Roy. Astron. Soc.}\
  }\textbf {\bibinfo {volume} {265}},\ \bibinfo {pages} {250} (\bibinfo {year}
  {1993})}\BibitemShut {NoStop}%
\bibitem [{\citenamefont {Hernquist}(1990)}]{Hernquist:1990be}%
  \BibitemOpen
  \bibfield  {author} {\bibinfo {author} {\bibfnamefont {L.}~\bibnamefont
  {Hernquist}},\ }\bibfield  {title} {\bibinfo {title} {{An Analytical Model
  for Spherical Galaxies and Bulges}},\ }\href {https://doi.org/10.1086/168845}
  {\bibfield  {journal} {\bibinfo  {journal} {Astrophys. J.}\ }\textbf
  {\bibinfo {volume} {356}},\ \bibinfo {pages} {359} (\bibinfo {year}
  {1990})}\BibitemShut {NoStop}%
\bibitem [{\citenamefont {Salucci}\ and\ \citenamefont
  {Burkert}(2000)}]{Salucci:2000ps}%
  \BibitemOpen
  \bibfield  {author} {\bibinfo {author} {\bibfnamefont {P.}~\bibnamefont
  {Salucci}}\ and\ \bibinfo {author} {\bibfnamefont {A.}~\bibnamefont
  {Burkert}},\ }\bibfield  {title} {\bibinfo {title} {{Dark matter scaling
  relations}},\ }\href {https://doi.org/10.1086/312747} {\bibfield  {journal}
  {\bibinfo  {journal} {Astrophys. J. Lett.}\ }\textbf {\bibinfo {volume}
  {537}},\ \bibinfo {pages} {L9} (\bibinfo {year} {2000})},\ \Eprint
  {https://arxiv.org/abs/astro-ph/0004397} {arXiv:astro-ph/0004397}
  \BibitemShut {NoStop}%
\bibitem [{\citenamefont {Dai}\ \emph {et~al.}(2022)\citenamefont {Dai},
  \citenamefont {Gong}, \citenamefont {Jiang},\ and\ \citenamefont
  {Liang}}]{Dai:2021olt}%
  \BibitemOpen
  \bibfield  {author} {\bibinfo {author} {\bibfnamefont {N.}~\bibnamefont
  {Dai}}, \bibinfo {author} {\bibfnamefont {Y.}~\bibnamefont {Gong}}, \bibinfo
  {author} {\bibfnamefont {T.}~\bibnamefont {Jiang}},\ and\ \bibinfo {author}
  {\bibfnamefont {D.}~\bibnamefont {Liang}},\ }\bibfield  {title} {\bibinfo
  {title} {{Intermediate mass-ratio inspirals with dark matter minispikes}},\
  }\href {https://doi.org/10.1103/PhysRevD.106.064003} {\bibfield  {journal}
  {\bibinfo  {journal} {Phys. Rev. D}\ }\textbf {\bibinfo {volume} {106}},\
  \bibinfo {pages} {064003} (\bibinfo {year} {2022})},\ \Eprint
  {https://arxiv.org/abs/2111.13514} {arXiv:2111.13514 [gr-qc]} \BibitemShut
  {NoStop}%
\bibitem [{\citenamefont {Cole}\ \emph {et~al.}(2023)\citenamefont {Cole},
  \citenamefont {Bertone}, \citenamefont {Coogan}, \citenamefont {Gaggero},
  \citenamefont {Karydas}, \citenamefont {Kavanagh}, \citenamefont {Spieksma},\
  and\ \citenamefont {Tomaselli}}]{Cole:2022yzw}%
  \BibitemOpen
  \bibfield  {author} {\bibinfo {author} {\bibfnamefont {P.~S.}\ \bibnamefont
  {Cole}}, \bibinfo {author} {\bibfnamefont {G.}~\bibnamefont {Bertone}},
  \bibinfo {author} {\bibfnamefont {A.}~\bibnamefont {Coogan}}, \bibinfo
  {author} {\bibfnamefont {D.}~\bibnamefont {Gaggero}}, \bibinfo {author}
  {\bibfnamefont {T.}~\bibnamefont {Karydas}}, \bibinfo {author} {\bibfnamefont
  {B.~J.}\ \bibnamefont {Kavanagh}}, \bibinfo {author} {\bibfnamefont
  {T.~F.~M.}\ \bibnamefont {Spieksma}},\ and\ \bibinfo {author} {\bibfnamefont
  {G.~M.}\ \bibnamefont {Tomaselli}},\ }\bibfield  {title} {\bibinfo {title}
  {{Distinguishing environmental effects on binary black hole gravitational
  waveforms}},\ }\href {https://doi.org/10.1038/s41550-023-01990-2} {\bibfield
  {journal} {\bibinfo  {journal} {Nature Astron.}\ }\textbf {\bibinfo {volume}
  {7}},\ \bibinfo {pages} {943} (\bibinfo {year} {2023})},\ \Eprint
  {https://arxiv.org/abs/2211.01362} {arXiv:2211.01362 [gr-qc]} \BibitemShut
  {NoStop}%
\bibitem [{\citenamefont {Kavanagh}\ \emph {et~al.}(2020)\citenamefont
  {Kavanagh}, \citenamefont {Nichols}, \citenamefont {Bertone},\ and\
  \citenamefont {Gaggero}}]{Kavanagh:2020cfn}%
  \BibitemOpen
  \bibfield  {author} {\bibinfo {author} {\bibfnamefont {B.~J.}\ \bibnamefont
  {Kavanagh}}, \bibinfo {author} {\bibfnamefont {D.~A.}\ \bibnamefont
  {Nichols}}, \bibinfo {author} {\bibfnamefont {G.}~\bibnamefont {Bertone}},\
  and\ \bibinfo {author} {\bibfnamefont {D.}~\bibnamefont {Gaggero}},\
  }\bibfield  {title} {\bibinfo {title} {{Detecting dark matter around black
  holes with gravitational waves: Effects of dark-matter dynamics on the
  gravitational waveform}},\ }\href
  {https://doi.org/10.1103/PhysRevD.102.083006} {\bibfield  {journal} {\bibinfo
   {journal} {Phys. Rev. D}\ }\textbf {\bibinfo {volume} {102}},\ \bibinfo
  {pages} {083006} (\bibinfo {year} {2020})},\ \Eprint
  {https://arxiv.org/abs/2002.12811} {arXiv:2002.12811 [gr-qc]} \BibitemShut
  {NoStop}%
\bibitem [{\citenamefont {Barsanti}\ \emph {et~al.}(2022)\citenamefont
  {Barsanti}, \citenamefont {Franchini}, \citenamefont {Gualtieri},
  \citenamefont {Maselli},\ and\ \citenamefont {Sotiriou}}]{Barsanti:2022ana}%
  \BibitemOpen
  \bibfield  {author} {\bibinfo {author} {\bibfnamefont {S.}~\bibnamefont
  {Barsanti}}, \bibinfo {author} {\bibfnamefont {N.}~\bibnamefont {Franchini}},
  \bibinfo {author} {\bibfnamefont {L.}~\bibnamefont {Gualtieri}}, \bibinfo
  {author} {\bibfnamefont {A.}~\bibnamefont {Maselli}},\ and\ \bibinfo {author}
  {\bibfnamefont {T.~P.}\ \bibnamefont {Sotiriou}},\ }\bibfield  {title}
  {\bibinfo {title} {{Extreme mass-ratio inspirals as probes of scalar fields:
  Eccentric equatorial orbits around Kerr black holes}},\ }\href
  {https://doi.org/10.1103/PhysRevD.106.044029} {\bibfield  {journal} {\bibinfo
   {journal} {Phys. Rev. D}\ }\textbf {\bibinfo {volume} {106}},\ \bibinfo
  {pages} {044029} (\bibinfo {year} {2022})},\ \Eprint
  {https://arxiv.org/abs/2203.05003} {arXiv:2203.05003 [gr-qc]} \BibitemShut
  {NoStop}%
\bibitem [{\citenamefont {Figueiredo}\ \emph {et~al.}(2023)\citenamefont
  {Figueiredo}, \citenamefont {Maselli},\ and\ \citenamefont
  {Cardoso}}]{Figueiredo:2023gas}%
  \BibitemOpen
  \bibfield  {author} {\bibinfo {author} {\bibfnamefont {E.}~\bibnamefont
  {Figueiredo}}, \bibinfo {author} {\bibfnamefont {A.}~\bibnamefont
  {Maselli}},\ and\ \bibinfo {author} {\bibfnamefont {V.}~\bibnamefont
  {Cardoso}},\ }\bibfield  {title} {\bibinfo {title} {{Black holes surrounded
  by generic dark matter profiles: Appearance and gravitational-wave
  emission}},\ }\href {https://doi.org/10.1103/PhysRevD.107.104033} {\bibfield
  {journal} {\bibinfo  {journal} {Phys. Rev. D}\ }\textbf {\bibinfo {volume}
  {107}},\ \bibinfo {pages} {104033} (\bibinfo {year} {2023})},\ \Eprint
  {https://arxiv.org/abs/2303.08183} {arXiv:2303.08183 [gr-qc]} \BibitemShut
  {NoStop}%
\bibitem [{\citenamefont {Zhou}\ \emph {et~al.}(2024)\citenamefont {Zhou},
  \citenamefont {Jin}, \citenamefont {Qiao},\ and\ \citenamefont
  {Wu}}]{Zhou:2024vhk}%
  \BibitemOpen
  \bibfield  {author} {\bibinfo {author} {\bibfnamefont {Y.-C.}\ \bibnamefont
  {Zhou}}, \bibinfo {author} {\bibfnamefont {H.-B.}\ \bibnamefont {Jin}},
  \bibinfo {author} {\bibfnamefont {C.-F.}\ \bibnamefont {Qiao}},\ and\
  \bibinfo {author} {\bibfnamefont {Y.-L.}\ \bibnamefont {Wu}},\ }\bibfield
  {title} {\bibinfo {title} {{Intermediate-mass-ratio inspirals with general
  dynamical friction in dark matter minispikes}},\ }\href
  {https://doi.org/10.3847/1538-4357/adddbc} {\bibfield  {journal} {\bibinfo
  {journal} {Astrophys. J.}\ }\textbf {\bibinfo {volume} {986}},\ \bibinfo
  {pages} {196} (\bibinfo {year} {2024})},\ \Eprint
  {https://arxiv.org/abs/2405.19240} {arXiv:2405.19240 [astro-ph.HE]}
  \BibitemShut {NoStop}%
\bibitem [{\citenamefont {Zhang}\ \emph {et~al.}(2024)\citenamefont {Zhang},
  \citenamefont {Fu},\ and\ \citenamefont {Dai}}]{Zhang:2024ugv}%
  \BibitemOpen
  \bibfield  {author} {\bibinfo {author} {\bibfnamefont {C.}~\bibnamefont
  {Zhang}}, \bibinfo {author} {\bibfnamefont {G.}~\bibnamefont {Fu}},\ and\
  \bibinfo {author} {\bibfnamefont {N.}~\bibnamefont {Dai}},\ }\bibfield
  {title} {\bibinfo {title} {{Detecting dark matter halos with extreme
  mass-ratio inspirals}},\ }\href
  {https://doi.org/10.1088/1475-7516/2024/04/088} {\bibfield  {journal}
  {\bibinfo  {journal} {JCAP}\ }\textbf {\bibinfo {volume} {04}},\ \bibinfo
  {pages} {088}},\ \Eprint {https://arxiv.org/abs/2401.04467} {arXiv:2401.04467
  [gr-qc]} \BibitemShut {NoStop}%
\bibitem [{\citenamefont {Shadykul}\ \emph {et~al.}(2025)\citenamefont
  {Shadykul}, \citenamefont {Chakrabarty},\ and\ \citenamefont
  {Malafarina}}]{Shadykul:2024ehz}%
  \BibitemOpen
  \bibfield  {author} {\bibinfo {author} {\bibfnamefont {D.}~\bibnamefont
  {Shadykul}}, \bibinfo {author} {\bibfnamefont {H.}~\bibnamefont
  {Chakrabarty}},\ and\ \bibinfo {author} {\bibfnamefont {D.}~\bibnamefont
  {Malafarina}},\ }\bibfield  {title} {\bibinfo {title} {{Intermediate mass
  ratio inspirals in dark matter halos}},\ }\href
  {https://doi.org/10.1103/PhysRevD.111.104003} {\bibfield  {journal} {\bibinfo
   {journal} {Phys. Rev. D}\ }\textbf {\bibinfo {volume} {111}},\ \bibinfo
  {pages} {104003} (\bibinfo {year} {2025})},\ \Eprint
  {https://arxiv.org/abs/2410.18657} {arXiv:2410.18657 [gr-qc]} \BibitemShut
  {NoStop}%
\bibitem [{\citenamefont {Kadota}\ \emph {et~al.}(2024)\citenamefont {Kadota},
  \citenamefont {Kim}, \citenamefont {Ko},\ and\ \citenamefont
  {Yang}}]{Kadota:2023wlm}%
  \BibitemOpen
  \bibfield  {author} {\bibinfo {author} {\bibfnamefont {K.}~\bibnamefont
  {Kadota}}, \bibinfo {author} {\bibfnamefont {J.~H.}\ \bibnamefont {Kim}},
  \bibinfo {author} {\bibfnamefont {P.}~\bibnamefont {Ko}},\ and\ \bibinfo
  {author} {\bibfnamefont {X.-Y.}\ \bibnamefont {Yang}},\ }\bibfield  {title}
  {\bibinfo {title} {{Gravitational wave probes on self-interacting dark matter
  surrounding an intermediate mass black hole}},\ }\href
  {https://doi.org/10.1103/PhysRevD.109.015022} {\bibfield  {journal} {\bibinfo
   {journal} {Phys. Rev. D}\ }\textbf {\bibinfo {volume} {109}},\ \bibinfo
  {pages} {015022} (\bibinfo {year} {2024})},\ \Eprint
  {https://arxiv.org/abs/2306.10828} {arXiv:2306.10828 [hep-ph]} \BibitemShut
  {NoStop}%
\bibitem [{\citenamefont {Kim}\ and\ \citenamefont {Yang}(2025)}]{Kim:2024rgf}%
  \BibitemOpen
  \bibfield  {author} {\bibinfo {author} {\bibfnamefont {J.~H.}\ \bibnamefont
  {Kim}}\ and\ \bibinfo {author} {\bibfnamefont {X.-Y.}\ \bibnamefont {Yang}},\
  }\bibfield  {title} {\bibinfo {title} {{Gravitational wave duet by resonating
  binary black holes within ultralight dark matter}},\ }\href
  {https://doi.org/10.1103/ybtp-fzwl} {\bibfield  {journal} {\bibinfo
  {journal} {Phys. Rev. D}\ }\textbf {\bibinfo {volume} {112}},\ \bibinfo
  {pages} {083040} (\bibinfo {year} {2025})},\ \Eprint
  {https://arxiv.org/abs/2407.14604} {arXiv:2407.14604 [astro-ph.CO]}
  \BibitemShut {NoStop}%
\bibitem [{\citenamefont {Ding}\ \emph {et~al.}(2025)\citenamefont {Ding},
  \citenamefont {He},\ and\ \citenamefont {Zhu}}]{Ding:2025hqf}%
  \BibitemOpen
  \bibfield  {author} {\bibinfo {author} {\bibfnamefont {Q.}~\bibnamefont
  {Ding}}, \bibinfo {author} {\bibfnamefont {M.}~\bibnamefont {He}},\ and\
  \bibinfo {author} {\bibfnamefont {H.-Y.}\ \bibnamefont {Zhu}},\ }\bibfield
  {title} {\bibinfo {title} {{Extracting Properties of Dark Dense Environments
  around Black Holes from Gravitational Waves}},\ }\href@noop {} {\  (\bibinfo
  {year} {2025})},\ \Eprint {https://arxiv.org/abs/2510.27424}
  {arXiv:2510.27424 [gr-qc]} \BibitemShut {NoStop}%
\bibitem [{\citenamefont {Chandrasekhar}(1943)}]{Chandrasekhar:1943ys}%
  \BibitemOpen
  \bibfield  {author} {\bibinfo {author} {\bibfnamefont {S.}~\bibnamefont
  {Chandrasekhar}},\ }\bibfield  {title} {\bibinfo {title} {{Dynamical
  Friction. I. General Considerations: the Coefficient of Dynamical
  Friction}},\ }\href {https://doi.org/10.1086/144517} {\bibfield  {journal}
  {\bibinfo  {journal} {Astrophys. J.}\ }\textbf {\bibinfo {volume} {97}},\
  \bibinfo {pages} {255} (\bibinfo {year} {1943})}\BibitemShut {NoStop}%
\bibitem [{\citenamefont {Cardoso}\ \emph {et~al.}(2021)\citenamefont
  {Cardoso}, \citenamefont {Macedo},\ and\ \citenamefont
  {Vicente}}]{Cardoso:2020iji}%
  \BibitemOpen
  \bibfield  {author} {\bibinfo {author} {\bibfnamefont {V.}~\bibnamefont
  {Cardoso}}, \bibinfo {author} {\bibfnamefont {C.~F.~B.}\ \bibnamefont
  {Macedo}},\ and\ \bibinfo {author} {\bibfnamefont {R.}~\bibnamefont
  {Vicente}},\ }\bibfield  {title} {\bibinfo {title} {{Eccentricity evolution
  of compact binaries and applications to gravitational-wave physics}},\ }\href
  {https://doi.org/10.1103/PhysRevD.103.023015} {\bibfield  {journal} {\bibinfo
   {journal} {Phys. Rev. D}\ }\textbf {\bibinfo {volume} {103}},\ \bibinfo
  {pages} {023015} (\bibinfo {year} {2021})},\ \Eprint
  {https://arxiv.org/abs/2010.15151} {arXiv:2010.15151 [gr-qc]} \BibitemShut
  {NoStop}%
\bibitem [{\citenamefont {Bondi}(1952)}]{Bondi:1952ni}%
  \BibitemOpen
  \bibfield  {author} {\bibinfo {author} {\bibfnamefont {H.}~\bibnamefont
  {Bondi}},\ }\bibfield  {title} {\bibinfo {title} {{On spherically symmetrical
  accretion}},\ }\href {https://doi.org/10.1093/mnras/112.2.195} {\bibfield
  {journal} {\bibinfo  {journal} {Mon. Not. Roy. Astron. Soc.}\ }\textbf
  {\bibinfo {volume} {112}},\ \bibinfo {pages} {195} (\bibinfo {year}
  {1952})}\BibitemShut {NoStop}%
\bibitem [{\citenamefont {Macedo}\ \emph {et~al.}(2013)\citenamefont {Macedo},
  \citenamefont {Pani}, \citenamefont {Cardoso},\ and\ \citenamefont
  {Crispino}}]{Macedo:2013qea}%
  \BibitemOpen
  \bibfield  {author} {\bibinfo {author} {\bibfnamefont {C.~F.~B.}\
  \bibnamefont {Macedo}}, \bibinfo {author} {\bibfnamefont {P.}~\bibnamefont
  {Pani}}, \bibinfo {author} {\bibfnamefont {V.}~\bibnamefont {Cardoso}},\ and\
  \bibinfo {author} {\bibfnamefont {L.~C.~B.}\ \bibnamefont {Crispino}},\
  }\bibfield  {title} {\bibinfo {title} {{Into the lair: gravitational-wave
  signatures of dark matter}},\ }\href
  {https://doi.org/10.1088/0004-637X/774/1/48} {\bibfield  {journal} {\bibinfo
  {journal} {Astrophys. J.}\ }\textbf {\bibinfo {volume} {774}},\ \bibinfo
  {pages} {48} (\bibinfo {year} {2013})},\ \Eprint
  {https://arxiv.org/abs/1302.2646} {arXiv:1302.2646 [gr-qc]} \BibitemShut
  {NoStop}%
\bibitem [{\citenamefont {Bondi}\ and\ \citenamefont
  {Hoyle}(1944)}]{Bondi:1944rnk}%
  \BibitemOpen
  \bibfield  {author} {\bibinfo {author} {\bibfnamefont {H.}~\bibnamefont
  {Bondi}}\ and\ \bibinfo {author} {\bibfnamefont {F.}~\bibnamefont {Hoyle}},\
  }\bibfield  {title} {\bibinfo {title} {{On the Mechanism of Accretion by
  Stars}},\ }\href {https://doi.org/10.1093/mnras/104.5.273} {\bibfield
  {journal} {\bibinfo  {journal} {Mon. Not. Roy. Astron. Soc.}\ }\textbf
  {\bibinfo {volume} {104}},\ \bibinfo {pages} {273} (\bibinfo {year}
  {1944})}\BibitemShut {NoStop}%
\bibitem [{\citenamefont {Edgar}(2004)}]{Edgar:2004mk}%
  \BibitemOpen
  \bibfield  {author} {\bibinfo {author} {\bibfnamefont {R.~G.}\ \bibnamefont
  {Edgar}},\ }\bibfield  {title} {\bibinfo {title} {{A Review of
  Bondi-Hoyle-Lyttleton accretion}},\ }\href
  {https://doi.org/10.1016/j.newar.2004.06.001} {\bibfield  {journal} {\bibinfo
   {journal} {New Astron. Rev.}\ }\textbf {\bibinfo {volume} {48}},\ \bibinfo
  {pages} {843} (\bibinfo {year} {2004})},\ \Eprint
  {https://arxiv.org/abs/astro-ph/0406166} {arXiv:astro-ph/0406166}
  \BibitemShut {NoStop}%
\bibitem [{\citenamefont {Rahman}\ \emph {et~al.}(2024)\citenamefont {Rahman},
  \citenamefont {Kumar},\ and\ \citenamefont {Bhattacharyya}}]{Rahman:2023sof}%
  \BibitemOpen
  \bibfield  {author} {\bibinfo {author} {\bibfnamefont {M.}~\bibnamefont
  {Rahman}}, \bibinfo {author} {\bibfnamefont {S.}~\bibnamefont {Kumar}},\ and\
  \bibinfo {author} {\bibfnamefont {A.}~\bibnamefont {Bhattacharyya}},\
  }\bibfield  {title} {\bibinfo {title} {{Probing astrophysical environment
  with eccentric extreme mass-ratio inspirals}},\ }\href
  {https://doi.org/10.1088/1475-7516/2024/01/035} {\bibfield  {journal}
  {\bibinfo  {journal} {JCAP}\ }\textbf {\bibinfo {volume} {01}},\ \bibinfo
  {pages} {035}},\ \Eprint {https://arxiv.org/abs/2306.14971} {arXiv:2306.14971
  [gr-qc]} \BibitemShut {NoStop}%
\bibitem [{\citenamefont {Zhao}\ \emph {et~al.}(2025)\citenamefont {Zhao},
  \citenamefont {Dai},\ and\ \citenamefont {Gong}}]{Zhao:2024bpp}%
  \BibitemOpen
  \bibfield  {author} {\bibinfo {author} {\bibfnamefont {Y.}~\bibnamefont
  {Zhao}}, \bibinfo {author} {\bibfnamefont {N.}~\bibnamefont {Dai}},\ and\
  \bibinfo {author} {\bibfnamefont {Y.}~\bibnamefont {Gong}},\ }\bibfield
  {title} {\bibinfo {title} {{Distinguishing dark matter halos with Extreme
  mass ratio inspirals}},\ }\href {https://doi.org/10.1093/mnras/staf1471}
  {\bibfield  {journal} {\bibinfo  {journal} {Mon. Not. Roy. Astron. Soc.}\
  }\textbf {\bibinfo {volume} {2326}},\ \bibinfo {pages} {2337} (\bibinfo
  {year} {2025})},\ \Eprint {https://arxiv.org/abs/2410.06882}
  {arXiv:2410.06882 [gr-qc]} \BibitemShut {NoStop}%
\bibitem [{\citenamefont {Gliorio}\ \emph {et~al.}(2025)\citenamefont
  {Gliorio}, \citenamefont {Berti}, \citenamefont {Maselli},\ and\
  \citenamefont {Speeney}}]{Gliorio:2025cbh}%
  \BibitemOpen
  \bibfield  {author} {\bibinfo {author} {\bibfnamefont {S.}~\bibnamefont
  {Gliorio}}, \bibinfo {author} {\bibfnamefont {E.}~\bibnamefont {Berti}},
  \bibinfo {author} {\bibfnamefont {A.}~\bibnamefont {Maselli}},\ and\ \bibinfo
  {author} {\bibfnamefont {N.}~\bibnamefont {Speeney}},\ }\bibfield  {title}
  {\bibinfo {title} {{Extreme mass ratio inspirals in dark matter halos:
  Dynamics and distinguishability of halo models}},\ }\href
  {https://doi.org/10.1103/dw6c-14pt} {\bibfield  {journal} {\bibinfo
  {journal} {Phys. Rev. D}\ }\textbf {\bibinfo {volume} {112}},\ \bibinfo
  {pages} {124050} (\bibinfo {year} {2025})},\ \Eprint
  {https://arxiv.org/abs/2503.16649} {arXiv:2503.16649 [gr-qc]} \BibitemShut
  {NoStop}%
\bibitem [{\citenamefont {Zhao}\ and\ \citenamefont
  {Gong}(2026)}]{Zhao:2026yis}%
  \BibitemOpen
  \bibfield  {author} {\bibinfo {author} {\bibfnamefont {Y.}~\bibnamefont
  {Zhao}}\ and\ \bibinfo {author} {\bibfnamefont {Y.}~\bibnamefont {Gong}},\
  }\bibfield  {title} {\bibinfo {title} {{Dark matter distributions around
  extreme mass ratio inspirals: effects of radial pressure and relativistic
  treatment}},\ }\href@noop {} {\  (\bibinfo {year} {2026})},\ \Eprint
  {https://arxiv.org/abs/2602.12022} {arXiv:2602.12022 [gr-qc]} \BibitemShut
  {NoStop}%
\bibitem [{\citenamefont {Dai}\ \emph {et~al.}(2024)\citenamefont {Dai},
  \citenamefont {Gong}, \citenamefont {Zhao},\ and\ \citenamefont
  {Jiang}}]{Dai:2023cft}%
  \BibitemOpen
  \bibfield  {author} {\bibinfo {author} {\bibfnamefont {N.}~\bibnamefont
  {Dai}}, \bibinfo {author} {\bibfnamefont {Y.}~\bibnamefont {Gong}}, \bibinfo
  {author} {\bibfnamefont {Y.}~\bibnamefont {Zhao}},\ and\ \bibinfo {author}
  {\bibfnamefont {T.}~\bibnamefont {Jiang}},\ }\bibfield  {title} {\bibinfo
  {title} {{Extreme mass ratio inspirals in galaxies with dark matter halos}},\
  }\href {https://doi.org/10.1103/PhysRevD.110.084080} {\bibfield  {journal}
  {\bibinfo  {journal} {Phys. Rev. D}\ }\textbf {\bibinfo {volume} {110}},\
  \bibinfo {pages} {084080} (\bibinfo {year} {2024})},\ \Eprint
  {https://arxiv.org/abs/2301.05088} {arXiv:2301.05088 [gr-qc]} \BibitemShut
  {NoStop}%
\bibitem [{\citenamefont {Li}\ \emph {et~al.}(2025{\natexlab{a}})\citenamefont
  {Li}, \citenamefont {Guo}, \citenamefont {Cao},\ and\ \citenamefont
  {Zhang}}]{Li:2025qtb}%
  \BibitemOpen
  \bibfield  {author} {\bibinfo {author} {\bibfnamefont {Z.}~\bibnamefont
  {Li}}, \bibinfo {author} {\bibfnamefont {X.}~\bibnamefont {Guo}}, \bibinfo
  {author} {\bibfnamefont {Z.}~\bibnamefont {Cao}},\ and\ \bibinfo {author}
  {\bibfnamefont {Y.-L.}\ \bibnamefont {Zhang}},\ }\bibfield  {title} {\bibinfo
  {title} {{Detectability of dark matter density distribution via gravitational
  waves from binary black holes in the Galactic Center}},\ }\href
  {https://doi.org/10.1103/zr7l-7y5c} {\bibfield  {journal} {\bibinfo
  {journal} {Phys. Rev. D}\ }\textbf {\bibinfo {volume} {112}},\ \bibinfo
  {pages} {063055} (\bibinfo {year} {2025}{\natexlab{a}})},\ \Eprint
  {https://arxiv.org/abs/2506.19327} {arXiv:2506.19327 [astro-ph.HE]}
  \BibitemShut {NoStop}%
\bibitem [{\citenamefont {Li}\ \emph {et~al.}(2025{\natexlab{b}})\citenamefont
  {Li}, \citenamefont {Qiao},\ and\ \citenamefont {Tao}}]{Li:2025eln}%
  \BibitemOpen
  \bibfield  {author} {\bibinfo {author} {\bibfnamefont {G.-H.}\ \bibnamefont
  {Li}}, \bibinfo {author} {\bibfnamefont {C.-K.}\ \bibnamefont {Qiao}},\ and\
  \bibinfo {author} {\bibfnamefont {J.}~\bibnamefont {Tao}},\ }\bibfield
  {title} {\bibinfo {title} {{Periodic orbits and their gravitational waves in
  EMRIs: supermassive black hole affected by galactic dark matter halos}},\
  }\href@noop {} {\  (\bibinfo {year} {2025}{\natexlab{b}})},\ \Eprint
  {https://arxiv.org/abs/2510.24989} {arXiv:2510.24989 [gr-qc]} \BibitemShut
  {NoStop}%
\bibitem [{\citenamefont {Alloqulov}\ \emph {et~al.}(2025)\citenamefont
  {Alloqulov}, \citenamefont {Xamidov}, \citenamefont {Shaymatov},\ and\
  \citenamefont {Ahmedov}}]{Alloqulov:2025ucf}%
  \BibitemOpen
  \bibfield  {author} {\bibinfo {author} {\bibfnamefont {M.}~\bibnamefont
  {Alloqulov}}, \bibinfo {author} {\bibfnamefont {T.}~\bibnamefont {Xamidov}},
  \bibinfo {author} {\bibfnamefont {S.}~\bibnamefont {Shaymatov}},\ and\
  \bibinfo {author} {\bibfnamefont {B.}~\bibnamefont {Ahmedov}},\ }\bibfield
  {title} {\bibinfo {title} {{Gravitational waveforms from periodic orbits
  around a Schwarzschild black hole embedded in a Dehnen-type dark matter
  halo}},\ }\href {https://doi.org/10.1140/epjc/s10052-025-14529-8} {\bibfield
  {journal} {\bibinfo  {journal} {Eur. Phys. J. C}\ }\textbf {\bibinfo {volume}
  {85}},\ \bibinfo {pages} {798} (\bibinfo {year} {2025})},\ \Eprint
  {https://arxiv.org/abs/2504.05236} {arXiv:2504.05236 [gr-qc]} \BibitemShut
  {NoStop}%
\bibitem [{\citenamefont {Haroon}\ and\ \citenamefont
  {Zhu}(2025)}]{Haroon:2025rzx}%
  \BibitemOpen
  \bibfield  {author} {\bibinfo {author} {\bibfnamefont {S.}~\bibnamefont
  {Haroon}}\ and\ \bibinfo {author} {\bibfnamefont {T.}~\bibnamefont {Zhu}},\
  }\bibfield  {title} {\bibinfo {title} {{Periodic orbits and their
  gravitational wave radiations in a black hole with a dark matter halo}},\
  }\href {https://doi.org/10.1103/ckdt-wtsl} {\bibfield  {journal} {\bibinfo
  {journal} {Phys. Rev. D}\ }\textbf {\bibinfo {volume} {112}},\ \bibinfo
  {pages} {044046} (\bibinfo {year} {2025})},\ \Eprint
  {https://arxiv.org/abs/2502.09171} {arXiv:2502.09171 [gr-qc]} \BibitemShut
  {NoStop}%
\bibitem [{\citenamefont {Das}\ \emph {et~al.}(2025{\natexlab{a}})\citenamefont
  {Das}, \citenamefont {Dalui}, \citenamefont {Lee},\ and\ \citenamefont
  {Cai}}]{Das:2025vja}%
  \BibitemOpen
  \bibfield  {author} {\bibinfo {author} {\bibfnamefont {S.}~\bibnamefont
  {Das}}, \bibinfo {author} {\bibfnamefont {S.}~\bibnamefont {Dalui}}, \bibinfo
  {author} {\bibfnamefont {B.-H.}\ \bibnamefont {Lee}},\ and\ \bibinfo {author}
  {\bibfnamefont {Y.-F.}\ \bibnamefont {Cai}},\ }\bibfield  {title} {\bibinfo
  {title} {{Extreme-Mass-Ratio Inspirals Embedded in Dark Matter Halo I:
  Existence of Homoclinic Orbit and Near-Horizon Chaos}},\ }\href@noop {} {\
  (\bibinfo {year} {2025}{\natexlab{a}})},\ \Eprint
  {https://arxiv.org/abs/2511.03657} {arXiv:2511.03657 [gr-qc]} \BibitemShut
  {NoStop}%
\bibitem [{\citenamefont {Das}\ \emph {et~al.}(2025{\natexlab{b}})\citenamefont
  {Das}, \citenamefont {Dalui}, \citenamefont {Lee},\ and\ \citenamefont
  {Cai}}]{Das:2025eiv}%
  \BibitemOpen
  \bibfield  {author} {\bibinfo {author} {\bibfnamefont {S.}~\bibnamefont
  {Das}}, \bibinfo {author} {\bibfnamefont {S.}~\bibnamefont {Dalui}}, \bibinfo
  {author} {\bibfnamefont {B.-H.}\ \bibnamefont {Lee}},\ and\ \bibinfo {author}
  {\bibfnamefont {Y.-F.}\ \bibnamefont {Cai}},\ }\bibfield  {title} {\bibinfo
  {title} {{Extreme-Mass-Ratio Inspirals Embedded in Dark Matter Halo II:
  Chaotic Imprints in Gravitational Waves}},\ }\href@noop {} {\  (\bibinfo
  {year} {2025}{\natexlab{b}})},\ \Eprint {https://arxiv.org/abs/2512.04848}
  {arXiv:2512.04848 [gr-qc]} \BibitemShut {NoStop}%
\bibitem [{\citenamefont {Cardoso}\ \emph
  {et~al.}(2022{\natexlab{b}})\citenamefont {Cardoso}, \citenamefont
  {Destounis}, \citenamefont {Duque}, \citenamefont {Macedo},\ and\
  \citenamefont {Maselli}}]{Cardoso:2021wlq}%
  \BibitemOpen
  \bibfield  {author} {\bibinfo {author} {\bibfnamefont {V.}~\bibnamefont
  {Cardoso}}, \bibinfo {author} {\bibfnamefont {K.}~\bibnamefont {Destounis}},
  \bibinfo {author} {\bibfnamefont {F.}~\bibnamefont {Duque}}, \bibinfo
  {author} {\bibfnamefont {R.~P.}\ \bibnamefont {Macedo}},\ and\ \bibinfo
  {author} {\bibfnamefont {A.}~\bibnamefont {Maselli}},\ }\bibfield  {title}
  {\bibinfo {title} {{Black holes in galaxies: Environmental impact on
  gravitational-wave generation and propagation}},\ }\href
  {https://doi.org/10.1103/PhysRevD.105.L061501} {\bibfield  {journal}
  {\bibinfo  {journal} {Phys. Rev. D}\ }\textbf {\bibinfo {volume} {105}},\
  \bibinfo {pages} {L061501} (\bibinfo {year} {2022}{\natexlab{b}})},\ \Eprint
  {https://arxiv.org/abs/2109.00005} {arXiv:2109.00005 [gr-qc]} \BibitemShut
  {NoStop}%
\bibitem [{\citenamefont {Liu}\ \emph {et~al.}(2024{\natexlab{a}})\citenamefont
  {Liu}, \citenamefont {Yang},\ and\ \citenamefont {Long}}]{Liu:2024xcd}%
  \BibitemOpen
  \bibfield  {author} {\bibinfo {author} {\bibfnamefont {D.}~\bibnamefont
  {Liu}}, \bibinfo {author} {\bibfnamefont {Y.}~\bibnamefont {Yang}},\ and\
  \bibinfo {author} {\bibfnamefont {Z.-W.}\ \bibnamefont {Long}},\ }\bibfield
  {title} {\bibinfo {title} {{Probing black holes in a dark matter spike of M87
  using quasinormal modes}},\ }\href
  {https://doi.org/10.1140/epjc/s10052-024-13096-8} {\bibfield  {journal}
  {\bibinfo  {journal} {Eur. Phys. J. C}\ }\textbf {\bibinfo {volume} {84}},\
  \bibinfo {pages} {731} (\bibinfo {year} {2024}{\natexlab{a}})},\ \Eprint
  {https://arxiv.org/abs/2401.09182} {arXiv:2401.09182 [gr-qc]} \BibitemShut
  {NoStop}%
\bibitem [{\citenamefont {Liu}\ \emph {et~al.}(2024{\natexlab{b}})\citenamefont
  {Liu}, \citenamefont {Yang},\ and\ \citenamefont {Long}}]{Liu:2023vno}%
  \BibitemOpen
  \bibfield  {author} {\bibinfo {author} {\bibfnamefont {D.}~\bibnamefont
  {Liu}}, \bibinfo {author} {\bibfnamefont {Y.}~\bibnamefont {Yang}},\ and\
  \bibinfo {author} {\bibfnamefont {Z.-W.}\ \bibnamefont {Long}},\ }\bibfield
  {title} {\bibinfo {title} {{Probing the black holes in a dark matter halo of
  M87 using gravitational wave echoes}},\ }\href
  {https://doi.org/10.1140/epjc/s10052-024-13255-x} {\bibfield  {journal}
  {\bibinfo  {journal} {Eur. Phys. J. C}\ }\textbf {\bibinfo {volume} {84}},\
  \bibinfo {pages} {871} (\bibinfo {year} {2024}{\natexlab{b}})},\ \Eprint
  {https://arxiv.org/abs/2312.07074} {arXiv:2312.07074 [gr-qc]} \BibitemShut
  {NoStop}%
\bibitem [{\citenamefont {Zhao}\ \emph {et~al.}(2024)\citenamefont {Zhao},
  \citenamefont {Sun}, \citenamefont {Cao}, \citenamefont {Lin},\ and\
  \citenamefont {Qian}}]{Zhao:2023itk}%
  \BibitemOpen
  \bibfield  {author} {\bibinfo {author} {\bibfnamefont {Y.}~\bibnamefont
  {Zhao}}, \bibinfo {author} {\bibfnamefont {B.}~\bibnamefont {Sun}}, \bibinfo
  {author} {\bibfnamefont {Z.}~\bibnamefont {Cao}}, \bibinfo {author}
  {\bibfnamefont {K.}~\bibnamefont {Lin}},\ and\ \bibinfo {author}
  {\bibfnamefont {W.-L.}\ \bibnamefont {Qian}},\ }\bibfield  {title} {\bibinfo
  {title} {{Influence of dark matter equation of state on the axial
  gravitational ringing of supermassive black holes}},\ }\href
  {https://doi.org/10.1103/PhysRevD.109.044031} {\bibfield  {journal} {\bibinfo
   {journal} {Phys. Rev. D}\ }\textbf {\bibinfo {volume} {109}},\ \bibinfo
  {pages} {044031} (\bibinfo {year} {2024})},\ \Eprint
  {https://arxiv.org/abs/2308.15371} {arXiv:2308.15371 [gr-qc]} \BibitemShut
  {NoStop}%
\bibitem [{\citenamefont {Hughes}(2000)}]{Hughes:1999bq}%
  \BibitemOpen
  \bibfield  {author} {\bibinfo {author} {\bibfnamefont {S.~A.}\ \bibnamefont
  {Hughes}},\ }\bibfield  {title} {\bibinfo {title} {{The Evolution of
  circular, nonequatorial orbits of Kerr black holes due to gravitational wave
  emission}},\ }\href {https://doi.org/10.1103/PhysRevD.65.069902} {\bibfield
  {journal} {\bibinfo  {journal} {Phys. Rev. D}\ }\textbf {\bibinfo {volume}
  {61}},\ \bibinfo {pages} {084004} (\bibinfo {year} {2000})},\ \bibinfo {note}
  {[Erratum: Phys.Rev.D 63, 049902 (2001), Erratum: Phys.Rev.D 65, 069902
  (2002), Erratum: Phys.Rev.D 67, 089901 (2003), Erratum: Phys.Rev.D 78, 109902
  (2008), Erratum: Phys.Rev.D 90, 109904 (2014)]},\ \Eprint
  {https://arxiv.org/abs/gr-qc/9910091} {arXiv:gr-qc/9910091} \BibitemShut
  {NoStop}%
\bibitem [{\citenamefont {Hughes}(2001)}]{Hughes:2001jr}%
  \BibitemOpen
  \bibfield  {author} {\bibinfo {author} {\bibfnamefont {S.~A.}\ \bibnamefont
  {Hughes}},\ }\bibfield  {title} {\bibinfo {title} {{Evolution of circular,
  nonequatorial orbits of Kerr black holes due to gravitational wave emission.
  II. Inspiral trajectories and gravitational wave forms}},\ }\href
  {https://doi.org/10.1103/PhysRevD.64.064004} {\bibfield  {journal} {\bibinfo
  {journal} {Phys. Rev. D}\ }\textbf {\bibinfo {volume} {64}},\ \bibinfo
  {pages} {064004} (\bibinfo {year} {2001})},\ \bibinfo {note} {[Erratum:
  Phys.Rev.D 88, 109902 (2013)]},\ \Eprint
  {https://arxiv.org/abs/gr-qc/0104041} {arXiv:gr-qc/0104041} \BibitemShut
  {NoStop}%
\bibitem [{\citenamefont {Drasco}\ \emph {et~al.}(2005)\citenamefont {Drasco},
  \citenamefont {Flanagan},\ and\ \citenamefont {Hughes}}]{Drasco:2005is}%
  \BibitemOpen
  \bibfield  {author} {\bibinfo {author} {\bibfnamefont {S.}~\bibnamefont
  {Drasco}}, \bibinfo {author} {\bibfnamefont {E.~E.}\ \bibnamefont
  {Flanagan}},\ and\ \bibinfo {author} {\bibfnamefont {S.~A.}\ \bibnamefont
  {Hughes}},\ }\bibfield  {title} {\bibinfo {title} {{Computing inspirals in
  Kerr in the adiabatic regime. I. The Scalar case}},\ }\href
  {https://doi.org/10.1088/0264-9381/22/15/011} {\bibfield  {journal} {\bibinfo
   {journal} {Class. Quant. Grav.}\ }\textbf {\bibinfo {volume} {22}},\
  \bibinfo {pages} {S801} (\bibinfo {year} {2005})},\ \Eprint
  {https://arxiv.org/abs/gr-qc/0505075} {arXiv:gr-qc/0505075} \BibitemShut
  {NoStop}%
\bibitem [{\citenamefont {Drasco}\ and\ \citenamefont
  {Hughes}(2006)}]{Drasco:2005kz}%
  \BibitemOpen
  \bibfield  {author} {\bibinfo {author} {\bibfnamefont {S.}~\bibnamefont
  {Drasco}}\ and\ \bibinfo {author} {\bibfnamefont {S.~A.}\ \bibnamefont
  {Hughes}},\ }\bibfield  {title} {\bibinfo {title} {{Gravitational wave
  snapshots of generic extreme mass ratio inspirals}},\ }\href
  {https://doi.org/10.1103/PhysRevD.73.024027} {\bibfield  {journal} {\bibinfo
  {journal} {Phys. Rev. D}\ }\textbf {\bibinfo {volume} {73}},\ \bibinfo
  {pages} {024027} (\bibinfo {year} {2006})},\ \bibinfo {note} {[Erratum:
  Phys.Rev.D 88, 109905 (2013), Erratum: Phys.Rev.D 90, 109905 (2014)]},\
  \Eprint {https://arxiv.org/abs/gr-qc/0509101} {arXiv:gr-qc/0509101}
  \BibitemShut {NoStop}%
\bibitem [{\citenamefont {Sundararajan}\ \emph {et~al.}(2008)\citenamefont
  {Sundararajan}, \citenamefont {Khanna}, \citenamefont {Hughes},\ and\
  \citenamefont {Drasco}}]{Sundararajan:2008zm}%
  \BibitemOpen
  \bibfield  {author} {\bibinfo {author} {\bibfnamefont {P.~A.}\ \bibnamefont
  {Sundararajan}}, \bibinfo {author} {\bibfnamefont {G.}~\bibnamefont
  {Khanna}}, \bibinfo {author} {\bibfnamefont {S.~A.}\ \bibnamefont {Hughes}},\
  and\ \bibinfo {author} {\bibfnamefont {S.}~\bibnamefont {Drasco}},\
  }\bibfield  {title} {\bibinfo {title} {{Towards adiabatic waveforms for
  inspiral into Kerr black holes: II. Dynamical sources and generic orbits}},\
  }\href {https://doi.org/10.1103/PhysRevD.78.024022} {\bibfield  {journal}
  {\bibinfo  {journal} {Phys. Rev. D}\ }\textbf {\bibinfo {volume} {78}},\
  \bibinfo {pages} {024022} (\bibinfo {year} {2008})},\ \Eprint
  {https://arxiv.org/abs/0803.0317} {arXiv:0803.0317 [gr-qc]} \BibitemShut
  {NoStop}%
\bibitem [{\citenamefont {Isoyama}\ \emph {et~al.}(2022)\citenamefont
  {Isoyama}, \citenamefont {Fujita}, \citenamefont {Chua}, \citenamefont
  {Nakano}, \citenamefont {Pound},\ and\ \citenamefont
  {Sago}}]{Isoyama:2021jjd}%
  \BibitemOpen
  \bibfield  {author} {\bibinfo {author} {\bibfnamefont {S.}~\bibnamefont
  {Isoyama}}, \bibinfo {author} {\bibfnamefont {R.}~\bibnamefont {Fujita}},
  \bibinfo {author} {\bibfnamefont {A.~J.~K.}\ \bibnamefont {Chua}}, \bibinfo
  {author} {\bibfnamefont {H.}~\bibnamefont {Nakano}}, \bibinfo {author}
  {\bibfnamefont {A.}~\bibnamefont {Pound}},\ and\ \bibinfo {author}
  {\bibfnamefont {N.}~\bibnamefont {Sago}},\ }\bibfield  {title} {\bibinfo
  {title} {{Adiabatic Waveforms from Extreme-Mass-Ratio Inspirals: An
  Analytical Approach}},\ }\href
  {https://doi.org/10.1103/PhysRevLett.128.231101} {\bibfield  {journal}
  {\bibinfo  {journal} {Phys. Rev. Lett.}\ }\textbf {\bibinfo {volume} {128}},\
  \bibinfo {pages} {231101} (\bibinfo {year} {2022})},\ \Eprint
  {https://arxiv.org/abs/2111.05288} {arXiv:2111.05288 [gr-qc]} \BibitemShut
  {NoStop}%
\bibitem [{\citenamefont {Barack}\ and\ \citenamefont
  {Cutler}(2004)}]{Barack:2003fp}%
  \BibitemOpen
  \bibfield  {author} {\bibinfo {author} {\bibfnamefont {L.}~\bibnamefont
  {Barack}}\ and\ \bibinfo {author} {\bibfnamefont {C.}~\bibnamefont
  {Cutler}},\ }\bibfield  {title} {\bibinfo {title} {{LISA capture sources:
  Approximate waveforms, signal-to-noise ratios, and parameter estimation
  accuracy}},\ }\href {https://doi.org/10.1103/PhysRevD.69.082005} {\bibfield
  {journal} {\bibinfo  {journal} {Phys. Rev. D}\ }\textbf {\bibinfo {volume}
  {69}},\ \bibinfo {pages} {082005} (\bibinfo {year} {2004})},\ \Eprint
  {https://arxiv.org/abs/gr-qc/0310125} {arXiv:gr-qc/0310125} \BibitemShut
  {NoStop}%
\bibitem [{\citenamefont {Gair}\ \emph {et~al.}(2005)\citenamefont {Gair},
  \citenamefont {Kennefick},\ and\ \citenamefont {Larson}}]{Gair:2005is}%
  \BibitemOpen
  \bibfield  {author} {\bibinfo {author} {\bibfnamefont {J.~R.}\ \bibnamefont
  {Gair}}, \bibinfo {author} {\bibfnamefont {D.~J.}\ \bibnamefont
  {Kennefick}},\ and\ \bibinfo {author} {\bibfnamefont {S.~L.}\ \bibnamefont
  {Larson}},\ }\bibfield  {title} {\bibinfo {title} {{Semi-relativistic
  approximation to gravitational radiation from encounters with black holes}},\
  }\href {https://doi.org/10.1103/PhysRevD.74.109901} {\bibfield  {journal}
  {\bibinfo  {journal} {Phys. Rev. D}\ }\textbf {\bibinfo {volume} {72}},\
  \bibinfo {pages} {084009} (\bibinfo {year} {2005})},\ \bibinfo {note}
  {[Erratum: Phys.Rev.D 74, 109901 (2006)]},\ \Eprint
  {https://arxiv.org/abs/gr-qc/0508049} {arXiv:gr-qc/0508049} \BibitemShut
  {NoStop}%
\bibitem [{\citenamefont {Babak}\ \emph {et~al.}(2007)\citenamefont {Babak},
  \citenamefont {Fang}, \citenamefont {Gair}, \citenamefont {Glampedakis},\
  and\ \citenamefont {Hughes}}]{Babak:2006uv}%
  \BibitemOpen
  \bibfield  {author} {\bibinfo {author} {\bibfnamefont {S.}~\bibnamefont
  {Babak}}, \bibinfo {author} {\bibfnamefont {H.}~\bibnamefont {Fang}},
  \bibinfo {author} {\bibfnamefont {J.~R.}\ \bibnamefont {Gair}}, \bibinfo
  {author} {\bibfnamefont {K.}~\bibnamefont {Glampedakis}},\ and\ \bibinfo
  {author} {\bibfnamefont {S.~A.}\ \bibnamefont {Hughes}},\ }\bibfield  {title}
  {\bibinfo {title} {{'Kludge' gravitational waveforms for a test-body orbiting
  a Kerr black hole}},\ }\href {https://doi.org/10.1103/PhysRevD.75.024005}
  {\bibfield  {journal} {\bibinfo  {journal} {Phys. Rev. D}\ }\textbf {\bibinfo
  {volume} {75}},\ \bibinfo {pages} {024005} (\bibinfo {year} {2007})},\
  \bibinfo {note} {[Erratum: Phys.Rev.D 77, 04990 (2008)]},\ \Eprint
  {https://arxiv.org/abs/gr-qc/0607007} {arXiv:gr-qc/0607007} \BibitemShut
  {NoStop}%
\bibitem [{\citenamefont {Chua}\ \emph {et~al.}(2017)\citenamefont {Chua},
  \citenamefont {Moore},\ and\ \citenamefont {Gair}}]{Chua:2017ujo}%
  \BibitemOpen
  \bibfield  {author} {\bibinfo {author} {\bibfnamefont {A.~J.~K.}\
  \bibnamefont {Chua}}, \bibinfo {author} {\bibfnamefont {C.~J.}\ \bibnamefont
  {Moore}},\ and\ \bibinfo {author} {\bibfnamefont {J.~R.}\ \bibnamefont
  {Gair}},\ }\bibfield  {title} {\bibinfo {title} {{Augmented kludge waveforms
  for detecting extreme-mass-ratio inspirals}},\ }\href
  {https://doi.org/10.1103/PhysRevD.96.044005} {\bibfield  {journal} {\bibinfo
  {journal} {Phys. Rev. D}\ }\textbf {\bibinfo {volume} {96}},\ \bibinfo
  {pages} {044005} (\bibinfo {year} {2017})},\ \Eprint
  {https://arxiv.org/abs/1705.04259} {arXiv:1705.04259 [gr-qc]} \BibitemShut
  {NoStop}%
\bibitem [{\citenamefont {Xu}\ \emph {et~al.}(2018)\citenamefont {Xu},
  \citenamefont {Hou}, \citenamefont {Gong},\ and\ \citenamefont
  {Wang}}]{Xu:2018wow}%
  \BibitemOpen
  \bibfield  {author} {\bibinfo {author} {\bibfnamefont {Z.}~\bibnamefont
  {Xu}}, \bibinfo {author} {\bibfnamefont {X.}~\bibnamefont {Hou}}, \bibinfo
  {author} {\bibfnamefont {X.}~\bibnamefont {Gong}},\ and\ \bibinfo {author}
  {\bibfnamefont {J.}~\bibnamefont {Wang}},\ }\bibfield  {title} {\bibinfo
  {title} {{Black Hole Space-time In Dark Matter Halo}},\ }\href
  {https://doi.org/10.1088/1475-7516/2018/09/038} {\bibfield  {journal}
  {\bibinfo  {journal} {JCAP}\ }\textbf {\bibinfo {volume} {09}},\ \bibinfo
  {pages} {038}},\ \Eprint {https://arxiv.org/abs/1803.00767} {arXiv:1803.00767
  [gr-qc]} \BibitemShut {NoStop}%
\bibitem [{\citenamefont {Pantig}\ and\ \citenamefont
  {{\"O}vg{\"u}n}(2022)}]{Pantig:2022whj}%
  \BibitemOpen
  \bibfield  {author} {\bibinfo {author} {\bibfnamefont {R.~C.}\ \bibnamefont
  {Pantig}}\ and\ \bibinfo {author} {\bibfnamefont {A.}~\bibnamefont
  {{\"O}vg{\"u}n}},\ }\bibfield  {title} {\bibinfo {title} {{Dehnen halo effect
  on a black hole in an ultra-faint dwarf galaxy}},\ }\href
  {https://doi.org/10.1088/1475-7516/2022/08/056} {\bibfield  {journal}
  {\bibinfo  {journal} {JCAP}\ }\textbf {\bibinfo {volume} {08}}\bibfield
  {number} {\bibinfo  {number} { (08)},\ \bibinfo {pages} {056}},\ }\Eprint
  {https://arxiv.org/abs/2202.07404} {arXiv:2202.07404 [astro-ph.GA]}
  \BibitemShut {NoStop}%
\bibitem [{\citenamefont {Xu}\ \emph {et~al.}(2020)\citenamefont {Xu},
  \citenamefont {Gong},\ and\ \citenamefont {Zhang}}]{Xu:2020jpv}%
  \BibitemOpen
  \bibfield  {author} {\bibinfo {author} {\bibfnamefont {Z.}~\bibnamefont
  {Xu}}, \bibinfo {author} {\bibfnamefont {X.}~\bibnamefont {Gong}},\ and\
  \bibinfo {author} {\bibfnamefont {S.-N.}\ \bibnamefont {Zhang}},\ }\bibfield
  {title} {\bibinfo {title} {{Black hole immersed dark matter halo}},\ }\href
  {https://doi.org/10.1103/PhysRevD.101.024029} {\bibfield  {journal} {\bibinfo
   {journal} {Phys. Rev. D}\ }\textbf {\bibinfo {volume} {101}},\ \bibinfo
  {pages} {024029} (\bibinfo {year} {2020})}\BibitemShut {NoStop}%
\bibitem [{\citenamefont {Matos}\ \emph {et~al.}(2000)\citenamefont {Matos},
  \citenamefont {Guzman},\ and\ \citenamefont {Nunez}}]{Matos:2000ki}%
  \BibitemOpen
  \bibfield  {author} {\bibinfo {author} {\bibfnamefont {T.}~\bibnamefont
  {Matos}}, \bibinfo {author} {\bibfnamefont {F.~S.}\ \bibnamefont {Guzman}},\
  and\ \bibinfo {author} {\bibfnamefont {D.}~\bibnamefont {Nunez}},\ }\bibfield
   {title} {\bibinfo {title} {{Spherical scalar field halo in galaxies}},\
  }\href {https://doi.org/10.1103/PhysRevD.62.061301} {\bibfield  {journal}
  {\bibinfo  {journal} {Phys. Rev. D}\ }\textbf {\bibinfo {volume} {62}},\
  \bibinfo {pages} {061301} (\bibinfo {year} {2000})},\ \Eprint
  {https://arxiv.org/abs/astro-ph/0003398} {arXiv:astro-ph/0003398}
  \BibitemShut {NoStop}%
\bibitem [{\citenamefont {Qiao}\ and\ \citenamefont {Su}(2024)}]{Qiao:2024ehj}%
  \BibitemOpen
  \bibfield  {author} {\bibinfo {author} {\bibfnamefont {C.-K.}\ \bibnamefont
  {Qiao}}\ and\ \bibinfo {author} {\bibfnamefont {P.}~\bibnamefont {Su}},\
  }\bibfield  {title} {\bibinfo {title} {{Time delay of light in the
  gravitational lensing of supermassive black holes in dark matter halos}},\
  }\href {https://doi.org/10.1140/epjc/s10052-024-13403-3} {\bibfield
  {journal} {\bibinfo  {journal} {Eur. Phys. J. C}\ }\textbf {\bibinfo {volume}
  {84}},\ \bibinfo {pages} {1032} (\bibinfo {year} {2024})},\ \Eprint
  {https://arxiv.org/abs/2403.05682} {arXiv:2403.05682 [gr-qc]} \BibitemShut
  {NoStop}%
\bibitem [{\citenamefont {Liu}\ \emph {et~al.}(2024{\natexlab{c}})\citenamefont
  {Liu}, \citenamefont {Qiao},\ and\ \citenamefont {Tao}}]{Liu:2023xtb}%
  \BibitemOpen
  \bibfield  {author} {\bibinfo {author} {\bibfnamefont {Y.-G.}\ \bibnamefont
  {Liu}}, \bibinfo {author} {\bibfnamefont {C.-K.}\ \bibnamefont {Qiao}},\ and\
  \bibinfo {author} {\bibfnamefont {J.}~\bibnamefont {Tao}},\ }\bibfield
  {title} {\bibinfo {title} {{Gravitational lensing of spherically symmetric
  black holes in dark matter halos}},\ }\href
  {https://doi.org/10.1088/1475-7516/2024/10/075} {\bibfield  {journal}
  {\bibinfo  {journal} {JCAP}\ }\textbf {\bibinfo {volume} {10}},\ \bibinfo
  {pages} {075}},\ \Eprint {https://arxiv.org/abs/2312.15760} {arXiv:2312.15760
  [gr-qc]} \BibitemShut {NoStop}%
\bibitem [{\citenamefont {Fukushige}\ \emph {et~al.}(2004)\citenamefont
  {Fukushige}, \citenamefont {Kawai},\ and\ \citenamefont
  {Makino}}]{Fukushige:2003xc}%
  \BibitemOpen
  \bibfield  {author} {\bibinfo {author} {\bibfnamefont {T.}~\bibnamefont
  {Fukushige}}, \bibinfo {author} {\bibfnamefont {A.}~\bibnamefont {Kawai}},\
  and\ \bibinfo {author} {\bibfnamefont {J.}~\bibnamefont {Makino}},\
  }\bibfield  {title} {\bibinfo {title} {{Structure of dark matter halos from
  hierarchical clustering. 3. Shallowing of the Inner cusp}},\ }\href
  {https://doi.org/10.1086/383192} {\bibfield  {journal} {\bibinfo  {journal}
  {Astrophys. J.}\ }\textbf {\bibinfo {volume} {606}},\ \bibinfo {pages} {625}
  (\bibinfo {year} {2004})},\ \Eprint {https://arxiv.org/abs/astro-ph/0306203}
  {arXiv:astro-ph/0306203} \BibitemShut {NoStop}%
\bibitem [{\citenamefont {Ishihara}\ \emph {et~al.}(2016)\citenamefont
  {Ishihara}, \citenamefont {Suzuki}, \citenamefont {Ono}, \citenamefont
  {Kitamura},\ and\ \citenamefont {Asada}}]{Ishihara:2016vdc}%
  \BibitemOpen
  \bibfield  {author} {\bibinfo {author} {\bibfnamefont {A.}~\bibnamefont
  {Ishihara}}, \bibinfo {author} {\bibfnamefont {Y.}~\bibnamefont {Suzuki}},
  \bibinfo {author} {\bibfnamefont {T.}~\bibnamefont {Ono}}, \bibinfo {author}
  {\bibfnamefont {T.}~\bibnamefont {Kitamura}},\ and\ \bibinfo {author}
  {\bibfnamefont {H.}~\bibnamefont {Asada}},\ }\bibfield  {title} {\bibinfo
  {title} {{Gravitational bending angle of light for finite distance and the
  Gauss-Bonnet theorem}},\ }\href {https://doi.org/10.1103/PhysRevD.94.084015}
  {\bibfield  {journal} {\bibinfo  {journal} {Phys. Rev. D}\ }\textbf {\bibinfo
  {volume} {94}},\ \bibinfo {pages} {084015} (\bibinfo {year} {2016})},\
  \Eprint {https://arxiv.org/abs/1604.08308} {arXiv:1604.08308 [gr-qc]}
  \BibitemShut {NoStop}%
\bibitem [{\citenamefont {Takizawa}\ \emph {et~al.}(2020)\citenamefont
  {Takizawa}, \citenamefont {Ono},\ and\ \citenamefont
  {Asada}}]{Takizawa:2020egm}%
  \BibitemOpen
  \bibfield  {author} {\bibinfo {author} {\bibfnamefont {K.}~\bibnamefont
  {Takizawa}}, \bibinfo {author} {\bibfnamefont {T.}~\bibnamefont {Ono}},\ and\
  \bibinfo {author} {\bibfnamefont {H.}~\bibnamefont {Asada}},\ }\bibfield
  {title} {\bibinfo {title} {{Gravitational deflection angle of light:
  Definition by an observer and its application to an asymptotically nonflat
  spacetime}},\ }\href {https://doi.org/10.1103/PhysRevD.101.104032} {\bibfield
   {journal} {\bibinfo  {journal} {Phys. Rev. D}\ }\textbf {\bibinfo {volume}
  {101}},\ \bibinfo {pages} {104032} (\bibinfo {year} {2020})},\ \Eprint
  {https://arxiv.org/abs/2001.03290} {arXiv:2001.03290 [gr-qc]} \BibitemShut
  {NoStop}%
\bibitem [{\citenamefont {Li}\ \emph {et~al.}(2020)\citenamefont {Li},
  \citenamefont {Zhang},\ and\ \citenamefont {{\"O}vg{\"u}n}}]{Li:2020wvn}%
  \BibitemOpen
  \bibfield  {author} {\bibinfo {author} {\bibfnamefont {Z.}~\bibnamefont
  {Li}}, \bibinfo {author} {\bibfnamefont {G.}~\bibnamefont {Zhang}},\ and\
  \bibinfo {author} {\bibfnamefont {A.}~\bibnamefont {{\"O}vg{\"u}n}},\
  }\bibfield  {title} {\bibinfo {title} {{Circular Orbit of a Particle and Weak
  Gravitational Lensing}},\ }\href
  {https://doi.org/10.1103/PhysRevD.101.124058} {\bibfield  {journal} {\bibinfo
   {journal} {Phys. Rev. D}\ }\textbf {\bibinfo {volume} {101}},\ \bibinfo
  {pages} {124058} (\bibinfo {year} {2020})},\ \Eprint
  {https://arxiv.org/abs/2006.13047} {arXiv:2006.13047 [gr-qc]} \BibitemShut
  {NoStop}%
\bibitem [{\citenamefont {{\"O}vg{\"u}n}\ \emph {et~al.}(2018)\citenamefont
  {{\"O}vg{\"u}n}, \citenamefont {Sakall{\i}},\ and\ \citenamefont
  {Saavedra}}]{Ovgun:2018tua}%
  \BibitemOpen
  \bibfield  {author} {\bibinfo {author} {\bibfnamefont {A.}~\bibnamefont
  {{\"O}vg{\"u}n}}, \bibinfo {author} {\bibfnamefont {{\.I}.}~\bibnamefont
  {Sakall{\i}}},\ and\ \bibinfo {author} {\bibfnamefont {J.}~\bibnamefont
  {Saavedra}},\ }\bibfield  {title} {\bibinfo {title} {{Shadow cast and
  Deflection angle of Kerr-Newman-Kasuya spacetime}},\ }\href
  {https://doi.org/10.1088/1475-7516/2018/10/041} {\bibfield  {journal}
  {\bibinfo  {journal} {JCAP}\ }\textbf {\bibinfo {volume} {10}},\ \bibinfo
  {pages} {041}},\ \Eprint {https://arxiv.org/abs/1807.00388} {arXiv:1807.00388
  [gr-qc]} \BibitemShut {NoStop}%
\bibitem [{\citenamefont {Jusufi}\ \emph {et~al.}(2018)\citenamefont {Jusufi},
  \citenamefont {{\"O}vg{\"u}n}, \citenamefont {Saavedra}, \citenamefont
  {V{\'a}squez},\ and\ \citenamefont {Gonz{\'a}lez}}]{Jusufi:2018jof}%
  \BibitemOpen
  \bibfield  {author} {\bibinfo {author} {\bibfnamefont {K.}~\bibnamefont
  {Jusufi}}, \bibinfo {author} {\bibfnamefont {A.}~\bibnamefont
  {{\"O}vg{\"u}n}}, \bibinfo {author} {\bibfnamefont {J.}~\bibnamefont
  {Saavedra}}, \bibinfo {author} {\bibfnamefont {Y.}~\bibnamefont
  {V{\'a}squez}},\ and\ \bibinfo {author} {\bibfnamefont {P.~A.}\ \bibnamefont
  {Gonz{\'a}lez}},\ }\bibfield  {title} {\bibinfo {title} {{Deflection of light
  by rotating regular black holes using the Gauss-Bonnet theorem}},\ }\href
  {https://doi.org/10.1103/PhysRevD.97.124024} {\bibfield  {journal} {\bibinfo
  {journal} {Phys. Rev. D}\ }\textbf {\bibinfo {volume} {97}},\ \bibinfo
  {pages} {124024} (\bibinfo {year} {2018})},\ \Eprint
  {https://arxiv.org/abs/1804.00643} {arXiv:1804.00643 [gr-qc]} \BibitemShut
  {NoStop}%
\bibitem [{\citenamefont {Eda}\ \emph {et~al.}(2015)\citenamefont {Eda},
  \citenamefont {Itoh}, \citenamefont {Kuroyanagi},\ and\ \citenamefont
  {Silk}}]{Eda:2014kra}%
  \BibitemOpen
  \bibfield  {author} {\bibinfo {author} {\bibfnamefont {K.}~\bibnamefont
  {Eda}}, \bibinfo {author} {\bibfnamefont {Y.}~\bibnamefont {Itoh}}, \bibinfo
  {author} {\bibfnamefont {S.}~\bibnamefont {Kuroyanagi}},\ and\ \bibinfo
  {author} {\bibfnamefont {J.}~\bibnamefont {Silk}},\ }\bibfield  {title}
  {\bibinfo {title} {{Gravitational waves as a probe of dark matter
  minispikes}},\ }\href {https://doi.org/10.1103/PhysRevD.91.044045} {\bibfield
   {journal} {\bibinfo  {journal} {Phys. Rev. D}\ }\textbf {\bibinfo {volume}
  {91}},\ \bibinfo {pages} {044045} (\bibinfo {year} {2015})},\ \Eprint
  {https://arxiv.org/abs/1408.3534} {arXiv:1408.3534 [gr-qc]} \BibitemShut
  {NoStop}%
\bibitem [{\citenamefont {Hughes}(2019)}]{Hughes:2018qxz}%
  \BibitemOpen
  \bibfield  {author} {\bibinfo {author} {\bibfnamefont {S.~A.}\ \bibnamefont
  {Hughes}},\ }\bibfield  {title} {\bibinfo {title} {{Bound orbits of a slowly
  evolving black hole}},\ }\href {https://doi.org/10.1103/PhysRevD.100.064001}
  {\bibfield  {journal} {\bibinfo  {journal} {Phys. Rev. D}\ }\textbf {\bibinfo
  {volume} {100}},\ \bibinfo {pages} {064001} (\bibinfo {year} {2019})},\
  \Eprint {https://arxiv.org/abs/1806.09022} {arXiv:1806.09022 [gr-qc]}
  \BibitemShut {NoStop}%
\bibitem [{\citenamefont {Blachier}\ \emph {et~al.}(2024)\citenamefont
  {Blachier}, \citenamefont {Barrau}, \citenamefont {Martineau},\ and\
  \citenamefont {Renevey}}]{Blachier:2023ygh}%
  \BibitemOpen
  \bibfield  {author} {\bibinfo {author} {\bibfnamefont {B.}~\bibnamefont
  {Blachier}}, \bibinfo {author} {\bibfnamefont {A.}~\bibnamefont {Barrau}},
  \bibinfo {author} {\bibfnamefont {K.}~\bibnamefont {Martineau}},\ and\
  \bibinfo {author} {\bibfnamefont {C.}~\bibnamefont {Renevey}},\ }\bibfield
  {title} {\bibinfo {title} {{Competitive effects between gravitational
  radiation and mass variation for two-body systems in circular orbits}},\
  }\href {https://doi.org/10.1007/s10714-024-03201-3} {\bibfield  {journal}
  {\bibinfo  {journal} {Gen. Rel. Grav.}\ }\textbf {\bibinfo {volume} {56}},\
  \bibinfo {pages} {20} (\bibinfo {year} {2024})},\ \Eprint
  {https://arxiv.org/abs/2306.09069} {arXiv:2306.09069 [gr-qc]} \BibitemShut
  {NoStop}%
\bibitem [{\citenamefont {Peters}\ and\ \citenamefont
  {Mathews}(1963)}]{Peters:1963ux}%
  \BibitemOpen
  \bibfield  {author} {\bibinfo {author} {\bibfnamefont {P.~C.}\ \bibnamefont
  {Peters}}\ and\ \bibinfo {author} {\bibfnamefont {J.}~\bibnamefont
  {Mathews}},\ }\bibfield  {title} {\bibinfo {title} {{Gravitational radiation
  from point masses in a Keplerian orbit}},\ }\href
  {https://doi.org/10.1103/PhysRev.131.435} {\bibfield  {journal} {\bibinfo
  {journal} {Phys. Rev.}\ }\textbf {\bibinfo {volume} {131}},\ \bibinfo {pages}
  {435} (\bibinfo {year} {1963})}\BibitemShut {NoStop}%
\bibitem [{\citenamefont {Ashoorioon}\ \emph {et~al.}(2025)\citenamefont
  {Ashoorioon}, \citenamefont {Casadio}, \citenamefont {Jafarzade},
  \citenamefont {Jahani~Poshteh},\ and\ \citenamefont
  {Luongo}}]{Ashoorioon:2025ezk}%
  \BibitemOpen
  \bibfield  {author} {\bibinfo {author} {\bibfnamefont {A.}~\bibnamefont
  {Ashoorioon}}, \bibinfo {author} {\bibfnamefont {R.}~\bibnamefont {Casadio}},
  \bibinfo {author} {\bibfnamefont {K.}~\bibnamefont {Jafarzade}}, \bibinfo
  {author} {\bibfnamefont {M.~B.}\ \bibnamefont {Jahani~Poshteh}},\ and\
  \bibinfo {author} {\bibfnamefont {O.}~\bibnamefont {Luongo}},\ }\bibfield
  {title} {\bibinfo {title} {{Gravitational radiation reaction around a static
  black hole surrounded by a Dehnen type dark matter halo}},\ }\href@noop {} {\
   (\bibinfo {year} {2025})},\ \Eprint {https://arxiv.org/abs/2509.08569}
  {arXiv:2509.08569 [gr-qc]} \BibitemShut {NoStop}%
\bibitem [{\citenamefont {Peters}(1964)}]{Peters:1964zz}%
  \BibitemOpen
  \bibfield  {author} {\bibinfo {author} {\bibfnamefont {P.~C.}\ \bibnamefont
  {Peters}},\ }\bibfield  {title} {\bibinfo {title} {{Gravitational Radiation
  and the Motion of Two Point Masses}},\ }\href
  {https://doi.org/10.1103/PhysRev.136.B1224} {\bibfield  {journal} {\bibinfo
  {journal} {Phys. Rev.}\ }\textbf {\bibinfo {volume} {136}},\ \bibinfo {pages}
  {B1224} (\bibinfo {year} {1964})}\BibitemShut {NoStop}%
\bibitem [{\citenamefont {Cutler}\ \emph {et~al.}(1994)\citenamefont {Cutler},
  \citenamefont {Kennefick},\ and\ \citenamefont {Poisson}}]{Cutler:1994pb}%
  \BibitemOpen
  \bibfield  {author} {\bibinfo {author} {\bibfnamefont {C.}~\bibnamefont
  {Cutler}}, \bibinfo {author} {\bibfnamefont {D.}~\bibnamefont {Kennefick}},\
  and\ \bibinfo {author} {\bibfnamefont {E.}~\bibnamefont {Poisson}},\
  }\bibfield  {title} {\bibinfo {title} {{Gravitational radiation reaction for
  bound motion around a Schwarzschild black hole}},\ }\href
  {https://doi.org/10.1103/PhysRevD.50.3816} {\bibfield  {journal} {\bibinfo
  {journal} {Phys. Rev. D}\ }\textbf {\bibinfo {volume} {50}},\ \bibinfo
  {pages} {3816} (\bibinfo {year} {1994})}\BibitemShut {NoStop}%
\bibitem [{\citenamefont {Tu}\ \emph {et~al.}(2023)\citenamefont {Tu},
  \citenamefont {Zhu},\ and\ \citenamefont {Wang}}]{Tu:2023xab}%
  \BibitemOpen
  \bibfield  {author} {\bibinfo {author} {\bibfnamefont {Z.-Y.}\ \bibnamefont
  {Tu}}, \bibinfo {author} {\bibfnamefont {T.}~\bibnamefont {Zhu}},\ and\
  \bibinfo {author} {\bibfnamefont {A.}~\bibnamefont {Wang}},\ }\bibfield
  {title} {\bibinfo {title} {{Periodic orbits and their gravitational wave
  radiations in a polymer black hole in loop quantum gravity}},\ }\href
  {https://doi.org/10.1103/PhysRevD.108.024035} {\bibfield  {journal} {\bibinfo
   {journal} {Phys. Rev. D}\ }\textbf {\bibinfo {volume} {108}},\ \bibinfo
  {pages} {024035} (\bibinfo {year} {2023})},\ \Eprint
  {https://arxiv.org/abs/2304.14160} {arXiv:2304.14160 [gr-qc]} \BibitemShut
  {NoStop}%
\bibitem [{\citenamefont {Will}(2016)}]{Will:2016sgx}%
  \BibitemOpen
  \bibfield  {author} {\bibinfo {author} {\bibfnamefont {C.~M.}\ \bibnamefont
  {Will}},\ }\bibinfo {title} {{Gravity: Newtonian, Post-Newtonian, and General
  Relativistic}},\ in\ \href {https://doi.org/10.1007/978-3-319-20224-2_2}
  {\emph {\bibinfo {booktitle} {{Gravity: Where Do We Stand?}}}},\ \bibinfo
  {editor} {edited by\ \bibinfo {editor} {\bibfnamefont {R.}~\bibnamefont
  {Peron}}, \bibinfo {editor} {\bibfnamefont {M.}~\bibnamefont {Colpi}},
  \bibinfo {editor} {\bibfnamefont {V.}~\bibnamefont {Gorini}},\ and\ \bibinfo
  {editor} {\bibfnamefont {U.}~\bibnamefont {Moschella}}}\ (\bibinfo {year}
  {2016})\ pp.\ \bibinfo {pages} {9--72}\BibitemShut {NoStop}%
\bibitem [{\citenamefont {Maselli}\ \emph {et~al.}(2022)\citenamefont
  {Maselli}, \citenamefont {Franchini}, \citenamefont {Gualtieri},
  \citenamefont {Sotiriou}, \citenamefont {Barsanti},\ and\ \citenamefont
  {Pani}}]{Maselli:2021men}%
  \BibitemOpen
  \bibfield  {author} {\bibinfo {author} {\bibfnamefont {A.}~\bibnamefont
  {Maselli}}, \bibinfo {author} {\bibfnamefont {N.}~\bibnamefont {Franchini}},
  \bibinfo {author} {\bibfnamefont {L.}~\bibnamefont {Gualtieri}}, \bibinfo
  {author} {\bibfnamefont {T.~P.}\ \bibnamefont {Sotiriou}}, \bibinfo {author}
  {\bibfnamefont {S.}~\bibnamefont {Barsanti}},\ and\ \bibinfo {author}
  {\bibfnamefont {P.}~\bibnamefont {Pani}},\ }\bibfield  {title} {\bibinfo
  {title} {{Detecting fundamental fields with LISA observations of
  gravitational waves from extreme mass-ratio inspirals}},\ }\href
  {https://doi.org/10.1038/s41550-021-01589-5} {\bibfield  {journal} {\bibinfo
  {journal} {Nature Astron.}\ }\textbf {\bibinfo {volume} {6}},\ \bibinfo
  {pages} {464} (\bibinfo {year} {2022})},\ \Eprint
  {https://arxiv.org/abs/2106.11325} {arXiv:2106.11325 [gr-qc]} \BibitemShut
  {NoStop}%
\bibitem [{\citenamefont {Liang}\ \emph {et~al.}(2023)\citenamefont {Liang},
  \citenamefont {Xu}, \citenamefont {Mai},\ and\ \citenamefont
  {Shao}}]{Liang:2022gdk}%
  \BibitemOpen
  \bibfield  {author} {\bibinfo {author} {\bibfnamefont {D.}~\bibnamefont
  {Liang}}, \bibinfo {author} {\bibfnamefont {R.}~\bibnamefont {Xu}}, \bibinfo
  {author} {\bibfnamefont {Z.-F.}\ \bibnamefont {Mai}},\ and\ \bibinfo {author}
  {\bibfnamefont {L.}~\bibnamefont {Shao}},\ }\bibfield  {title} {\bibinfo
  {title} {{Probing vector hair of black holes with extreme-mass-ratio
  inspirals}},\ }\href {https://doi.org/10.1103/PhysRevD.107.044053} {\bibfield
   {journal} {\bibinfo  {journal} {Phys. Rev. D}\ }\textbf {\bibinfo {volume}
  {107}},\ \bibinfo {pages} {044053} (\bibinfo {year} {2023})},\ \Eprint
  {https://arxiv.org/abs/2212.09346} {arXiv:2212.09346 [gr-qc]} \BibitemShut
  {NoStop}%
\bibitem [{\citenamefont {Berti}\ \emph {et~al.}(2005)\citenamefont {Berti},
  \citenamefont {Buonanno},\ and\ \citenamefont {Will}}]{Berti:2004bd}%
  \BibitemOpen
  \bibfield  {author} {\bibinfo {author} {\bibfnamefont {E.}~\bibnamefont
  {Berti}}, \bibinfo {author} {\bibfnamefont {A.}~\bibnamefont {Buonanno}},\
  and\ \bibinfo {author} {\bibfnamefont {C.~M.}\ \bibnamefont {Will}},\
  }\bibfield  {title} {\bibinfo {title} {{Estimating spinning binary parameters
  and testing alternative theories of gravity with LISA}},\ }\href
  {https://doi.org/10.1103/PhysRevD.71.084025} {\bibfield  {journal} {\bibinfo
  {journal} {Phys. Rev. D}\ }\textbf {\bibinfo {volume} {71}},\ \bibinfo
  {pages} {084025} (\bibinfo {year} {2005})},\ \Eprint
  {https://arxiv.org/abs/gr-qc/0411129} {arXiv:gr-qc/0411129} \BibitemShut
  {NoStop}%
\bibitem [{\citenamefont {Maselli}\ \emph {et~al.}(2020)\citenamefont
  {Maselli}, \citenamefont {Franchini}, \citenamefont {Gualtieri},\ and\
  \citenamefont {Sotiriou}}]{Maselli:2020zgv}%
  \BibitemOpen
  \bibfield  {author} {\bibinfo {author} {\bibfnamefont {A.}~\bibnamefont
  {Maselli}}, \bibinfo {author} {\bibfnamefont {N.}~\bibnamefont {Franchini}},
  \bibinfo {author} {\bibfnamefont {L.}~\bibnamefont {Gualtieri}},\ and\
  \bibinfo {author} {\bibfnamefont {T.~P.}\ \bibnamefont {Sotiriou}},\
  }\bibfield  {title} {\bibinfo {title} {{Detecting scalar fields with Extreme
  Mass Ratio Inspirals}},\ }\href
  {https://doi.org/10.1103/PhysRevLett.125.141101} {\bibfield  {journal}
  {\bibinfo  {journal} {Phys. Rev. Lett.}\ }\textbf {\bibinfo {volume} {125}},\
  \bibinfo {pages} {141101} (\bibinfo {year} {2020})},\ \Eprint
  {https://arxiv.org/abs/2004.11895} {arXiv:2004.11895 [gr-qc]} \BibitemShut
  {NoStop}%
\end{thebibliography}%

	
\end{document}